\documentclass[aoas]{imsart}
\pdfoutput=1 
\RequirePackage[OT1]{fontenc}
\RequirePackage{amsthm,amsmath}
\RequirePackage{natbib}
\RequirePackage[colorlinks,citecolor=blue,urlcolor=blue]{hyperref}
\usepackage{graphicx}
\usepackage{graphics}
\usepackage{amssymb}
\usepackage{rotating}
\makeatletter
\def\journal@name{}
\makeatother
\pubyear{}
\volume{}
\issue{}
\firstpage{}  
\lastpage{}

\begin{document}

\begin{frontmatter}
\title{{The role of swabs in modeling the COVID-19 outbreak in the most affected regions of Italy}}

\begin{aug}
\author{\fnms{Claudia} \snm{Furlan}
\ead[label=e1]{furlan@stat.unipd.it}}
\and
\author{\fnms{Cinzia} \snm{Mortarino}
\ead[label=e2]{mortarino@stat.unipd.it}}

\address{Department of Statistical Sciences\\
University of Padova, Italy \\
\printead{e1,e2}}


\affiliation{University of Padova, Italy}

\end{aug}

\begin{abstract}
The daily fluctuations in the released number of Covid-19 cases played a big
role both at the beginning and in the most critical weeks of the outbreak,
when local authorities in Italy had to decide whether to  impose a lockdown
and at which level. Public opinion was focused on this information as well,
to understand how quickly the epidemic was spreading. When an increase/decrease
was communicated, especially a large one, it was not easy to understand if it was
due to a change in the epidemic evolution or if it was a fluctuation due to other
reasons, such as an increase in the number of swabs or a delay in the swab processing.

In this work, we propose a nonlinear asymmetric diffusion model, which includes
information on the daily number of swabs, to describe daily fluctuations in the number of
confirmed cases in addition to the main trend of the outbreak evolution.
The class of diffusion models has been selected to develop our proposal, as
it also allows estimation of the total number of confirmed cases at the end of
the outbreak.

The proposed model is compared to six existing models, among which the logistic and
the SIRD models are used as benchmarks, in the five most affected Italian regions.

\end{abstract}



\tableofcontents
\end{frontmatter}

\thispagestyle{empty}

\section{Introduction} \label{sec:intro}
Italy was the first nation to be affected by Covid-19 after China, and the epidemic
has mainly been located in Nothern Italy.
The first case was detected on February 21st, 2020, in
the municipality of Vo', a village near Padua in the Veneto Region of Northeast Italy.
On the same day, an infected patient was detected
in the small town of Codogno, which is located in the bordering Lombardy region.
From that time on, the epidemic followed a completely
different evolution in the two regions, and it quickly spread in northern Italy.

In Veneto, the local authorities imposed a lockdown on the whole municipality for two
weeks; both at the beginning and at the end of the two weeks, the population was tested
for the virus through nasopharyngeal swabs, and this approach
gave rise to the first epidemiological survey on  Covid-19 for understanding its
transmission dynamics \citep{lavezzo:20}.
Moreover, the hospital where the first
diagnosis occurred was closed, and  people who had previously accessed the facility were tested.

Among the Italian regions, Lombardy is the most affected by the epidemic, with a death toll three
times greater than that in China \citep{indolfi:20}. It is apparent  that, in Italy,
the regional autonomy regarding health policy has resulted in services with
different levels of quality \citep{indolfi:20}, such as the number of beds and the capacity of processing swabs.
With regard to the number of beds in Italy, the forecasts of hospitalisations was faced by \cite{gregori:20} for the Veneto region, while \cite{farcomeni:20} modeled the intensive care unit occupancy.

The capacity of processing swabs is of particular importance for detecting the
state of the epidemic, measuring the lockdown effects and, most importantly, reducing
the outbreak. In fact, only a swab outcome enacts the procedure to eventually
isolate the infected individual together with his/her close contacts. Consistent
delays in this
process will lead to failure in controlling the outbreak. Moreover, public attention
 is
focused every day on the released number of confirmed cases as a measure of
the state of the epidemic, especially at the beginning of the outbreak and during
the lockdown to evaluate its effect. When an increase/decrease is communicated,
especially a large one, it is not easy to understand if it is due to a change in
the epidemic evolution or if it is a fluctuation due to other reasons, such as
delays in the swab process or in the laboratory organization. The effect, however,
is relevant both to public opinion, spreading inaccurate optimistic/pessimistic views
of the situation, and to authorities who must decide whether to adopt restrictions and at which level.

Our opinion is that it is necessary to include the number of swabs to describe the
local
fluctuations in the epidemic evolution in addition to detecting the main trend. To the best of
our knowledge, no models are present in literature
with this characteristic. In fact, at the beginning of the outbreak,
 the curve of confirmed cases was usually modeled through an exponential
 \citep{remuzzi:20} or a logistic growth model \citep{batista:20, shen:20}. When the
 data collection window became long enough, the models were usually
 of two types: the compartmental and ARIMA models. The compartmental models
 represent the more blooming
 field of research and are based on modeling the infection, recovery and mortality rates by using the times
 series of the actually positive, recovered and dead cases \citep{anastassopoulou:20, batista:20sir, caccavo:20, fanelli:20, ivorra:20, iwata:20, lavezzo:20, liu:20, postnikov:20, wangping:20}.
 In the field of spatial statistics,  \cite{guliyev:20} contributed a spatial panel data model on
 confirmed, recovered and dead cases. Meanwhile, \cite{benvenuto:20} and
 \cite{chintalapudi:20} represent two contributions based on an ARIMA model.

In this paper, we made an effort to describe
the cumulative number of confirmed cases in the five most affected Italian regions,
based on the combination of a nonlinear model
and the number of completed swabs. In the class of growth models, we propose a new version of the dynamic potential model \citep{guseo:09},
where the novelty consists of the formulation of a new intervention function
with the
number of daily swabs as an explanatory variable. The model is particularly parsimonious
since the intervention function has only one {additional} parameter.
The base of the dynamic potential model was chosen since a) it has an
asymmetric shape and makes it possible  to model a `saddle', which is a rather common
nonlinear pattern; b) it gives an estimate of the total number of confirmed cases at the end of
the epidemic; and c) the total
number of confirmed cases is not fixed throughout the outbreak,
but it is allowed to change over time. Since the capability of processing swabs
increased over time and, consequently, the meeting criteria for people for being
tested were enlarged with the aim of detecting a larger number of asymptomatic
positive subjects, it is sensible to suppose that the number of  diagnosed cases
increases with time.

The proposed model was compared, in each region, with five alternative
growth models: the logistic model was used as a benchmark; the Generalized Bass model
\citep{bass:1994}, eventually including a parameter accounting for asymmetry
\citep{bemmaor:94}, with fixed market potential;
and the classic dynamic potential model \citep{guseo:09}, eventually including
a seasonal component \citep{guidolin:14}, Moreover, a SIRD model was used as a second
benchmark by summing the predictions of actually positive, recovered and dead
cases. Three-week forecasts of the spreading dynamics were provided for each model
as well. The models were compared in terms of R$^2$ and BIC values, for the
cumulative values. The squared linear correlation coefficient between observed and
fitted daily values was evaluated as well.

The rest of the paper proceeds as follows: In Section \ref{sec:data}, we provide a
description of the available data. In Section \ref{sec:models},  we describe the
proposed model and the 6 competing models. The results obtained in the 5 Italian
regions are illustrated in Section \ref{sec:applications}. Some concluding remarks
follow in Section \ref{sec:conclusions}.

\section{Data}\label{sec:data}
The data of the five Italian regions most affected by the epidemic, namely
Veneto,
Lombardy,
Piedmont,
Tuscany
and Emilia--Romagna,  were downloaded from the \cite{protezione:20}. The data
collection period was from the 24th of February to the 3rd of May 2020, which is the
last day before the so--called {\it Phase 2,} when the lockdown was partially removed.
For each day, the data consist of currently infected patients, both hospitalised or home isolated,
cumulated recovered people, cumulated deaths, the total number of confirmed cases, which is given by
the sum of the latter three components, and the cumulative number of swabs.

Official data before the 24th of February are not available, but the first infected
 people were detected on the 21st of February in both Veneto and Lombardy.
 Since the first days are important to correctly estimate the spread of an epidemic, for the latter
 regions we integrated the official dataset with information released by the newspapers and/or the
 official websites of the Regions about cases for the  21st to 23rd of February.

The reconstruction of the number of swabs in Lombardy for the first three days was facilitated by
the Region press release on the 21st of February \citep{reg_lom:21} and by the news article
of \cite{eco:23} on the 23rd; the value of the 22nd was imputed through their mean. For the
Veneto region, the imputation was subjectively fixed at  200, 700 and 1500 for days 21--23,
respectively. Moreover, the swab time series were cleaned since the cumulative values were
not nondecreasing. This happened on the 25th of February for Lombardy, and the imputation
was made by the average of the former and latter values. For Emilia-Romagna, there was an
analogous problem for the 28th to 30th of March, and the imputation was performed based on the
information released in \cite{reg_em:28, modena:30}.

\begin{figure}[t]
\centering
\includegraphics[width=12cm]{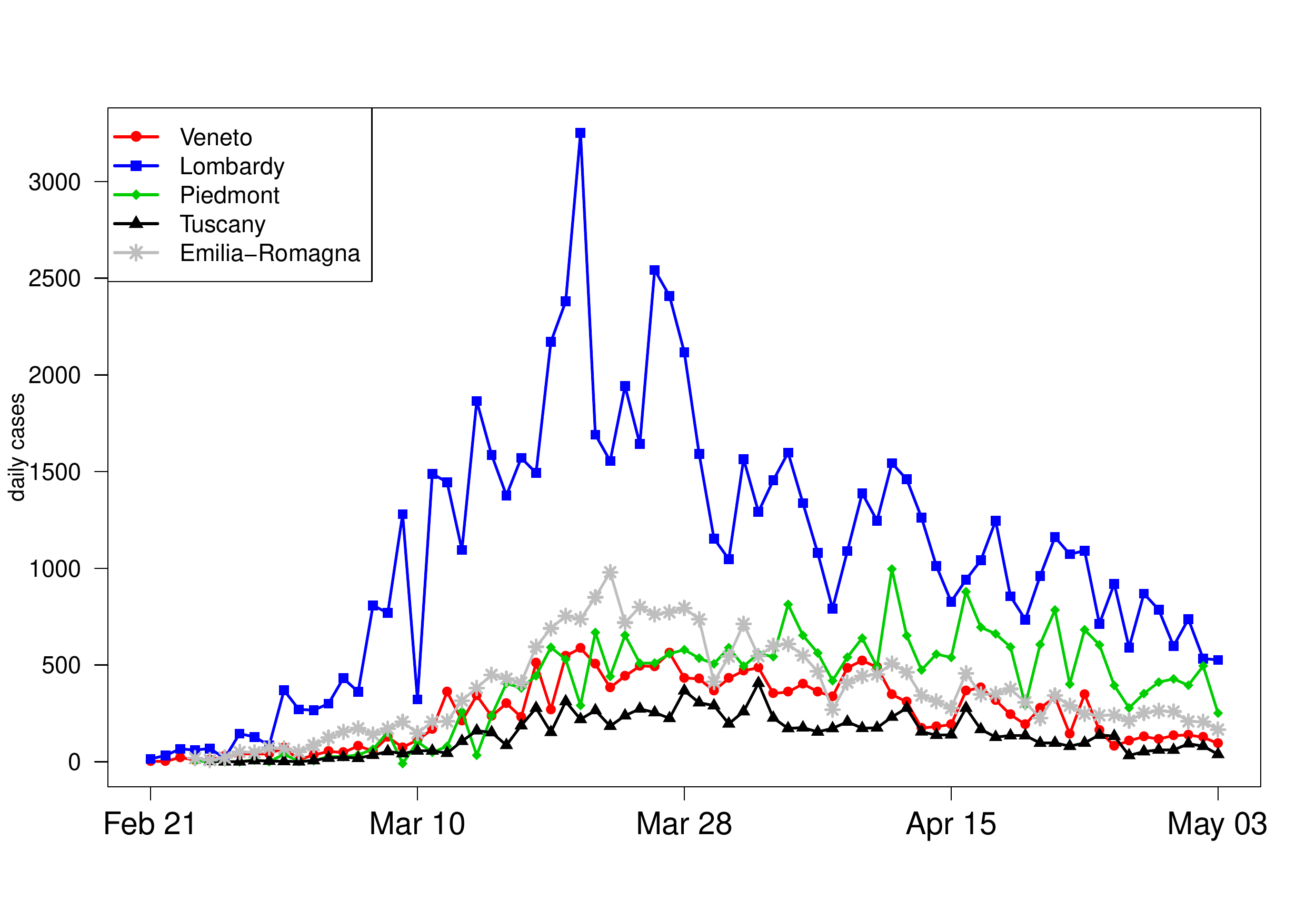} \vspace*{-4mm}
\caption{Daily number of confirmed Covid--19 cases in the five Italian regions}
\label{fig:data}
\end{figure}

\begin{figure}[t]
\centering
\begin{tabular}{cc}
\hspace{-10mm} (a) Veneto & \hspace{-8mm} (b) Lombardy \\
\hspace{-10mm}
\includegraphics[width=73mm]{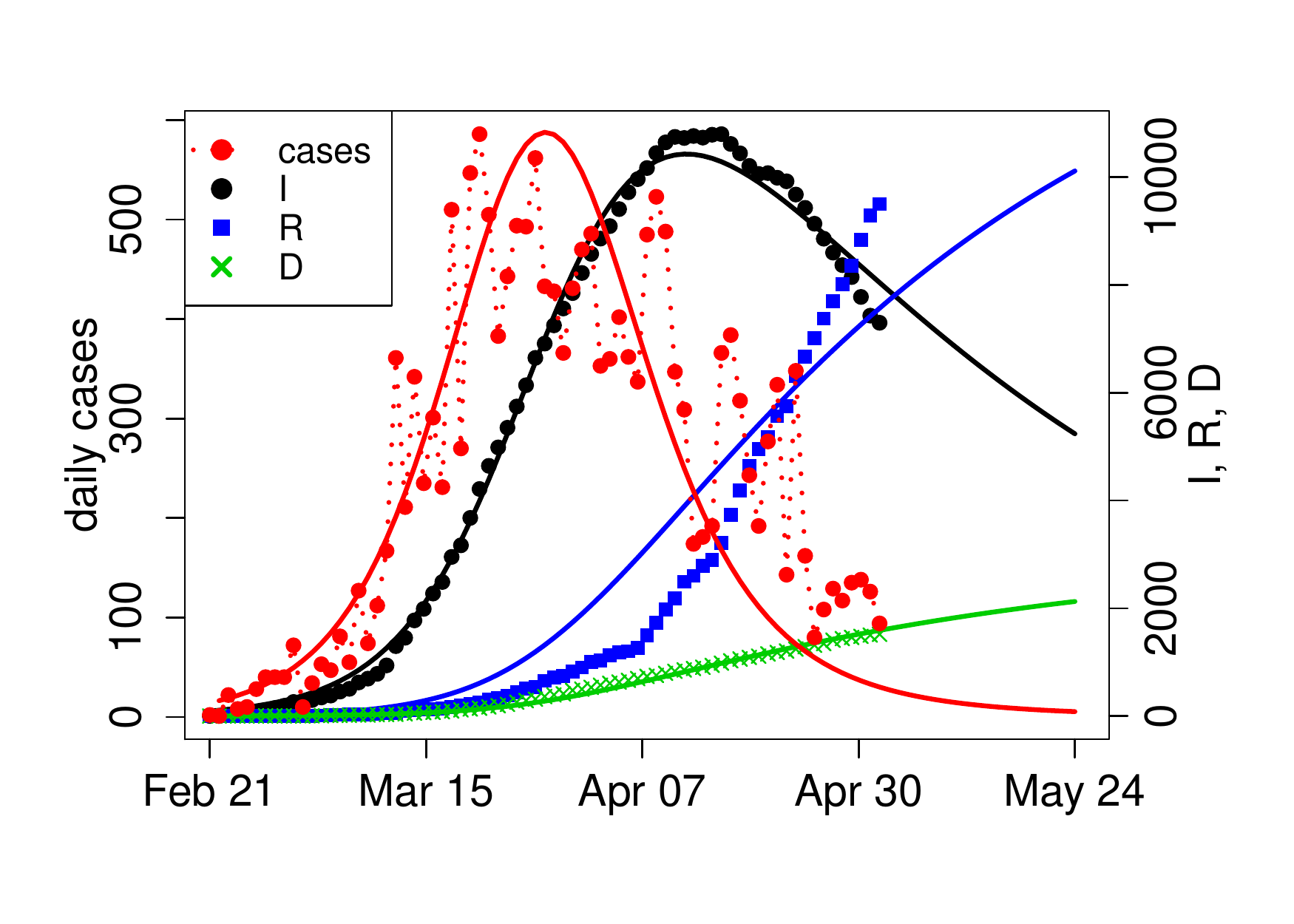} &
\hspace{-8mm}
\includegraphics[width=73mm]{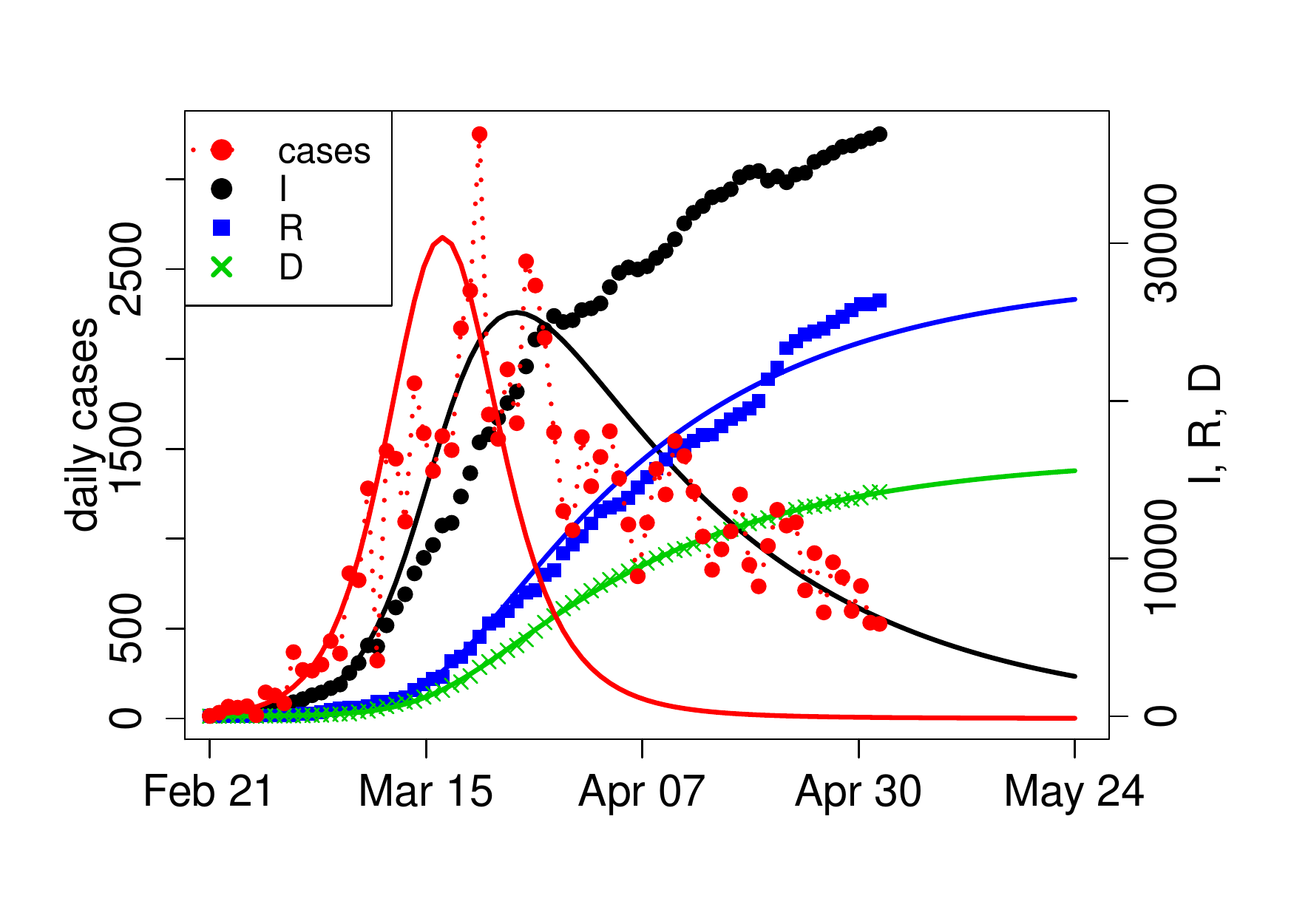}\\
\hspace{-10mm} (c) Piedmont & \hspace{-8mm} (d) Tuscany\\
\hspace{-10mm}
\includegraphics[width=73mm]{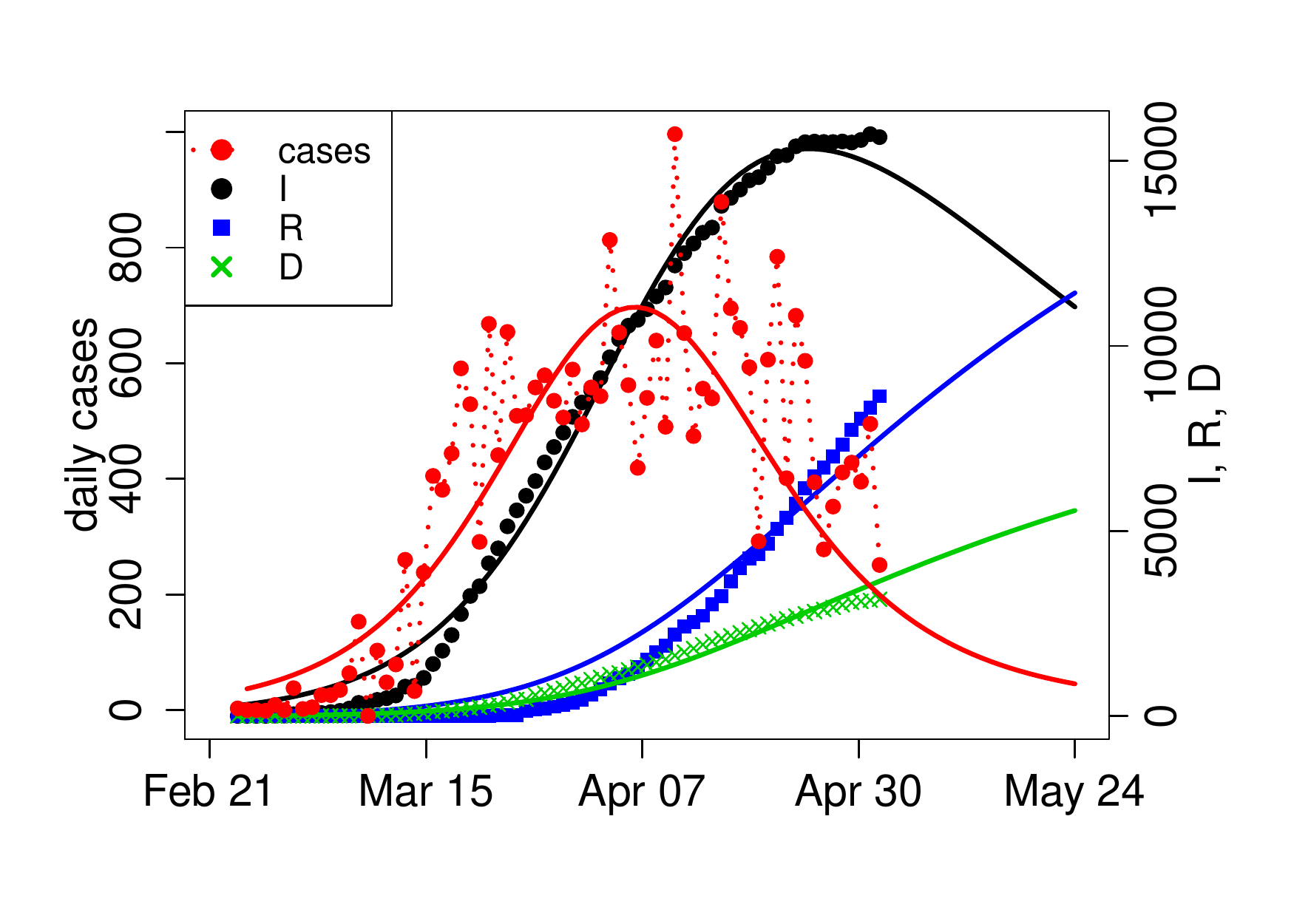} &
\hspace{-8mm}
\includegraphics[width=73mm]{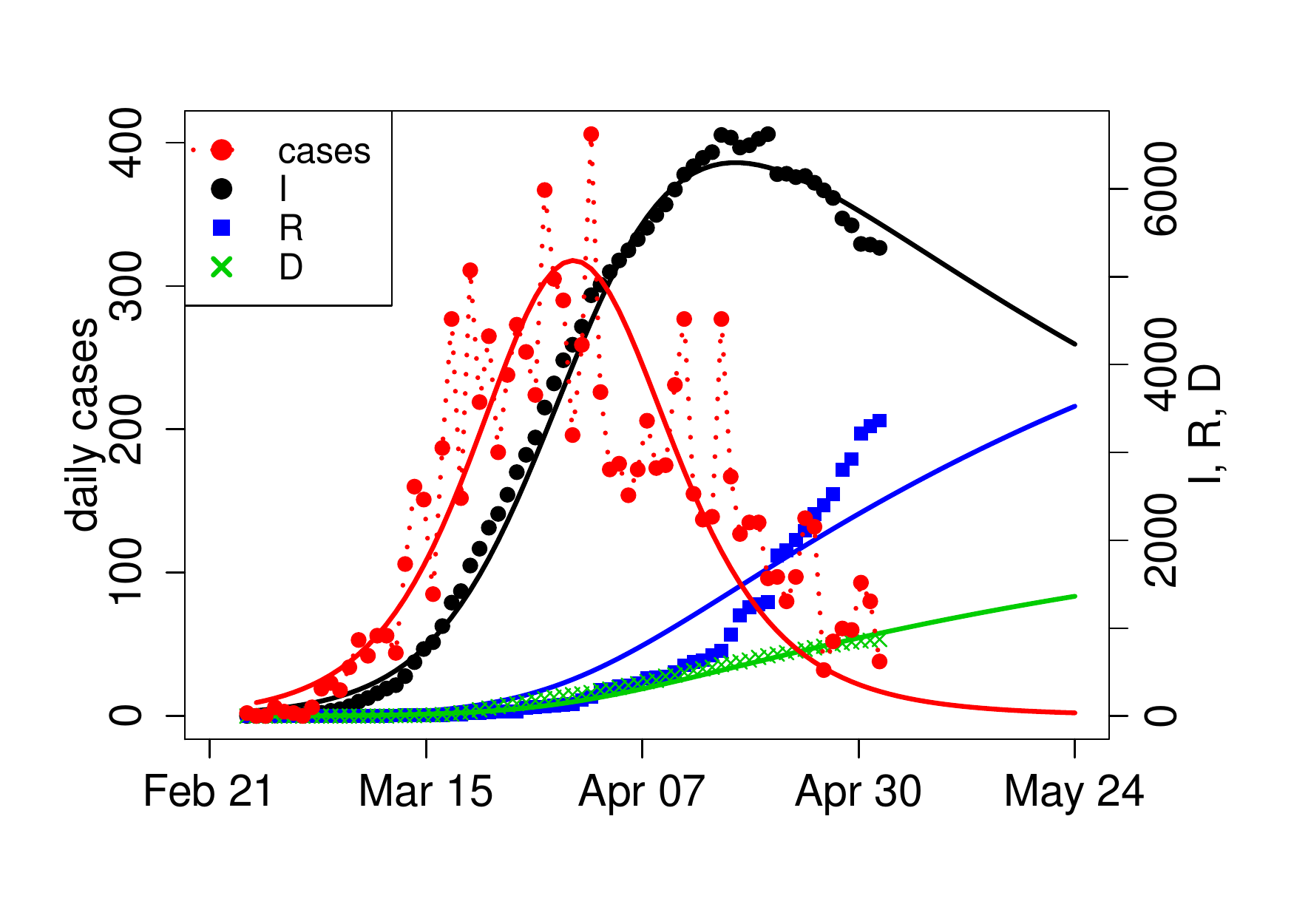}\\
\hspace{-10mm} (e) Emilia-Romagna & \\
\hspace{-10mm}
\includegraphics[width=73mm]{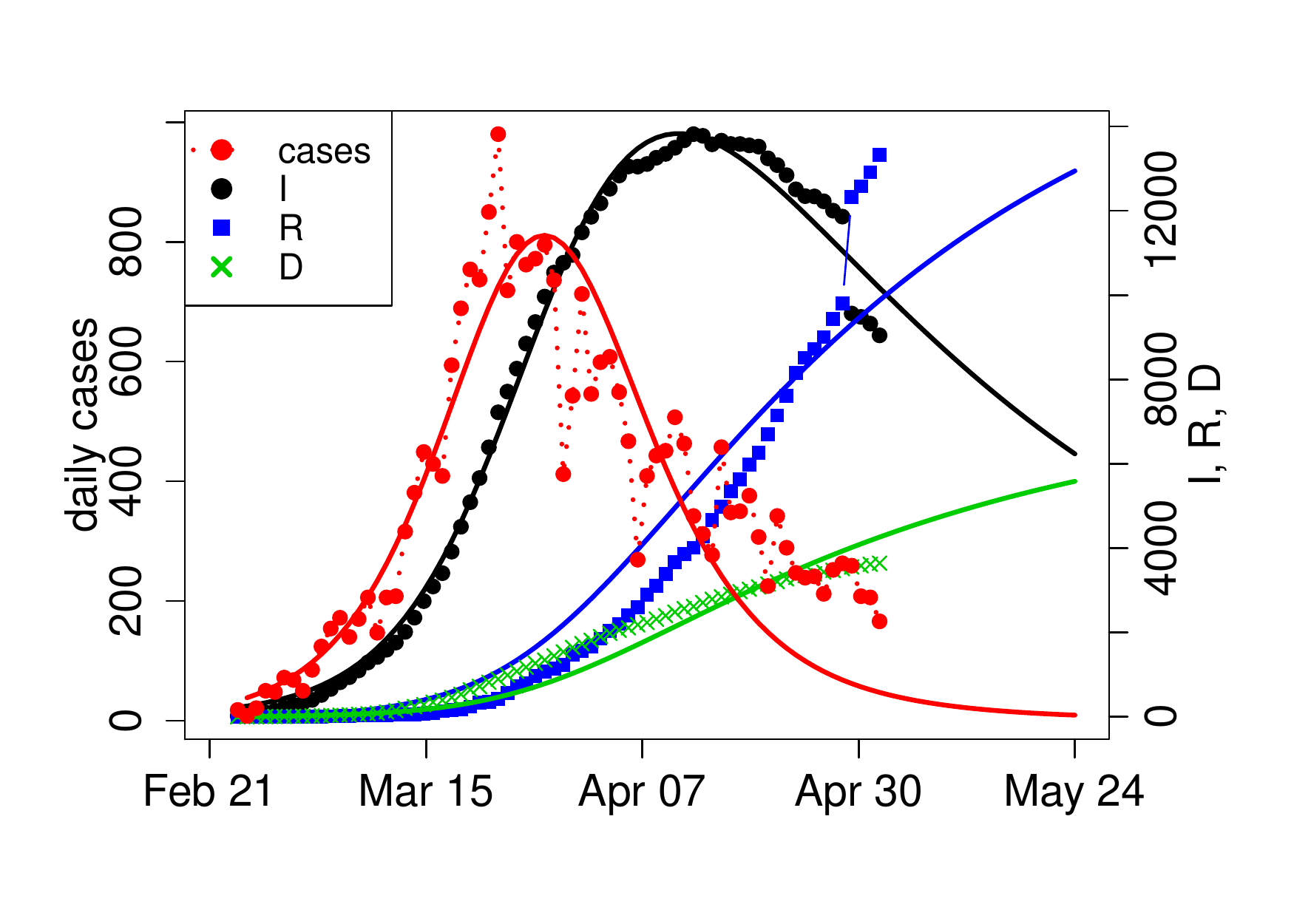} & \\
\end{tabular}
\caption{Data and forecasts based on the SIRD model {\rm (\ref{eq:SIRD})} in the five
regions. Left $y-$axis: daily number of confirmed cases.
Right $y-$axis: number of infected (I), recovered (R) and dead (D). Solid lines
correspond to fitted values.}
\label{fig:data_SIRD}
\end{figure}

Figure \ref{fig:data} shows the daily number of confirmed cases in the five regions.
The diffusion process peaked around the second half of March in the regions
where the epidemic started: Lombardy, Veneto and Emilia-Romagna. First, it is
worth noting the peculiarity of the epidemic in Lombardy, as the spread was much greater than in
the other regions. What is common is that the shape of the
 spread was asymmetric with a faster increase and a much slower decrease. Piedmont
 has  a  different  peculiarity,  with a
  rather flat spread in the second part of April because this region encountered problems
  with taking and
processing the swabs.

The capability of each  region in processing swabs changed
over time. Initially, swabs were essentially performed only
on symptomatic patients and their
strict contacts. Some regions, however, quickly increased  their capability to
process swabs,
and asymptomatic cases could be detected and isolated as well. Since only
people with a positive swab can be officially recorded as infected, and
as many infected people do not show symptoms, a large number of swabs is a
prerequisite for
diagnosis:
the more swabs taken, the more cases found. The relationship between confirmed cases and
the number of swabs is shown in
Figure \ref{fig:data_tamponi}, with daily values.
The  fluctuations
in the number of confirmed cases depend, to a certain extent, on the number of swabs processed, and
 for some regions, such as Lombardy, Veneto and Tuscany, the data show an almost regular weekly
  pattern, due perhaps to the organisation of the laboratories.

Figure \ref{fig:data_SIRD} depicts, for each region, the number of currently infected,
cumulated recovered people and cumulated deaths.
Fitted values obtained with
the SIRD model are also plotted (for details, see Section
\ref{sec:applications}).

\section{Models}\label{sec:models}
A general diffusion of innovations model can be defined through a nonlinear regression model as follows:
\begin{equation}\label{eq:generico}
 y(t) = z(t,\vartheta) +\varepsilon(t),
\end{equation}
where $y(t)$ are the cumulative sales of a product at time $t$ and $z(t,\vartheta)=z(t)$ is
a specific structure to be used to describe an evolution process.
Here, $\varepsilon_t$ are assumed to be i.i.d. Gaussian with variance $\sigma^2.$
The components of the parameter vector $\vartheta$ are jointly estimated using nonlinear least squares
(or, equivalently, likelihood estimation).
\clearpage

In this paper, we will compare the performance of alternative evolution structures.
The basic model is a logistic one (LOG):
\begin{equation}\label{eq:log}
 z(t) = m \frac{e^{\frac{t-\lambda}{\eta}}}{1+e^{\frac{t-\lambda}{\eta}}},
 \end{equation}
where $\lambda$ is the mode, median and average of the distribution, while $\eta$ is a shape parameter. Parameter $m$
is the market potential, which is the limiting value for $z(t),$ as $t$ goes to infinity.

The Generalized Bass Model \citep{bass:1994} is defined starting from the differential equation
\begin{equation}
 z^\prime (t) = \left[p+q\frac{z(t)}{m} \right][m-z(t)]w(t).
\end{equation}
Its solution, for the initial condition $z(0)=0,$ is
\begin{equation}\label{eq:gbm}
 z(t) = m \frac{1-e^{-(p+q)\int_0^t w(\tau)d\tau}}{1+\frac{q}{p} e^{-(p+q)\int_0^t w(\tau)d\tau}},
\end{equation}
where $m$ is the market potential, $p$ is the innovation coefficient,
$q$ is the imitation coefficient and $w(t)$ can be any integrable function. The effect of the intervention
function $w(t)$ is to accelerate or decrease diffusion with respect to a symmetric
unimodal path, which would arise in (\ref{eq:gbm}) for $w(t)=1$ for all $t$
values.
For $t$ values such as $w(t)>1$ diffusion is accelerated, while $w(t)<1$ corresponds to time periods with decreased diffusion speed.
Below, we examine the model (GBM$_{\rm RECT}$)
arising when $w(t)$ is specified by the so-called \textit{rectangular} shock:
\begin{equation}\label{eq:rect}
w_R(t)=1 +  c I_{a \leq t \leq b}.
\end{equation}
This allows us to describe the diffusion of a product for which we observe a constant shock
with intensity $c,$ either positive or negative, in the time interval $[a, b]$ \citep{guseo:05}.

Due to the asymmetric path observed for almost every region, we also examine a more
flexible model from \cite{bemmaor:94}:
\begin{equation}\label{eq:bemmaor}
 z(t) = m \frac{1-e^{-(p+q)t}}{[1+\frac{q}{p} e^{-(p+q)t}]^A},
\end{equation}
where $A$ is a further parameter allowing for asymmetry (positive asymmetry for $A>1,$
negative asymmetry for $A<1$).
If we insert a rectangular shock into it, we obtain the following model
(BeGBM$_{\rm RECT}$):
\begin{equation}\label{eq:bemrect}
 z(t) = m \frac{1-e^{-(p+q)\int_0^t w_R(\tau)d\tau}}{[1+\frac{q}{p} e^{-(p+q)\int_0^t w_R(\tau)d\tau}]^A},
\end{equation}
with function $w_R$ specified as in (\ref{eq:rect}).

A different way to provide flexibility to the evolutive structure can be obtained through a dynamic market potential
model \citep{guseo:09} (DMP):
\begin{equation}\label{eq:ggm}
 z(t) = m \sqrt{\frac{1-e^{-(p_c+q_c) t}}{1+\frac{q_c}{p_c} e^{-(p_c+q_c)t}}} \frac{1-e^{-(p+q) t}}{1+\frac{q}{p} e^{-(p+q)t}},
\end{equation}
where $p_c$ and $q_c$ are two parameters to describe how fast the dynamic market potential
approaches its maximum value, $m.$

Additionally, the DMP can be perturbed by shocks. For a general intervention function, $w(t),$
we obtain:
\begin{equation}\label{eq:ggmwt}
 z(t) = m \sqrt{\frac{1-e^{-(p_c+q_c) t}}{1+\frac{q_c}{p_c} e^{-(p_c+q_c)t}}}
 \frac{1-e^{-(p+q) \int_0^t w (\tau)d\tau}}{1+\frac{q}{p} e^{-(p+q)\int_0^t w (\tau)d\tau}}.
\end{equation}
If in model (\ref{eq:ggmwt}) we use, as proposed in \cite{guidolin:14}, the intervention function
\begin{equation}\label{eq:wstag}
w_s(t)= 1 +  \alpha_1 \cos \left( \frac{2 \pi t}{s} \right) +
\alpha_2 \sin \left( \frac{2 \pi t}{s} \right),
\end{equation}
we allow the model to incorporate cyclic seasonal fluctuations of width $\alpha_1$ and $\alpha_2$
with period $s$ (DMPseas).

Here, we propose to assess the usefulness of a dynamic market potential model as in
(\ref{eq:ggmwt}),
but with an intervention function depending upon the number of swabs analyzed at day $t,$
$B(t)$  (DMPsw). In particular, we suggest using
\begin{equation}\label{eq:wtamp}
w_B(t)= 1 +  \xi \left( \frac{B(t) - \mu_B}{\sigma_B} \right),
\end{equation}
where $\mu_B$ and $\sigma_B$ are the average and the standard deviation,
respectively, of the $B(t)$ values recorded during the observation period. It is easy to appreciate that such a structure accelerates,
with respect to an underlying trend described by a DMP, the number of cases
 whenever $B(t)$ exceeds its average, while
cases are reduced with a below-average number of swabs.

\begin{table}
\centering
\caption{Summary of the models analyzed.}\label{tab:listamodelli}
\hspace*{-1.2cm}
\begin{tabular}{lp{3.5cm}lclc}
\hline
 & \mbox{Model} & \mbox{Abbreviation} & Number of            &List of            & Equation\\
 &              &                     &           parameters &        parameters &         \\
\hline
1 & Logistic                                                 & LOG                & 3 & $(m, \lambda, \eta)$  & (\ref{eq:log})   \\
2 & GBM with rectangular$\;\;\;$ shock                       & GBM$_{\rm RECT}$   & 6 & $(m, p, q, a, b, c)$   & (\ref{eq:gbm})+(\ref{eq:rect})   \\
3 & Bemmaor GBM with$\;\;\;$ rectangular shock               & BeGBM$_{\rm RECT}$ & 7 & $(m, p, q, a, b, c, A)$  & (\ref{eq:bemrect})+(\ref{eq:rect})    \\
4 & Dynamic market$\;\;\;\;\;\;$ potential                   & DMP                & 5 & $(m, p_c, q_c, p, q)$  & (\ref{eq:ggm})   \\
5 & Dynamic market$\;\;\;\;\;\;\;$ potential+seasonal effect & DMPseas            & 8 & $(m, p_c, q_c, p, q, \alpha_1,  \alpha_2, s)$  & (\ref{eq:ggmwt})+(\ref{eq:wstag})   \\
6 & Dynamic market$\;\;\;\;\;\;\;$ potential+swabs           & DMPsw              & 6 & $(m, p_c, q_c, p, q, \xi)$  & (\ref{eq:ggmwt})+(\ref{eq:wtamp})   \\
7 
& Susceptibles, Infectives, Recovered, Deaths              & SIRD               & 5 & ($\beta$, $\gamma$, $\delta$, $N$, $I_0$) & (\ref{eq:SIRD}) \\
\hline
 \end{tabular}
\end{table}

A further benchmark is proposed in this work with the SIRD model, a compartmental model used for
describing and predicting the evolution of an infectious disease. Every individual of the population
may flow between the  compartments of `Susceptibles' ($S=S(t)$), `Infected' ($I=I(t)$),
`Recovered' ($R=R(t)$)  and `Deaths' ($D=D(t)$). We could apply this model using the data of
currently infected patients ($I$), cumulated recovered individuals ($R$) and cumulated deaths ($D$).
The forecasts of the confirmed cases are then calculated by summing the forecasts of $I$, $R$ and $D$.

This model uses the following system of differential equations:

\begin{equation}\label{eq:SIRD}
\left \{
\begin{array}{ll}
\displaystyle \frac{\partial S}{\partial t}$=$ -\displaystyle \frac{\beta I}{N}S\\[10pt]
\displaystyle \frac{\partial I}{\partial t}$=$ \displaystyle \frac{\beta I}{N}S -\gamma I -\delta I\\[10pt]
\displaystyle \frac{\partial R}{\partial t}$=$\gamma I \\[10pt]
\displaystyle \frac{\partial D}{\partial t}$=$\delta I\\[8pt]
N=S+I+R+D\\
S(0)=S_0, \ I(0)=I_0, \ R(0)=R_0, \ D(0)=D_0,\\
\end{array}
\right.
\end{equation}
where $N$ is the population size, while $\beta$, $\gamma$, and $\delta$ are
the rates of infection, recovery and mortality, respectively. The initial
conditions $R_0$ and $D_0$ correspond to the observed number of recovered and
dead individuals on the first day of the collection period. In this work, $I_0$
and $N$ are estimated to maximize the fitting, as the goal of this work is to
describe and predict the evolution of the total number of confirmed cases. The
last initial value to be defined is $S_0=N-I_0-R_0-D_0$. The parameter
set corresponds to ($\beta$, $\gamma$, $\delta$, $N$, $I_0$).

Table \ref{tab:listamodelli} proposes a summary of all the models that will be
applied in this study.

\section{Applications}\label{sec:applications}


Models of Table \ref{tab:listamodelli} were applied to the data of the five considered regions,
and forecasts up to May 24th are provided (three weeks ahead for
each region).
The first six models were fitted to the cumulative confirmed cases using NLS estimation;
asymptotic standard errors
and 95\%  asymptotic marginal confidence intervals ($mCIs$) are provided.

The SIRD model was fitted to I, R and D time series, assuming errors to be $iid$ normal
distributed, and therefore using MLE estimation. $R$
was used with the {\it bbmle} library \citep{R:bbmle}. To ensure that $\beta,$  $\gamma,$ $\delta$
estimates lay in $(0,1)$, the model was reparametrized  using their logit transformations.
Moreover,  for computational aspects,
the natural logarithms of both $N$ and $I_0$ were used.
Standard errors and 95\% profile likelihood $mCIs$ (based on inverting a
spline fit to the profile likelihood) are given.
By summing the
fitted values of I, R and D, we also obtained the fitted values of the cumulative
confirmed cases: this is used as a benchmark for evaluating the performance of the
proposed models. For each region, Figure \ref{fig:data_SIRD} depicts the evolution
of I, R and D  on the right $y-$axis and of the daily confirmed cases on the left
$y-$axis; solid lines correspond to fitted values.

Table \ref{tab:R2} summarizes the values of  the determination index $R^2$ for all models:
the huge values of {$R^2$} are unsurprising, given that we are working with
cumulative data and any S-shaped fitting produces high determination indexes.
A standard approach advises
the use of the
$R^2$ measure for comparative purposes only \citep{mortarino:15, ranit:2017};
 we, therefore, provided also the BIC (evaluated with cumulative values) and the
 squared linear correlation coefficient $\rho^2$  between observed and fitted \emph{daily} values as well.

\begin{table}
\centering
\caption{$R^2$ of the nonlinear models and corresponding BIC (cumulative data as response variable) and
squared linear correlation coefficient, $\rho^2,$
between observed \emph{instantaneous} sales and fitted \emph{instantaneous} sales.}
\label{tab:R2}
 \hspace*{-6mm}
 \begin{tabular}{lcc@{$\!\;\;$}c@{$\!\;\;$}c@{$\!\;\;$}c@{$\!\;\;$}c@{$\!\;\;$}c@{$\!\;\;$}c@{$\!\;\;$}c@{$\!\;\;$}c}
&& LOG                &GBM$_{\rm RECT}$   &BeGBM$_{\rm RECT}$ &DMP                &DMPseas            &DMPsw               &SIRD \\
\hline
&  &  & \\[-3mm]
Veneto     & $R^2$         &0.996912  &0.999785  &0.999845  &0.999822  &0.999825  &0.999898    & 0.987174 \\
           & BIC           &877.3940  &695.9181  &675.9767  &677.8449  &689.2245  &641.4728    & 989.9329\\
           & $\rho^2$      &0.738015  &0.796939  &0.822816  &0.828927  &0.834856  &0.858459    & 0.707816\\[2mm]
Lombardy   & $R^2$         &0.993010  &0.999629  &0.999900  &0.999834  &0.999860  &0.999919    & 0.720144    \\
           & BIC           &1143.348  &941.8574  &850.6930  &878.7936  &879.2002  &830.5268    & 1421.278   \\
           & $\rho^2$      &0.593900  &0.690817  &0.826706  &0.803006  &0.820098  &0.902698    & 0.328468 \\[2mm]
Piedmont   & $R^2$         &0.995989  &0.999791  &0.999831  &0.999880  &0.999895  &0.999905    & 0.993763    \\
           & BIC           &908.2256  &714.3012  &703.6910  &671.2378  &674.8002  &658.7328    & 947.6302   \\
           & $\rho^2$      &0.697338  &0.782143  &0.794564  &0.814390  &0.834386  &0.843469    & 0.683373    \\[2mm]
Tuscany    & $R^2$         &0.996129  &0.999432  &0.999725  &0.999772  &0.999792  &0.999796    & 0.991991    \\
           & BIC           &757.0439  &637.2911  &591.3882  &570.1541  &576.5137  &566.8199    & 827.7638   \\
           & $\rho^2$      &0.702671  &0.728945  &0.793943  &0.799981  &0.842169  &0.778397    & 0.712199 \\[2mm]
Emilia--   & $R^2$         &0.995195  &0.999822  &0.999923  &0.999776  &0.999862  &0.999925    & 0.993920 \\
~~Romagna  & BIC           &919.6076  &701.7807  &647.0657  &713.4355  &692.4212  &641.2345    & 944.5885\\
           & $\rho^2$      &0.741966  &0.890328  &0.920939  &0.907700  &0.921988  &0.904318    & 0.796537\\
\hline
 \end{tabular}
\end{table}

\subsection{Veneto}
\begin{figure}[t]
\centering
\begin{tabular}{cc}
\hspace{-10mm} (a) & \hspace{-8mm} (b) \\
\hspace{-8mm} \includegraphics[width=73mm]{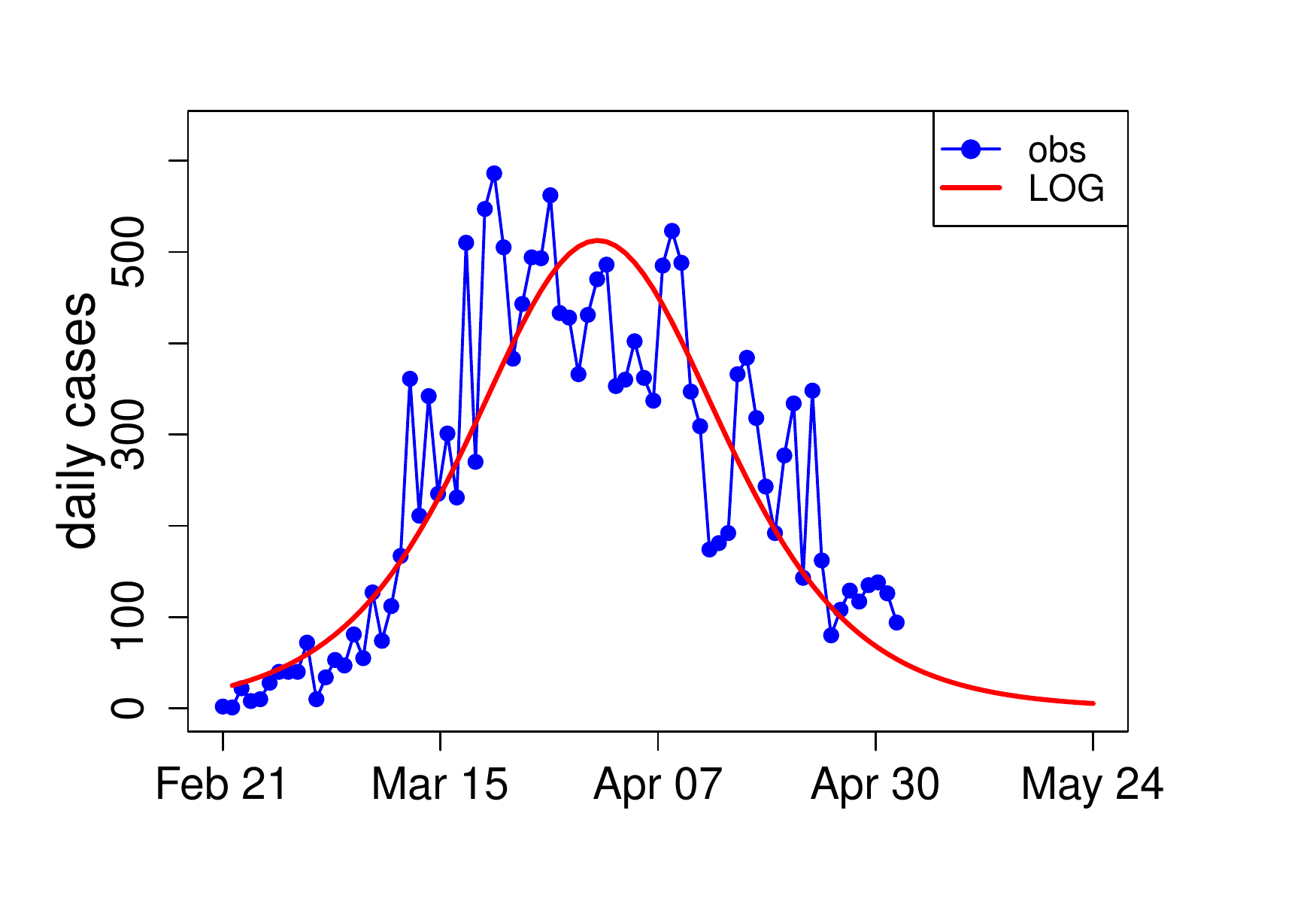} &
\hspace{-10mm} \includegraphics[width=73mm]{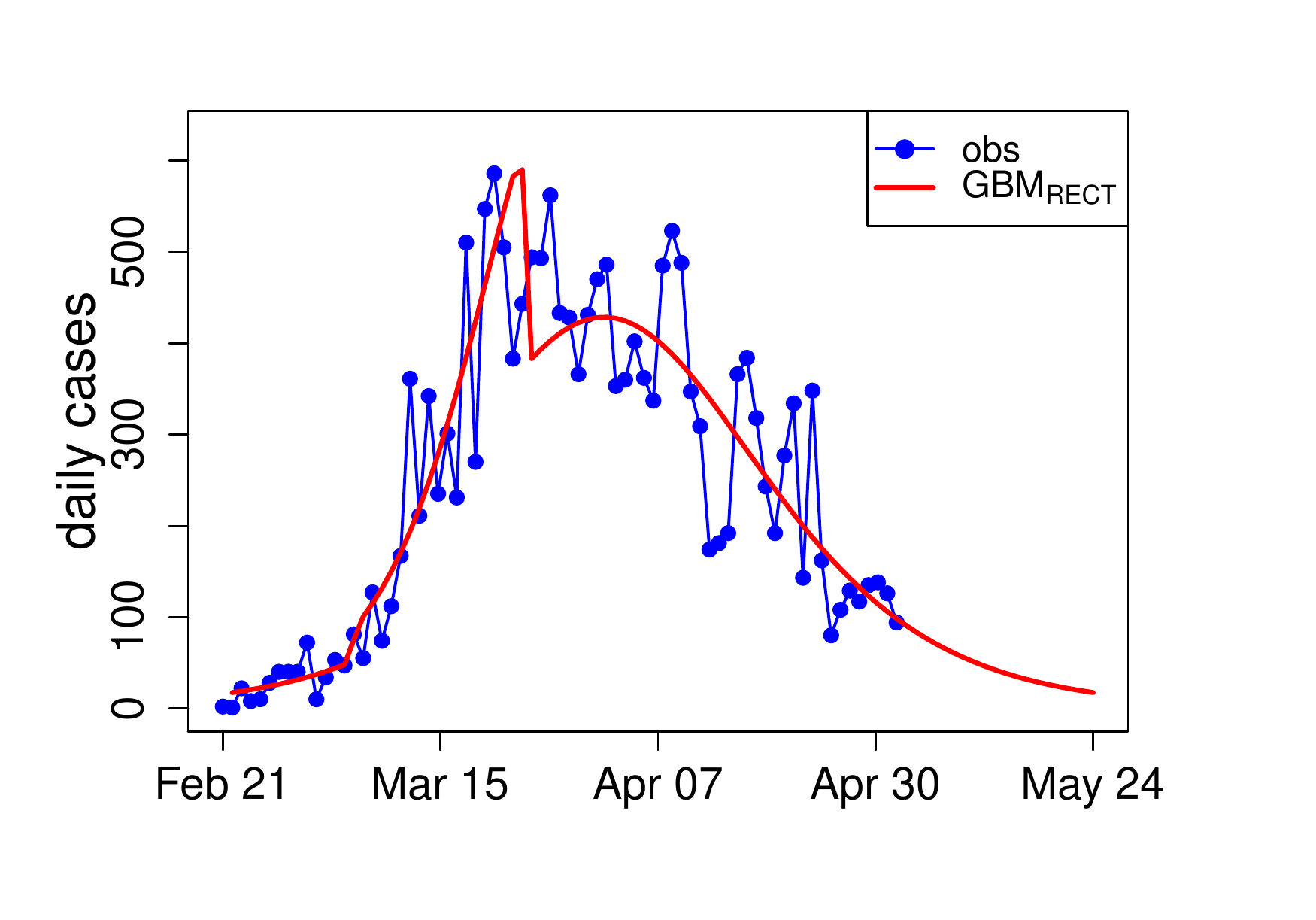}\\
\hspace{-8mm}(c) &  \hspace{-8mm} (d) \\
\hspace{-8mm} \includegraphics[width=73mm]{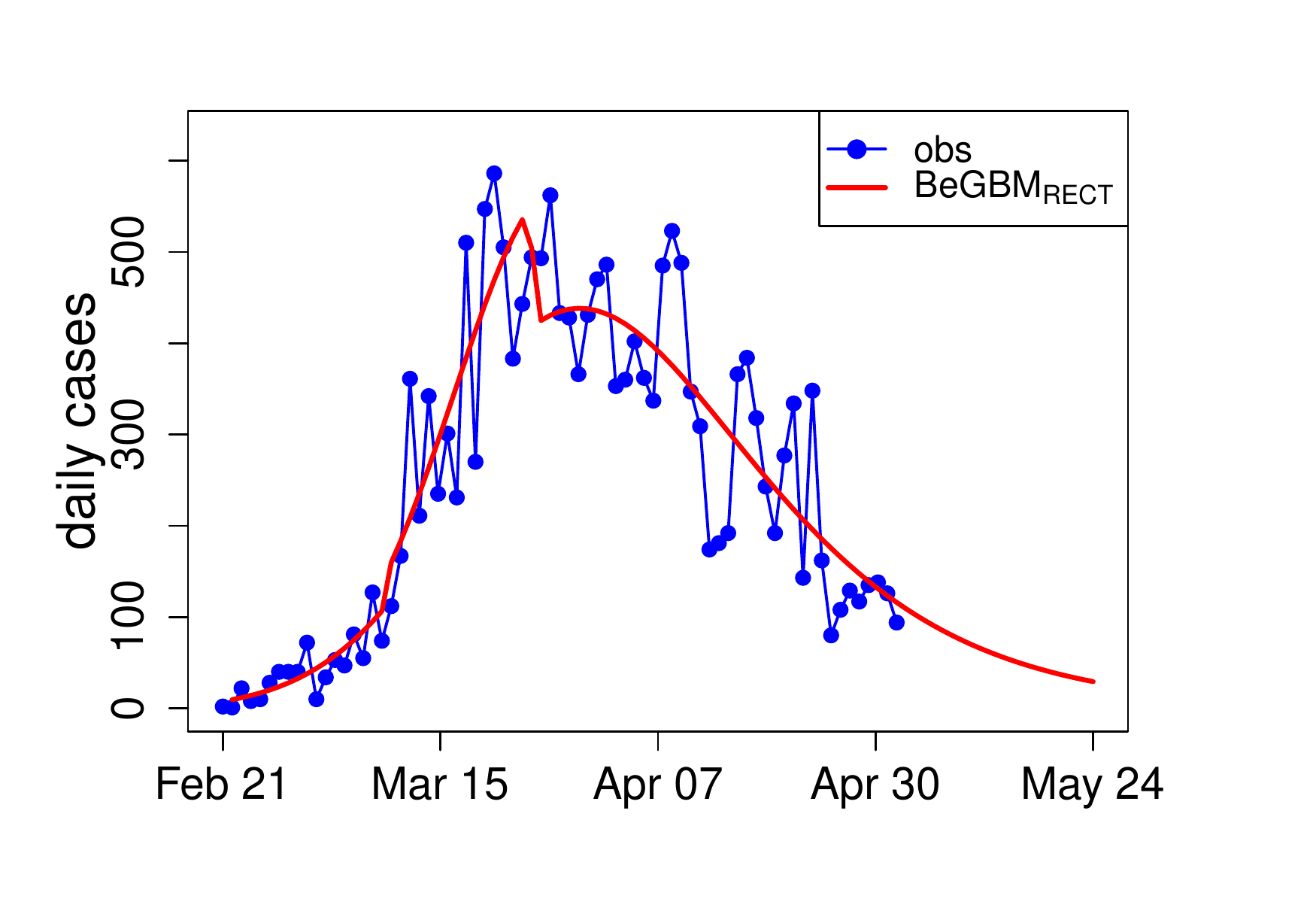} &
\hspace{-8mm}\includegraphics[width=73mm]{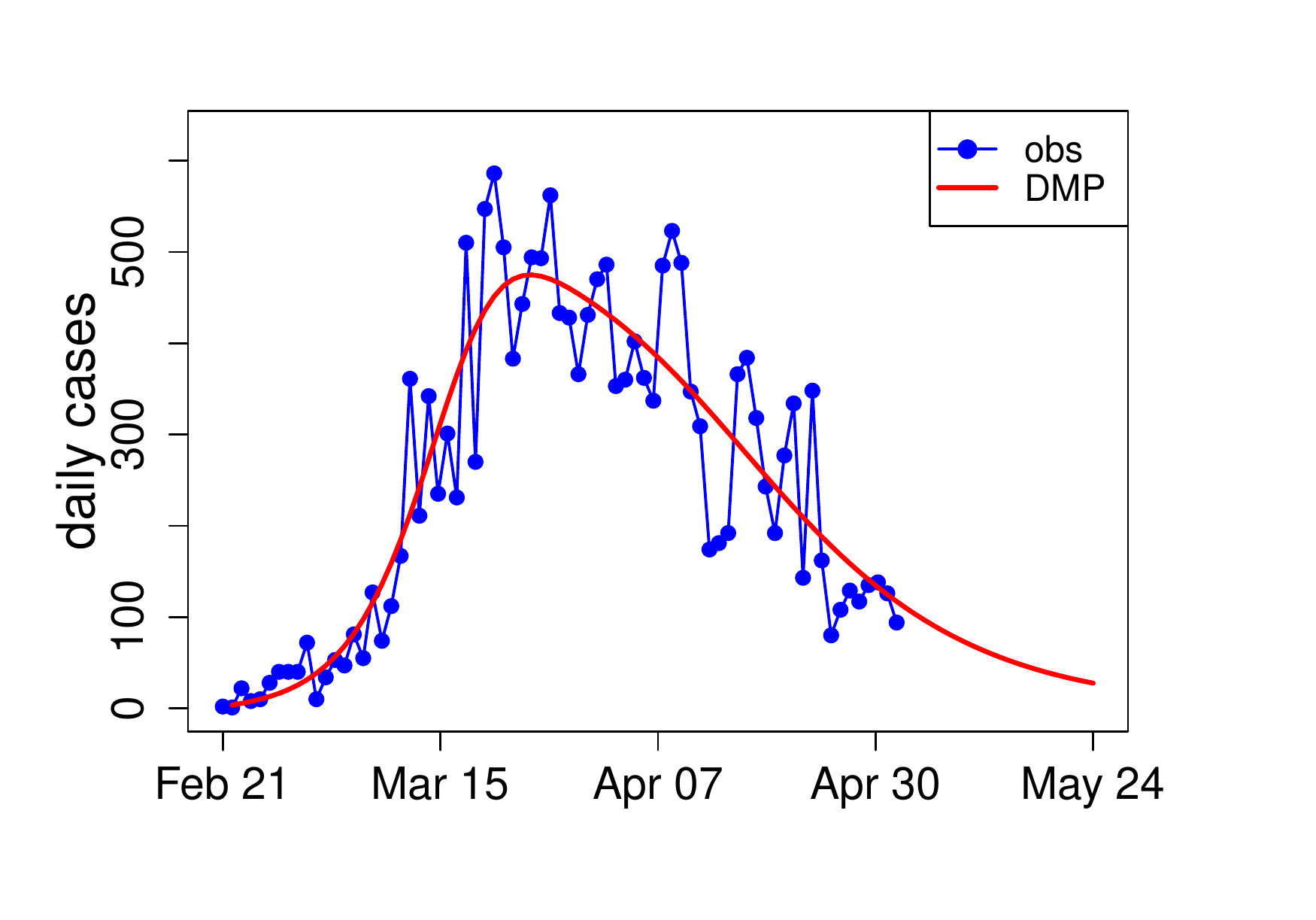}\\
\hspace{-10mm} (e) &  \hspace{-8mm} (f) \\
\hspace{-10mm} \includegraphics[width=73mm]{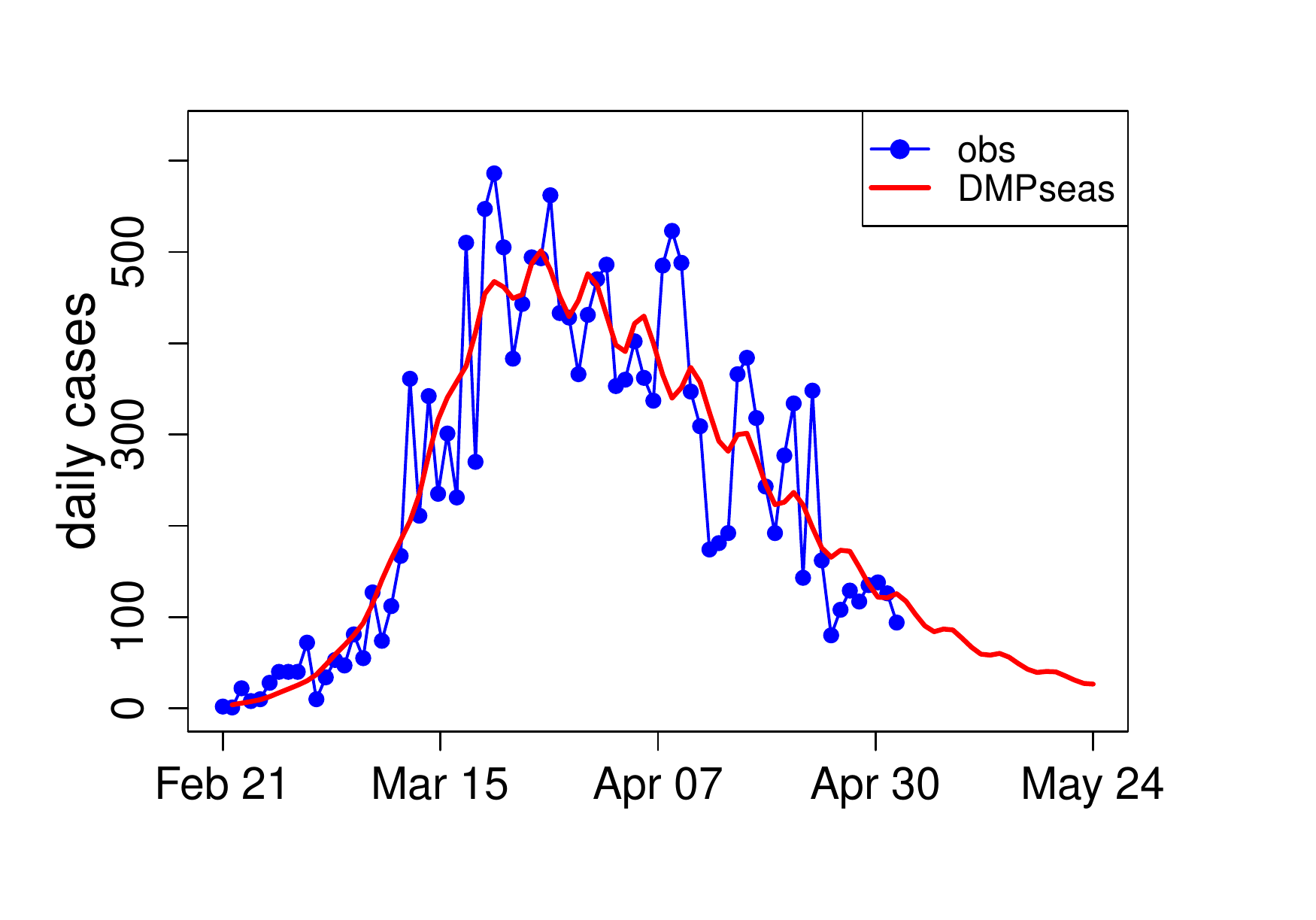} &
 \hspace{-8mm} \includegraphics[width=73mm]{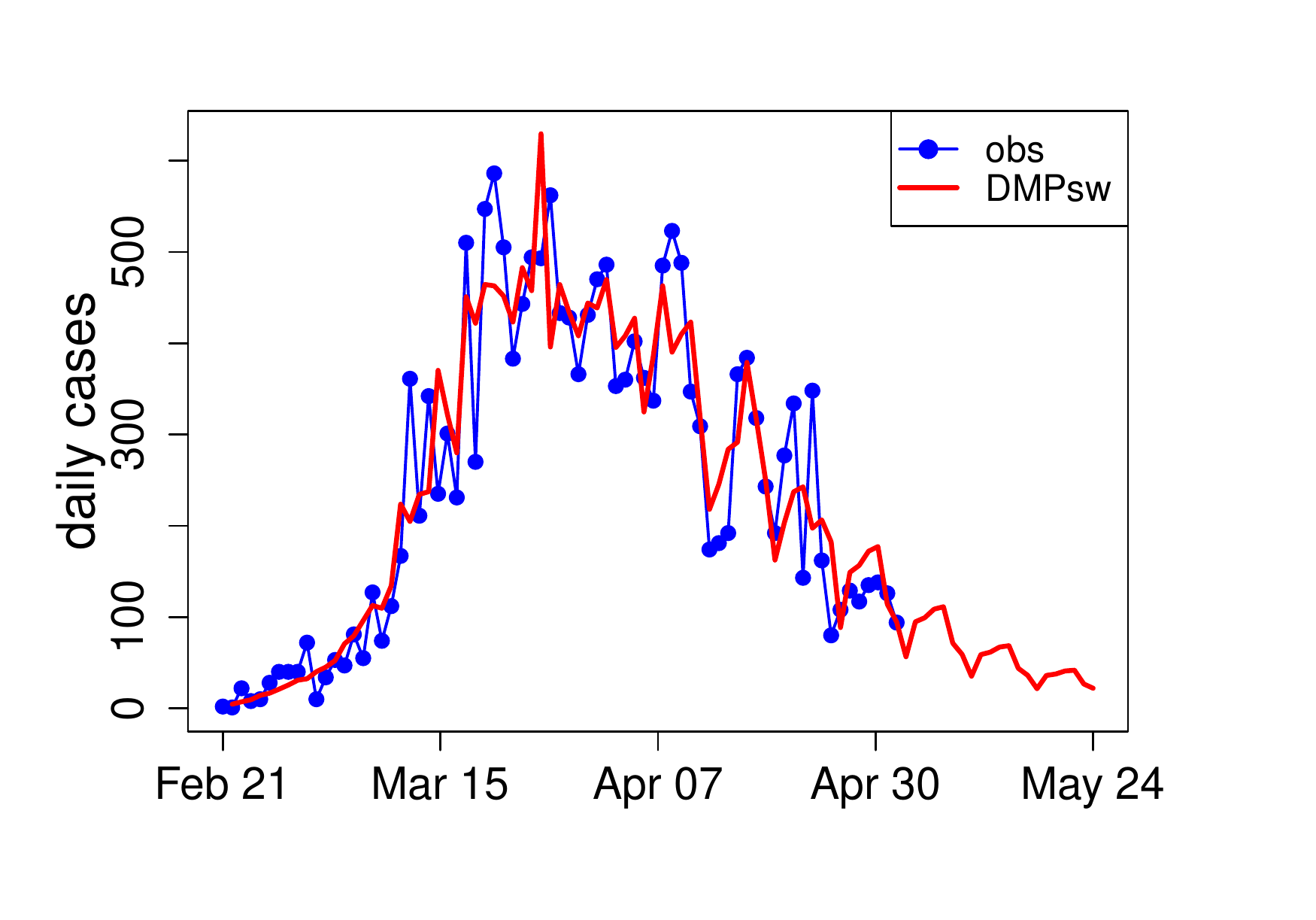}  \\
\end{tabular}
\caption{Veneto. Observed and fitted values with the alternative models.
{\rm (a)} Logistic                                                 (LOG);
{\rm (b)} GBM with rectangular shock                       (GBM$_{\rm RECT}$);
{\rm (c)} Bemmaor GBM with rectangular shock               (BeGBM$_{\rm RECT}$);
{\rm (d)} Dynamic market potential                   (DMP);
{\rm (e)} Dynamic market potential+seasonal effect (DMPseas);
{\rm (f)} Dynamic market potential+swabs           (DMPsw).}
\label{fig:fit veneto}
\end{figure}

The results for Veneto are displayed in Table \ref{tab:R2} ($R^2,$ BIC and $\rho^2$
between observed and fitted daily values), in Tables \ref{tab:venetoTMP},
\ref{tab:venetoLOG}--\ref{tab:venetoSIRD} (for parameter estimates for all the models fitted)
and in Figures \ref{fig:data_SIRD}(a) and \ref{fig:fit veneto},
where observed and fitted daily values are plotted.

From these results, we can infer that the logistic (Figure \ref{fig:fit veneto}(a)) and
the SIRD (Figure \ref{fig:data_SIRD}(a)) models, which both
represent  two commonly used benchmarks, are not adequate to describe the asymmetrical evolution of the epidemic, together with the
large fluctuations.

The results in Tables \ref{tab:venetoGBM} and \ref{tab:venetoBEGBM} show that a  positive ($\hat{c}>0$)
rectangular shock is significantly diagnosed both in the
GBM$_{\rm RECT}$ and the BeGBM$_{\rm RECT}$. In both cases, the shock denotes
an increase in cases starting around
March 5th ($t\simeq$ 14)
and
March 9th ($t\simeq$ 18), respectively, and ending around
March 23rd ($t\simeq$ 32) and
March 24th ($t\simeq$ 33), respectively.  Within these
models, the shock has the function of fitting the steep increase in cases in the first period
of the epidemic. Since the data for Veneto start from February 21st, both shocks end almost exactly two weeks later than the lockdown, established on March 8th.
This confirms that the lockdown policy was essential in reducing the spread of the epidemics, as the incubation period is up to 14 days.
The BeGBM$_{\rm RECT}$ suggests that the decrease after the peak is much
slower than the initial growth ($\hat{A}$=2.316), and, for this reason, we expect that the
subsequent models, which allow for asymmetry too, will also have good performance.

\begin{figure}[t]
\centering
\begin{tabular}{cc}
\hspace{-10mm} (a) Veneto: {$\mu_B=5250.03, \sigma_B=3517.83$}
& \hspace{-8mm} (b) Lombardy: {$\mu_B=5704.68, \sigma_B=3948.13$}   \\
\hspace{-10mm} \includegraphics[width=70mm]{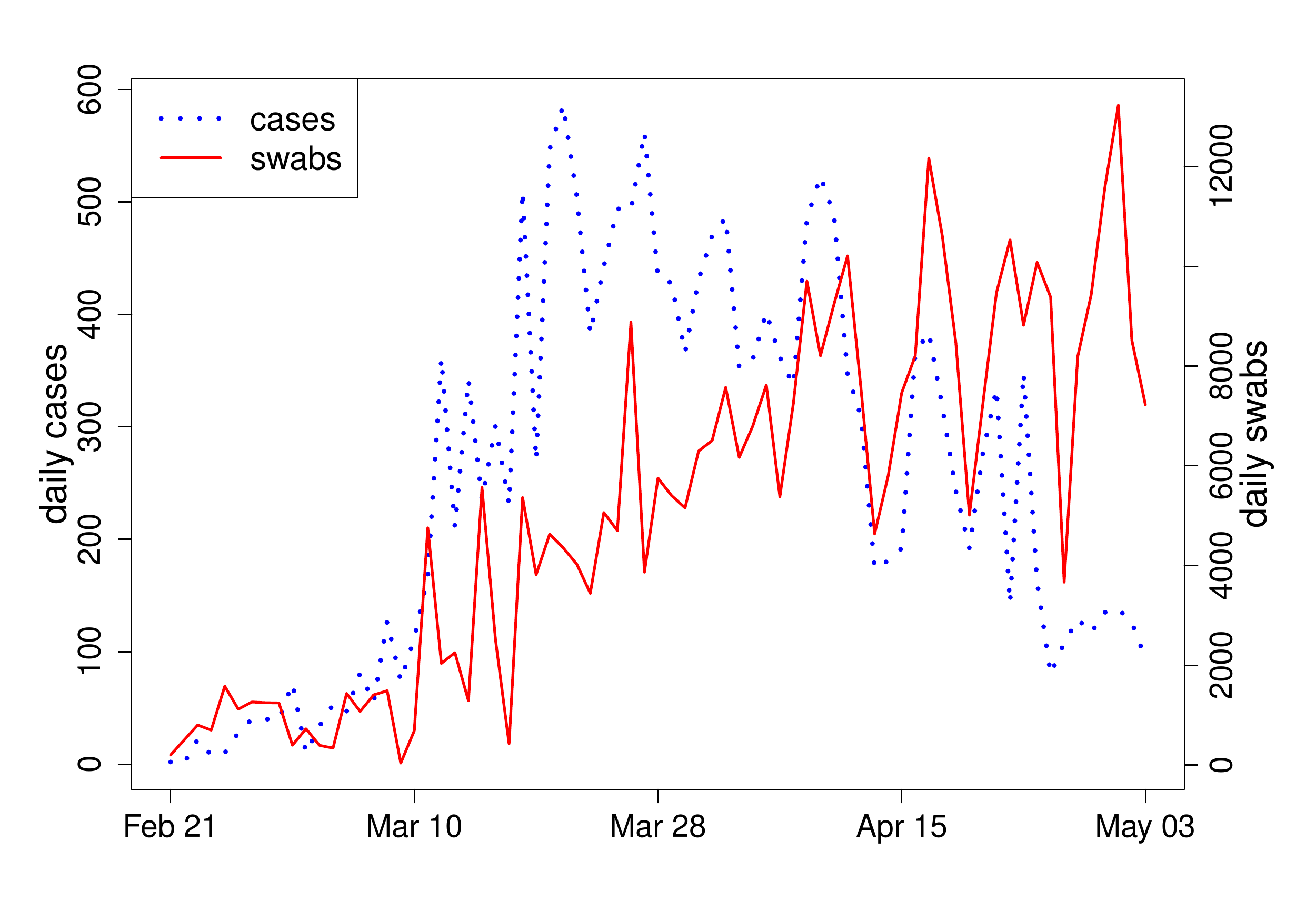} &
\hspace{-8mm} \includegraphics[width=70mm]{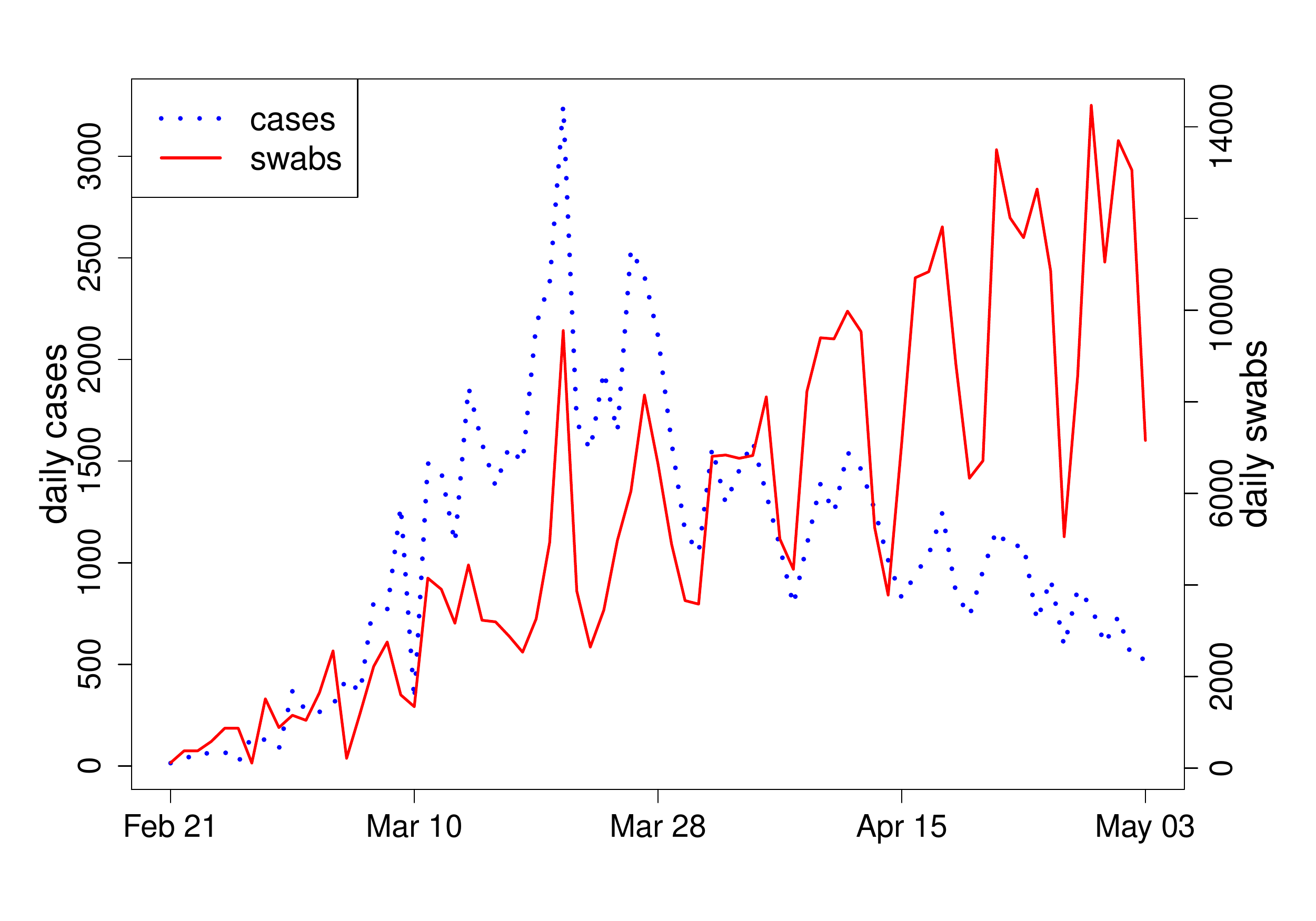}\\
\hspace{-10mm} (c) Piedmont: {$\mu_B=2493.72, \sigma_B=2169.88$}
& \hspace{-8mm} (d) Tuscany: {$\mu_B=2214.97, \sigma_B=1687.47$}  \\
\hspace{-10mm} \includegraphics[width=70mm]{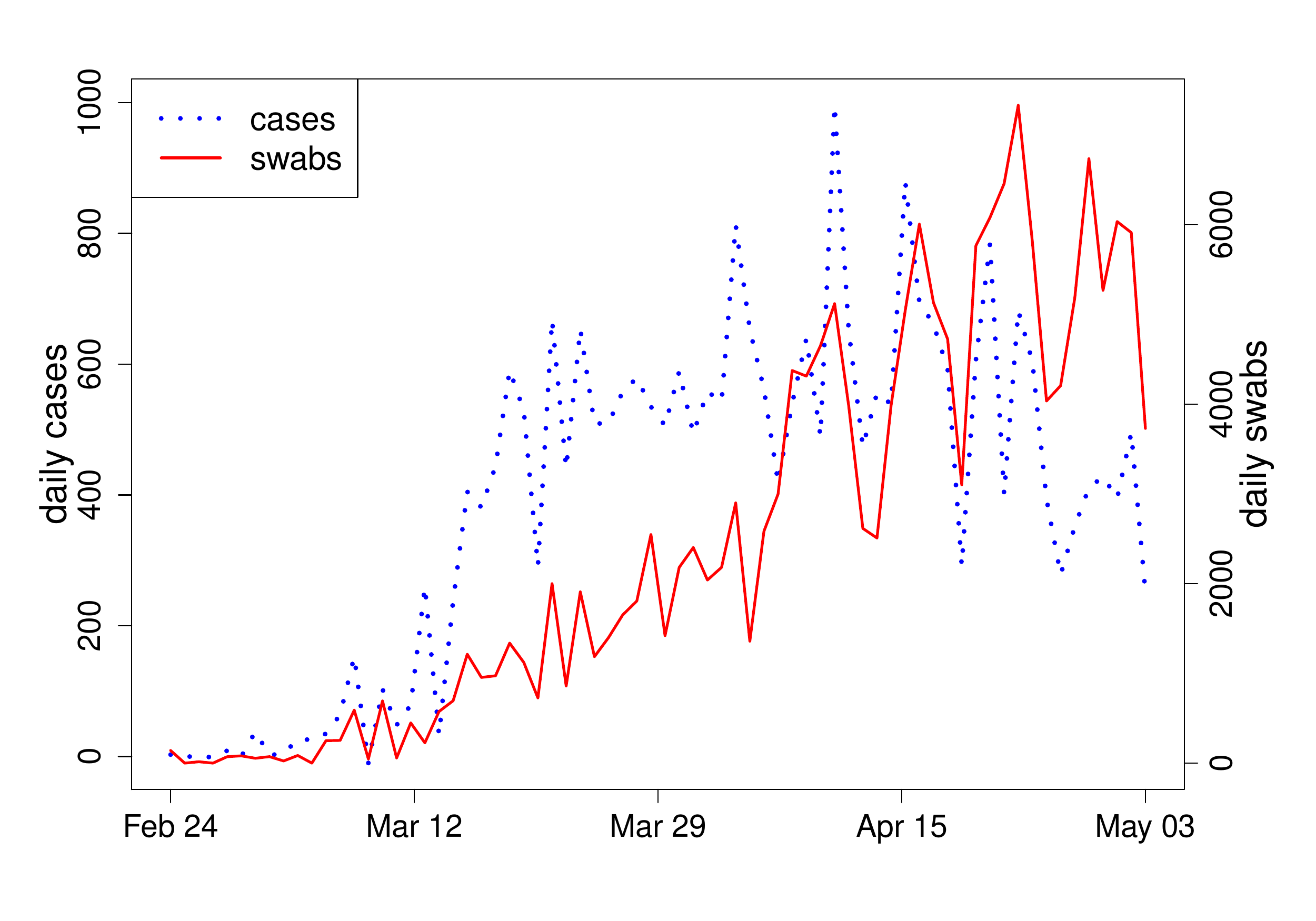} &
 \hspace{-8mm} \includegraphics[width=70mm]{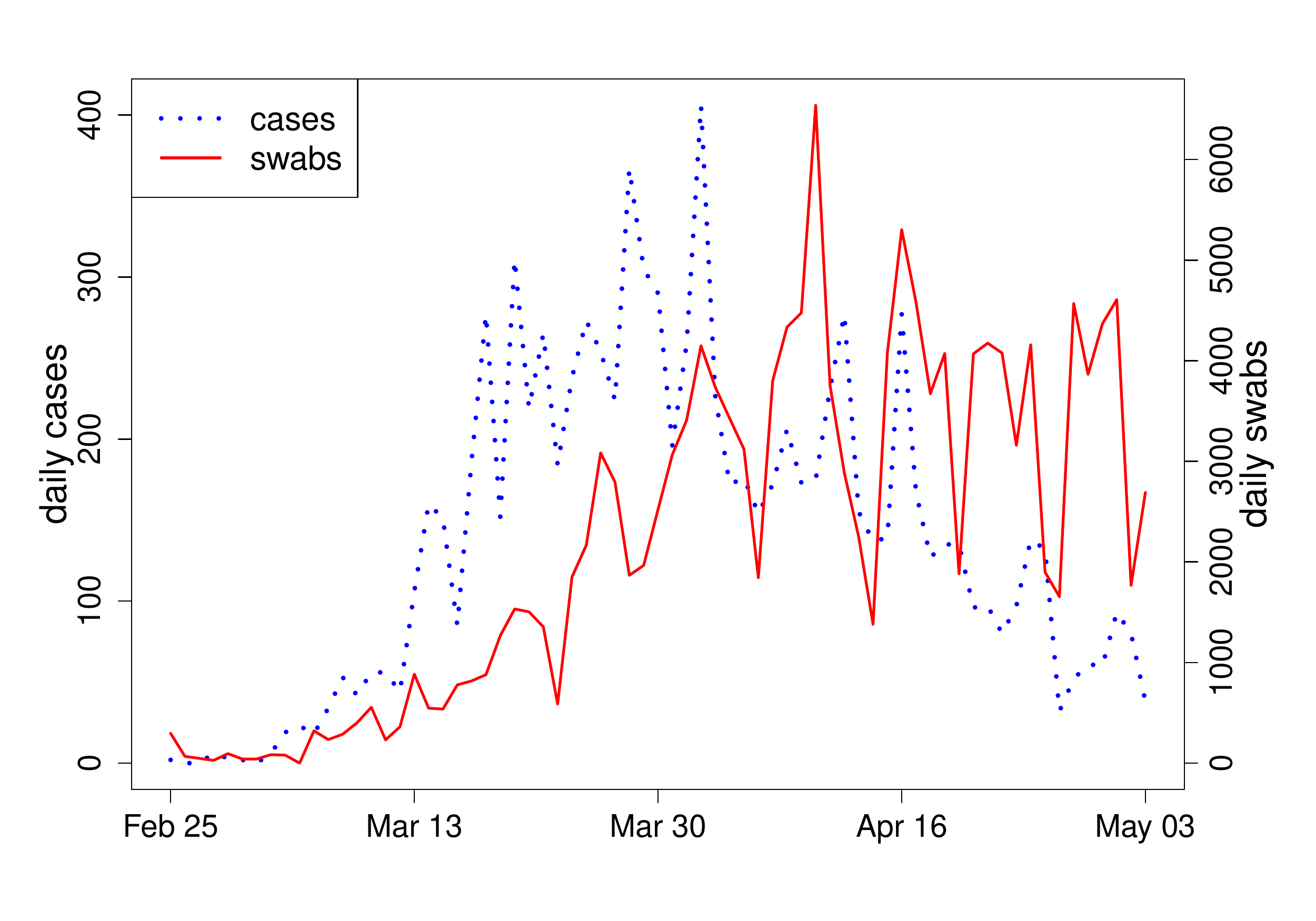}\\
\multicolumn{2}{l}{\hspace{-5mm} (e)  Emilia-Romagna: {$\mu_B=2854.01, \sigma_B=1975.41$}}  \\
\hspace{-10mm}\includegraphics[width=70mm]{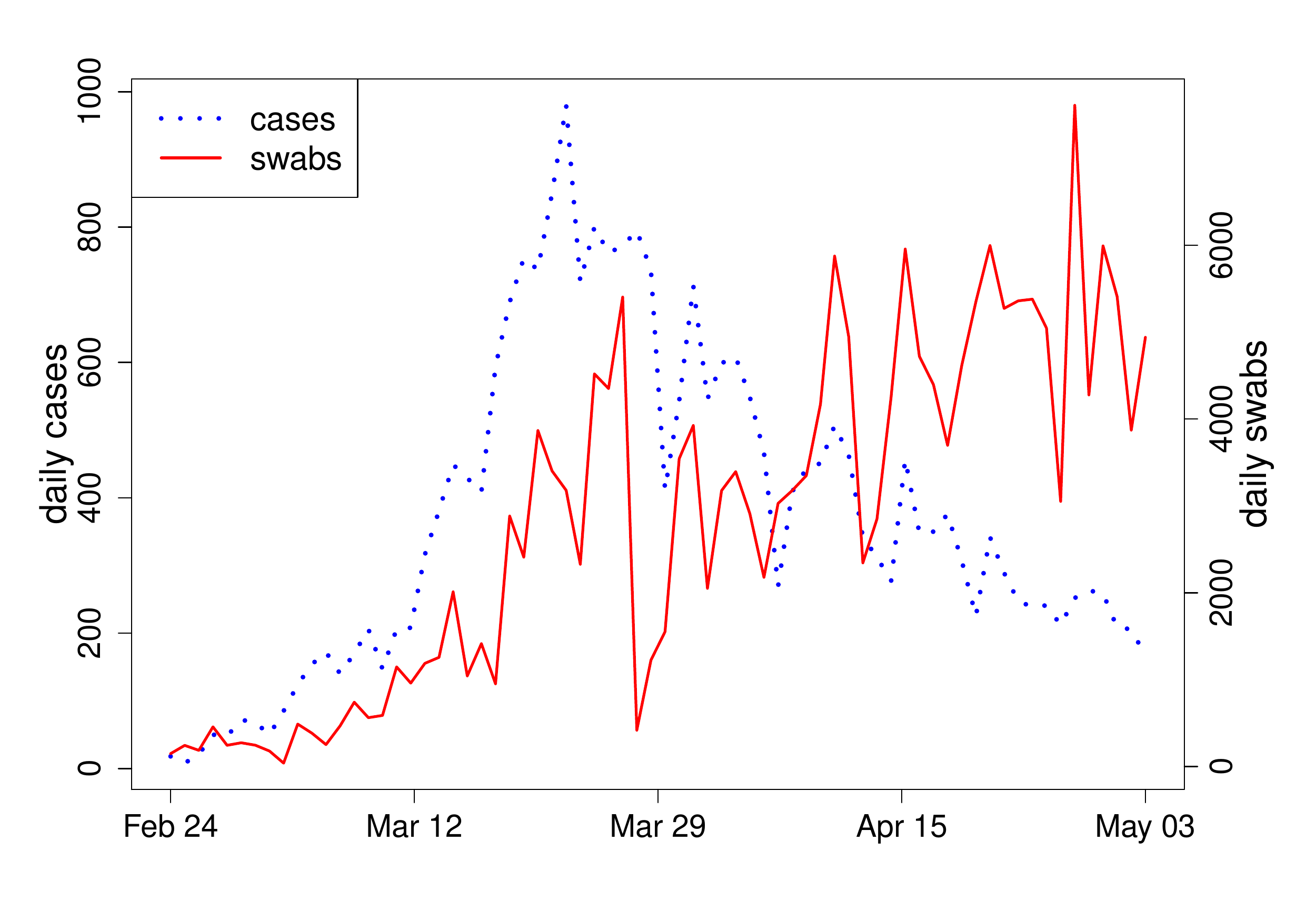} &
\end{tabular}
\caption{Daily number of confirmed cases and daily number of swabs processed in the five regions. $\mu_B$
and $\sigma_B$ correspond to the average and standard deviation, respectively, of the
swab time series (see {\rm Eq.~(\ref{eq:wtamp})}).}
\label{fig:data_tamponi}
\end{figure}
The DMP model (Figure \ref{fig:fit veneto}(d) and Table \ref{tab:venetoDMP}) summarizes the trend
of the series well, but the
$R^2$ and the BIC are slightly worse than the BeGBM$_{\rm RECT}.$ We also highlight that
a good fit with the DMP could not be attained in the early
\hfill phase
\hfill of
\hfill the
\hfill outbreak
\hfill with
\hfill a
\hfill smaller
\hfill number
\hfill of
\hfill observations,
\clearpage
\noindent
while the BeGBM$_{\rm RECT}$ could be correctly identified.

Among asymmetric models, however, only the DMPseas and DMPsw are able to take into account the
fluctuations around the main trend. The performance of the DMPseas
is very good ($R^2$=0.999825, $\rho^2$=0.834856). However,
the weekly cycle ($\hat{s}=7.003$ days, see Table \ref{tab:venetoSTAG})
in the DMPseas is still partially unsatisfactory because the range
of its fluctuations is not sufficiently large compared to the range of the observations (Figure \ref{fig:fit veneto}(e)).
Moreover, the high number of parameters in the DMPseas penalizes it
in terms of BIC (689.2245), whose value is larger than the corresponding values for
the BeGBM$_{\rm RECT}$ and DMP.

As highlighted in Section \ref{sec:data}, the complete time series with the daily number of
processed swabs is available. Figure \ref{fig:data_tamponi}(a) shows its values (right $y-$axis)
in relation to the number of daily confirmed cases (left $y-$axis). We notice
the good agreement between the paths of the two series, and the correspondence of their peaks
suggests that this relationship could be exploited.
The DMPsw (Figure \ref{fig:fit veneto}(f) and Table \ref{tab:venetoTMP}) performs
very well. In fact, we obtained  the largest values  for $R^2,$ 0.999898,
and $\rho^2,$ 0.858459, for this model as well as the lowest BIC value, 641.4728. The latter value, in particular, is reduced
by the small number of parameters of this model.
Clearly, the strict comparison between the BIC of this model
and the values obtained for the other models should take into account that,
in this model, the complete  series  of
 processed  daily swabs, $B(t),$  is
 used  as an input  to
the model. However, this information is  available, and  the performance of  the  model
suggests
it is  useful for achieving an accurate description. These results suggest that the daily number of
cases in Veneto followed an asymmetric trend, as
modelled by a DMP model, but large fluctuations around that trend can be observed as a
consequence of different numbers of swabs processed each day.
We underline that the forecasts displayed in Figure \ref{fig:fit veneto}(f) have been obtained
assuming that the number of swabs processed in the last week will be repeated in the subsequent
three weeks.
\begin{table}
\centering
\caption{Veneto.      Estimates, asymptotic standard errors and $95\%$ mCIs for the parameters of the           DMPsw  {\rm (\ref{eq:ggmwt})+(\ref{eq:wtamp})}.}
\label{tab:venetoTMP}
\begin{tabular}{crrr@{$\!\;\;$}l}
Parameter    &  Estimate  &  Standard Error  &    \multicolumn{2}{c}{Confidence Interval}    \\
\hline
$m$  &  19932.21  & 170.5066  &  (19591.88, & 20272.55)\\
$p_c$&  0.000321  & 0.000057  &  (0.000207, & 0.000435)\\
$q_c$&  0.235024  & 0.006705  &  (0.221640, & 0.248408)\\
$p$  &  0.016718  & 0.004224  &  (0.008286, & 0.025150)\\
$q$  &  0.033005  & 0.010126  &  (0.012794, & 0.053216)\\
$\xi$&  0.468809  & 0.088953  &  (0.291258, & 0.646360)\\
\hline
 \end{tabular}
\end{table}

With the fitted models, we also obtained the estimate
of the total number of infected people during the whole epidemic: that is, $N$ for
the SIRD model
and $m$ for the other models. Both the LOG and the SIRD models, which are the lower
performing models, provided smaller estimates ($\hat{m}=18270$ and $\hat{N}=17586$,
respectively).  The
estimates obtained for $m$ in the remaining models are very
similar,
ranging from 19432 (GBM$_{\rm RECT}$) to 20085 (BeGBM$_{\rm RECT}$).
In particular, for the DMPsw model, $\hat{m}$=19932.
Notice that this model predicts 1162 (+6.4\%) and 1846 (+10\%) more cases
 than the LOG and the SIRD models, respectively.
Were the lockdown policy of Phase 1 confirmed after May 3rd,
the estimate of $m$ given by the DMPsw model suggests that Veneto experienced, by May 3rd,
92\% of all expected cases (the total number of cases until May 3rd was 18318).


%
%
\subsection{Lombardy}
\begin{figure}[t]
\centering
\begin{tabular}{cc}
\hspace{-10mm} (a) & \hspace{-8mm} (b) \\
\hspace{-8mm} \includegraphics[width=73mm]{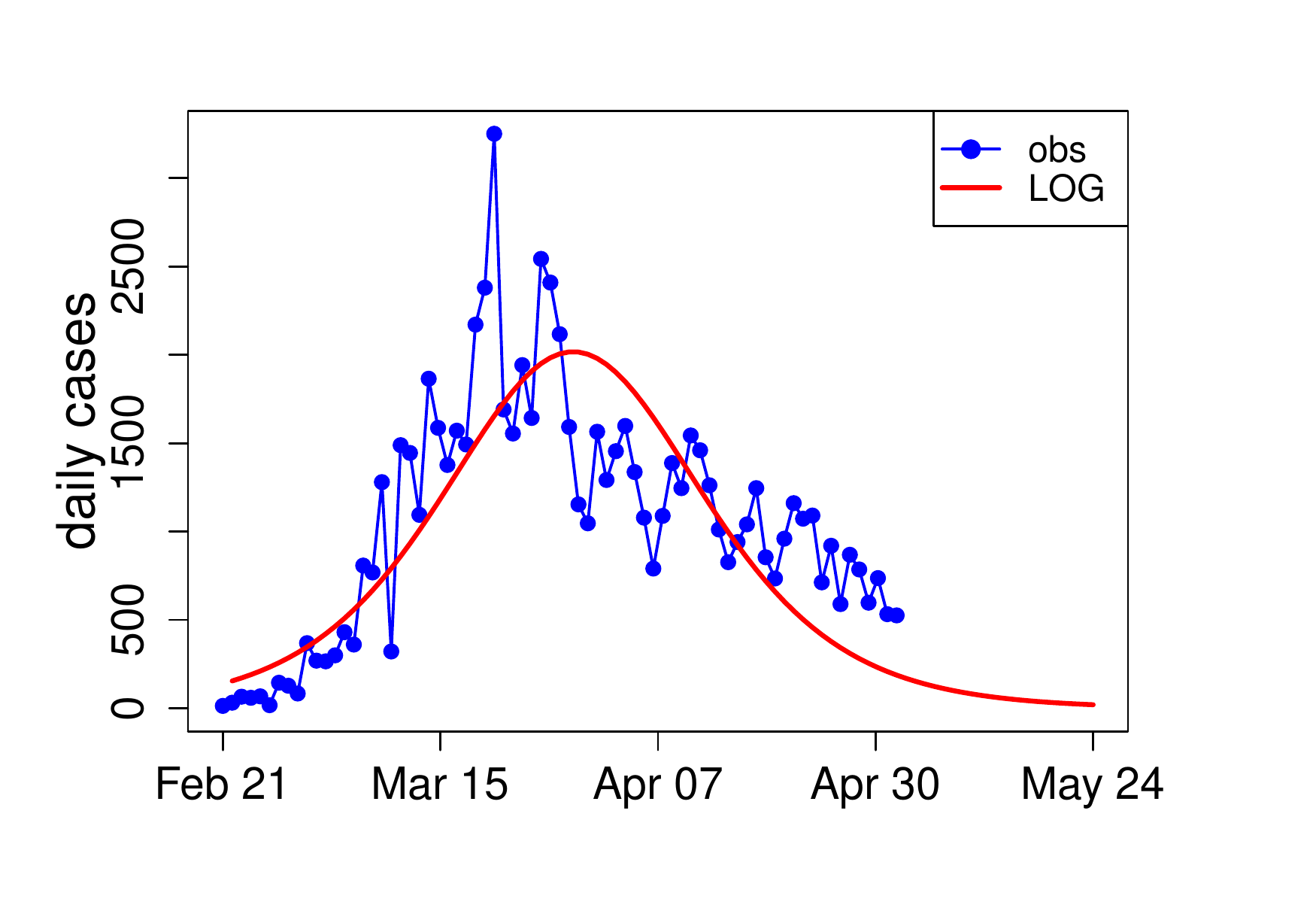} &
\hspace{-10mm} \includegraphics[width=73mm]{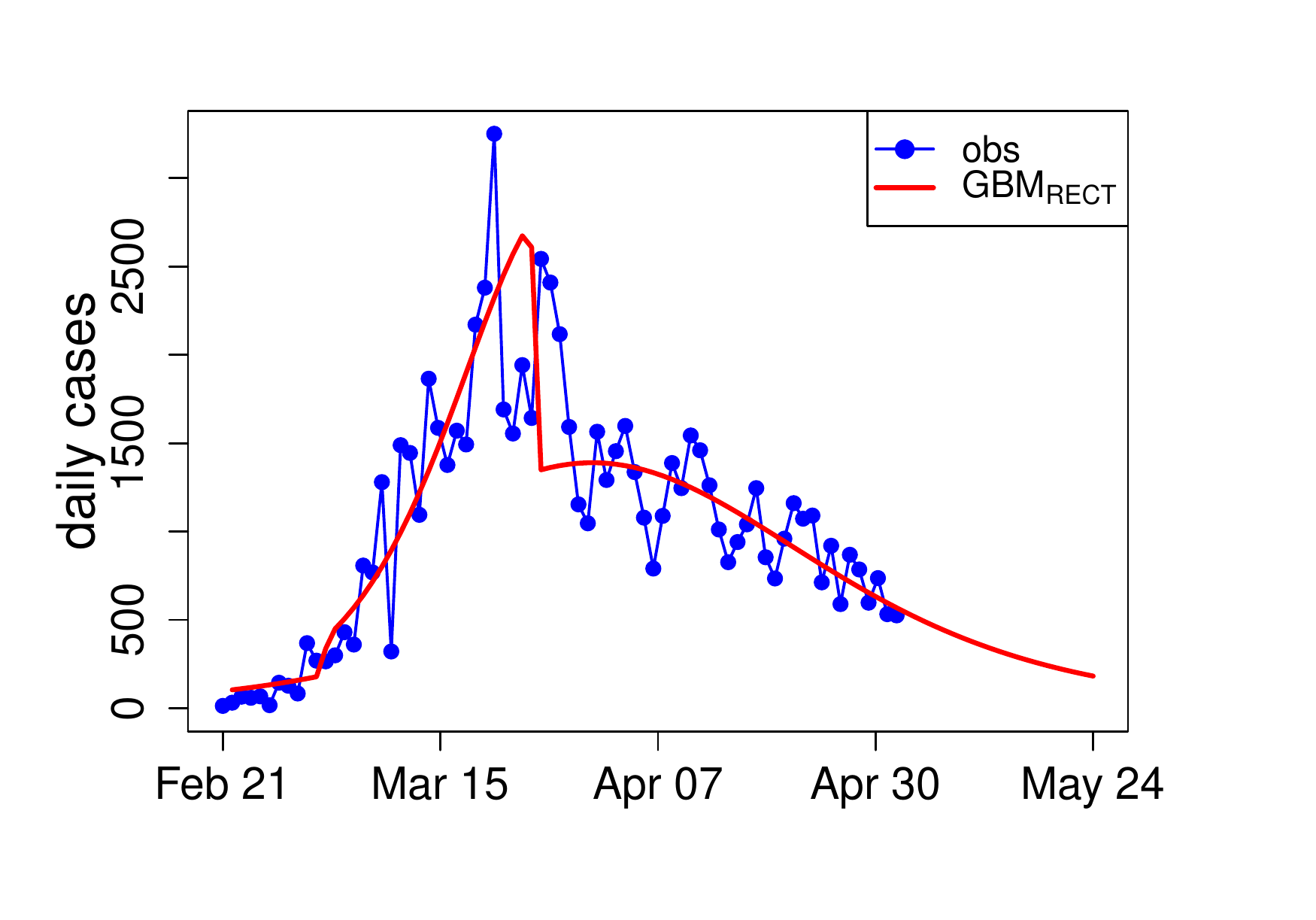}\\
\hspace{-8mm}(c) &  \hspace{-8mm} (d) \\
\hspace{-8mm} \includegraphics[width=73mm]{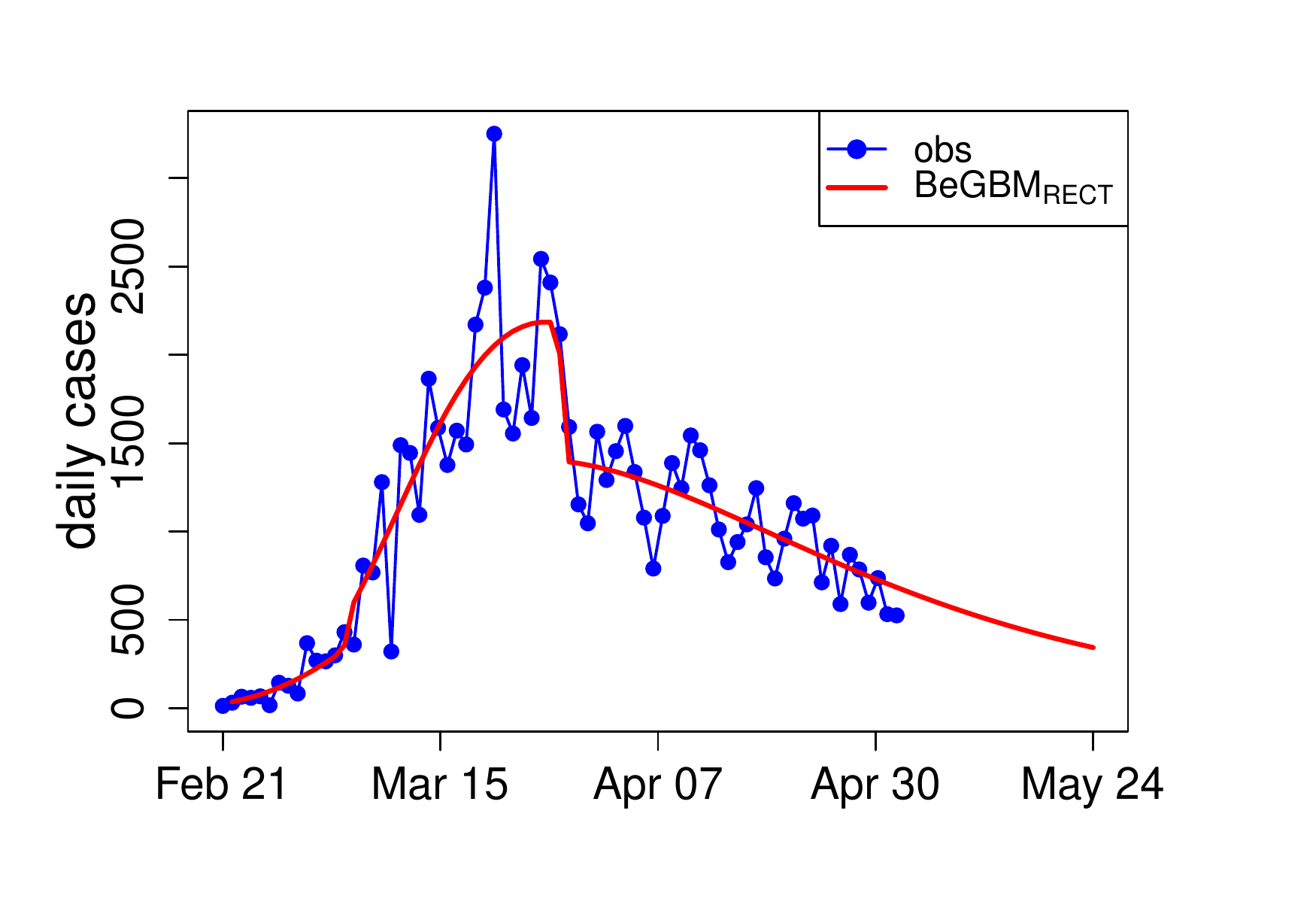} &
\hspace{-8mm}\includegraphics[width=73mm]{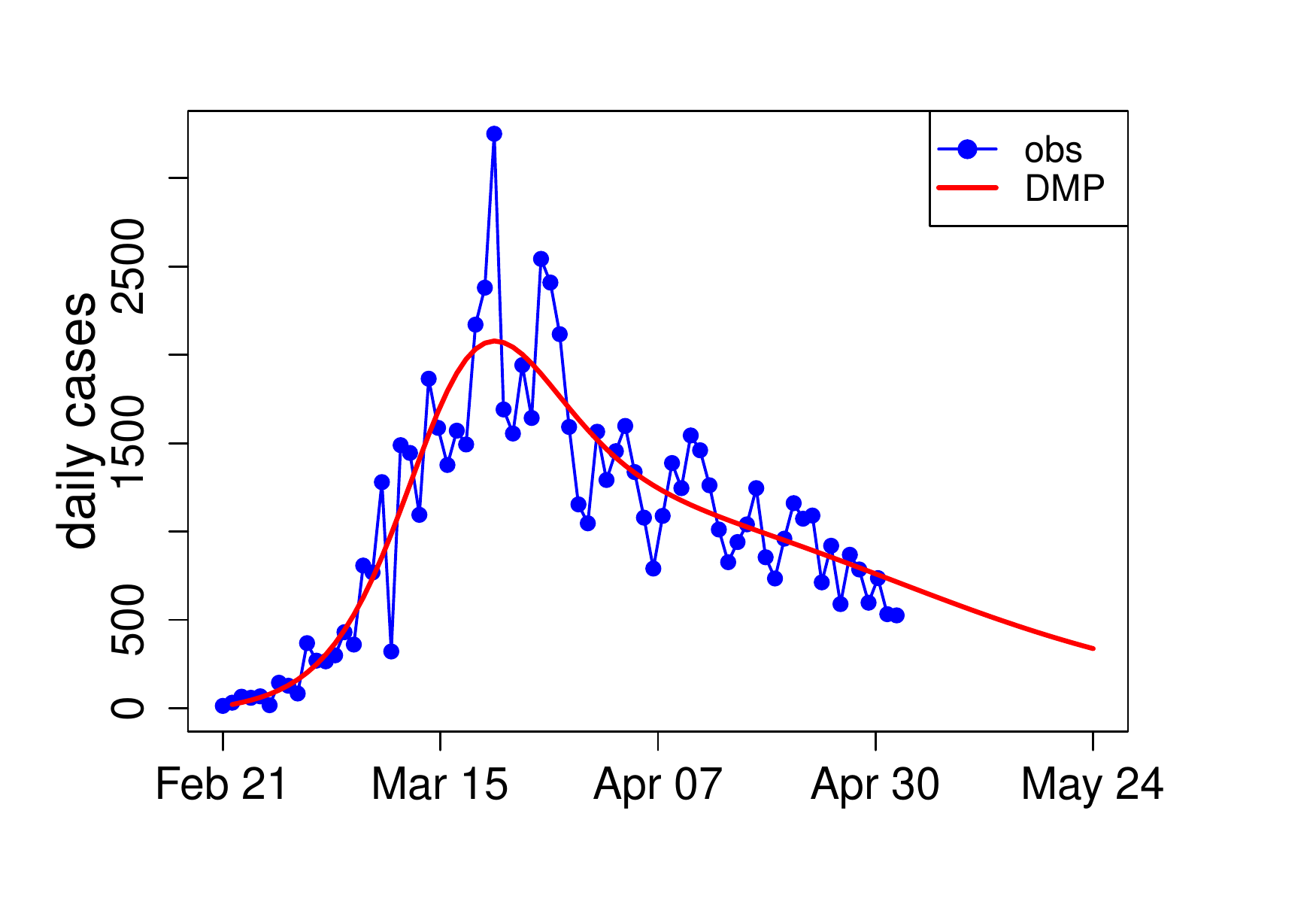}\\
\hspace{-10mm} (e) &  \hspace{-8mm} (f) \\
\hspace{-10mm} \includegraphics[width=73mm]{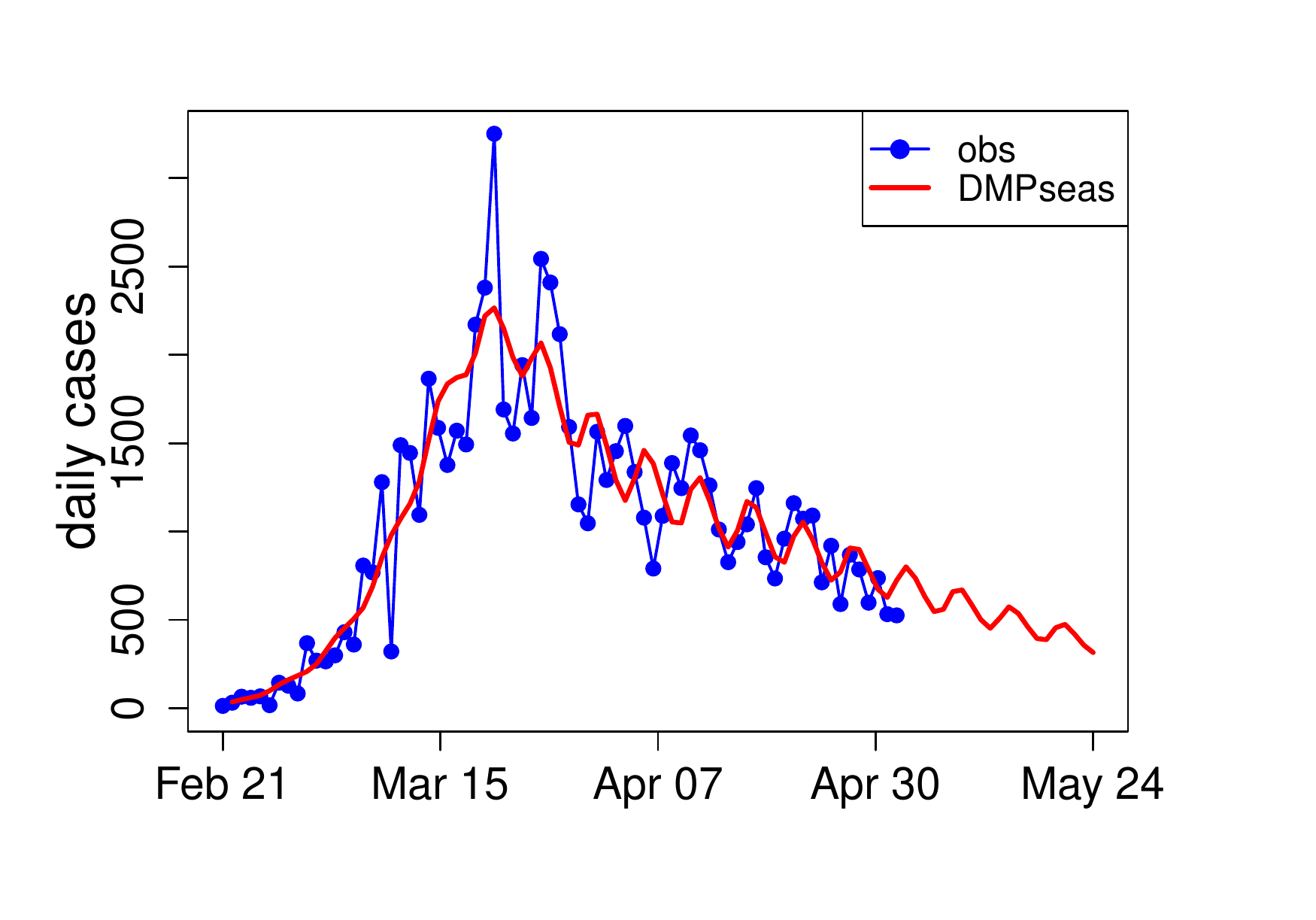} &
 \hspace{-8mm} \includegraphics[width=73mm]{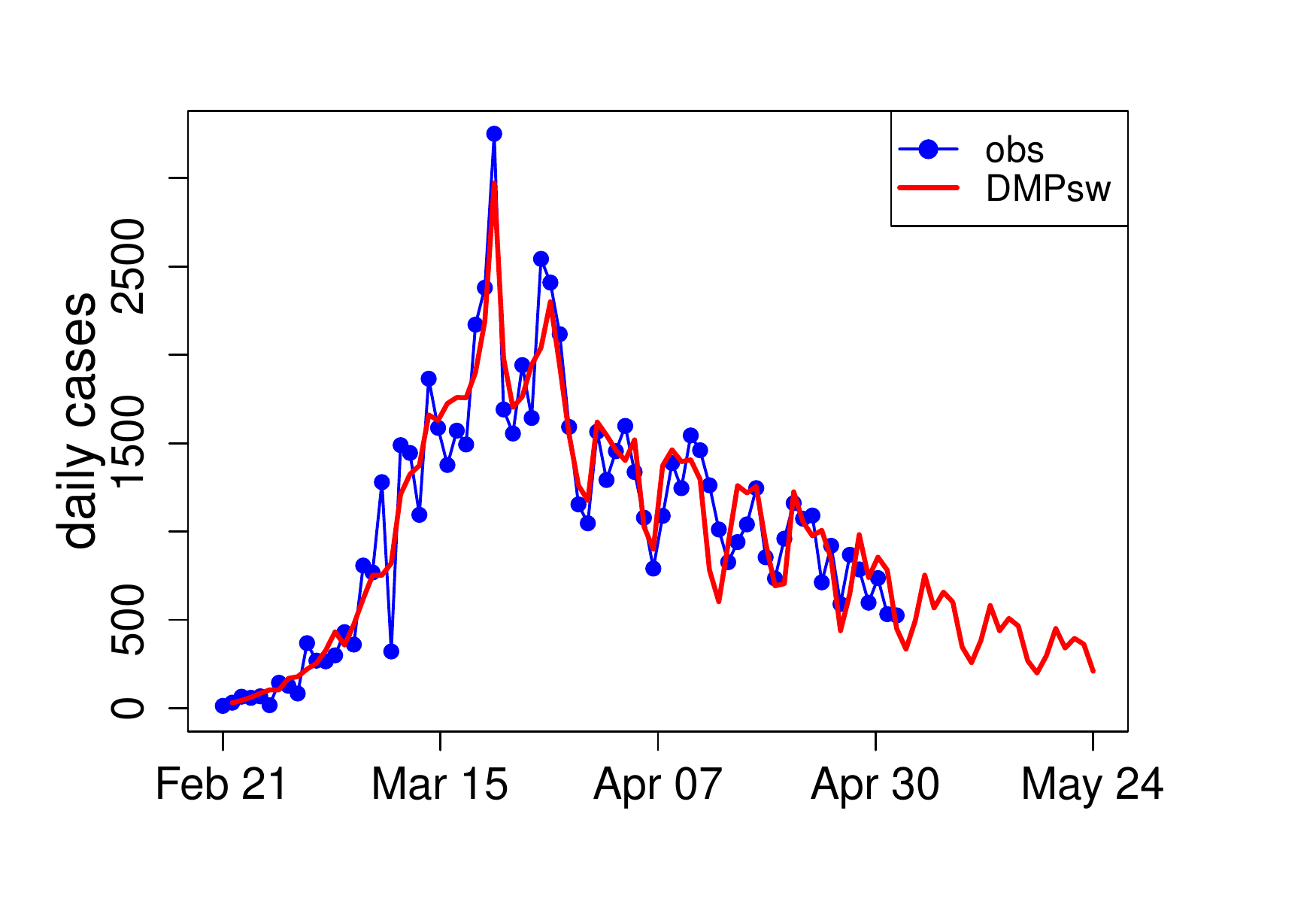}  \\
\end{tabular}
\caption{Lombardy. Observed and fitted values with the alternative models.
{\rm (a)} Logistic                                                 (LOG);
{\rm (b)} GBM with rectangular shock                       (GBM$_{\rm RECT}$);
{\rm (c)} Bemmaor GBM with rectangular shock               (BeGBM$_{\rm RECT}$);
{\rm (d)} Dynamic market potential                   (DMP);
{\rm (e)} Dynamic market potential+seasonal effect (DMPseas);
{\rm (f)} Dynamic market potential+swabs           (DMPsw).}
\label{fig:fit lombardia}
\end{figure}

Lombardy is the Italian region where COVID-19 spread in the most dramatic way. The total number of
infected people on May 3rd was 77528 with more than 14000 deaths (about half of the death toll up to that date in Italy as a whole).
The results for Lombardy are displayed in Table \ref{tab:R2} ($R^2,$ BIC and $\rho^2$), in Tables \ref{tab:lombardiaTMP},
\ref{tab:lombardiaLOG}--\ref{tab:lombardiaSIRD} (for parameter estimates for
all the models fitted) and in Figures \ref{fig:data_SIRD}(b) and \ref{fig:fit lombardia},
where observed and fitted daily values are plotted.

For this region, the logistic (Figure \ref{fig:fit lombardia}(a))
and  SIRD (Figure \ref{fig:data_SIRD}(b)) models are less effective
models in describing  the
asymmetrical evolution of the epidemic. For the SIRD model, in particular,
a good convergence point could not be attained.

The results in Tables \ref{tab:lombardiaGBM} and \ref{tab:lombardiaBEGBM} show that, as observed for
Veneto, a  positive ($\hat{c}>0$)
rectangular shock is significantly diagnosed at the beginning of the time series, both in the
GBM$_{\rm RECT}$ and the BeGBM$_{\rm RECT}$. The GBM$_{\rm RECT}$ estimates the end of the shock on
 March 25th ($t\simeq$ 34), but according to Figure \ref{fig:fit lombardia}(b),
this is not perfectly matching with the data. This is the reason why, for this model,
$\rho^2$ is particularly small (0.690817).\footnote{The $R^2$ is evaluated on cumulative
cases, which are the response variable. Since cumulative cases are measured on a much
larger scale, discrepancies between fitted and observed values are less relevant on
the $R^2$ than they are on the daily values. In this case, the lack-of-fit around the
peak heavily penalizes the $\rho^2.$}. \\
Conversely, the BeGBM$_{\rm RECT}$ better identifies
the end of the shock three days later,
on March 28th ($t\simeq$ 37), when we observe a relevant stable decrease.
For this region, the lockdown policy had a delayed effect compared to what happened in Veneto,
as the decrease in the number of cases was registered 20 days after March 8th, while the incubation period is
up to 14 days. One reason for such a wider interval could be possible delays in taking
  \clearpage
  \noindent
and processing the swabs; in fact, the health system of Lombardy
experienced an unexpected overload.

With the DMP model (Figure \ref{fig:fit lombardia}(d) and Table \ref{tab:lombardiaDMP}),
it is possibile to fully appreciate the asymmetrical shape of the outbreak, especially
the slow
 decrease in the number of cases in this region. However, its performance in terms of
$R^2,$ $\rho^2$ and BIC is worse than that of the BeGBM$_{\rm RECT}.$

The performance of the DMPseas, with a weekly cycle ($\hat{s}=7.005$ days)
(Table \ref{tab:lombardiaSTAG}), is not satisfactory, as it
 does not adequately capture
  the fluctuations (except for the very end of the series).
Here, too,
 the $R^2,$ $\rho^2$
 and BIC values are worse than those obtained with the BeGBM$_{\rm RECT}.$

Finally, the DMPsw (Figure \ref{fig:fit lombardia}(f) and Table \ref{tab:lombardiaTMP})
 performs very well. With this model, we obtained the largest values for $R^2,$ 0.999919,
 and $\rho^2,$ 0.902698.
The BIC value for this model, 830.5628, supports it with respect to the BeGBM$_{\rm RECT}$
(850.6930), which was the best model up to this point.
Figure
\ref{fig:data_tamponi}(b) shows the number of swabs (right $y-$axis)
in relation to daily cases (left $y-$axis). For this region, too, there is a
great agreement between the paths of the two series, with almost perfect correspondence
of their peaks. By comparing panels (a) and (b) of Figure \ref{fig:data_tamponi},
we can appreciate the differences in swab policies adopted by Veneto and Lombardy.
The latter region, which has about twice the number of inhabitants as Veneto,
processed on average 5705 swabs each day, which was not much more
than the average  in Veneto (5250),
even though Lombardy experienced more than four times the number of
\emph{officially} diagnosed people compared to Veneto.

\begin{table}
\centering
\caption{Lombardy.      Estimates, asymptotic standard errors and $95\%$ mCIs for the parameters of the           DMPsw  {\rm (\ref{eq:ggmwt})+(\ref{eq:wtamp})}.}
\label{tab:lombardiaTMP}
\begin{tabular}{crrr@{$\!\;\;$}l}
Parameter    &  Estimate  &  Standard Error  &    \multicolumn{2}{c}{Confidence Interval}    \\
\hline
$m$  &  95017.78  &  3787.978  &  (87456.94, & 102578.6.)\\
$p_c$&  0.000460  &  0.000079  &  (0.000302, & 0.000617)\\
$q_c$&  0.221508  &  0.006106  &  (0.209320, & 0.233696)\\
$p$  &  0.027625  &  0.002594  &  (0.022447, & 0.032804)\\
$q$  & -0.007438  &  0.008828  & (-0.025059, & 0.010184)\\
$\xi$&  0.537777  &  0.050836  &  (0.436308, & 0.639245)\\
\hline
 \end{tabular}
\end{table}

The estimates obtained for $m$ in
BeGBM$_{\rm RECT}$, DMP, DMPseas and DMPsw substantially agree, ranging from 95018 (DMPsw)
to 98723 (DMPseas). As underlined at the beginning of this subsection, the total number
of cases until May 3rd was 77528. The difference between this number and
$\hat{m}$=95018 of the DMPsw is quite large, confirming that this region, by May 3rd,
started Phase 2 in a riskier context than Veneto,
having experienced only 82\% of all expected cases.

\begin{figure}[t]
\centering
\begin{tabular}{cc}
\hspace{-10mm} (a) & \hspace{-8mm} (b) \\
\hspace{-8mm} \includegraphics[width=73mm]{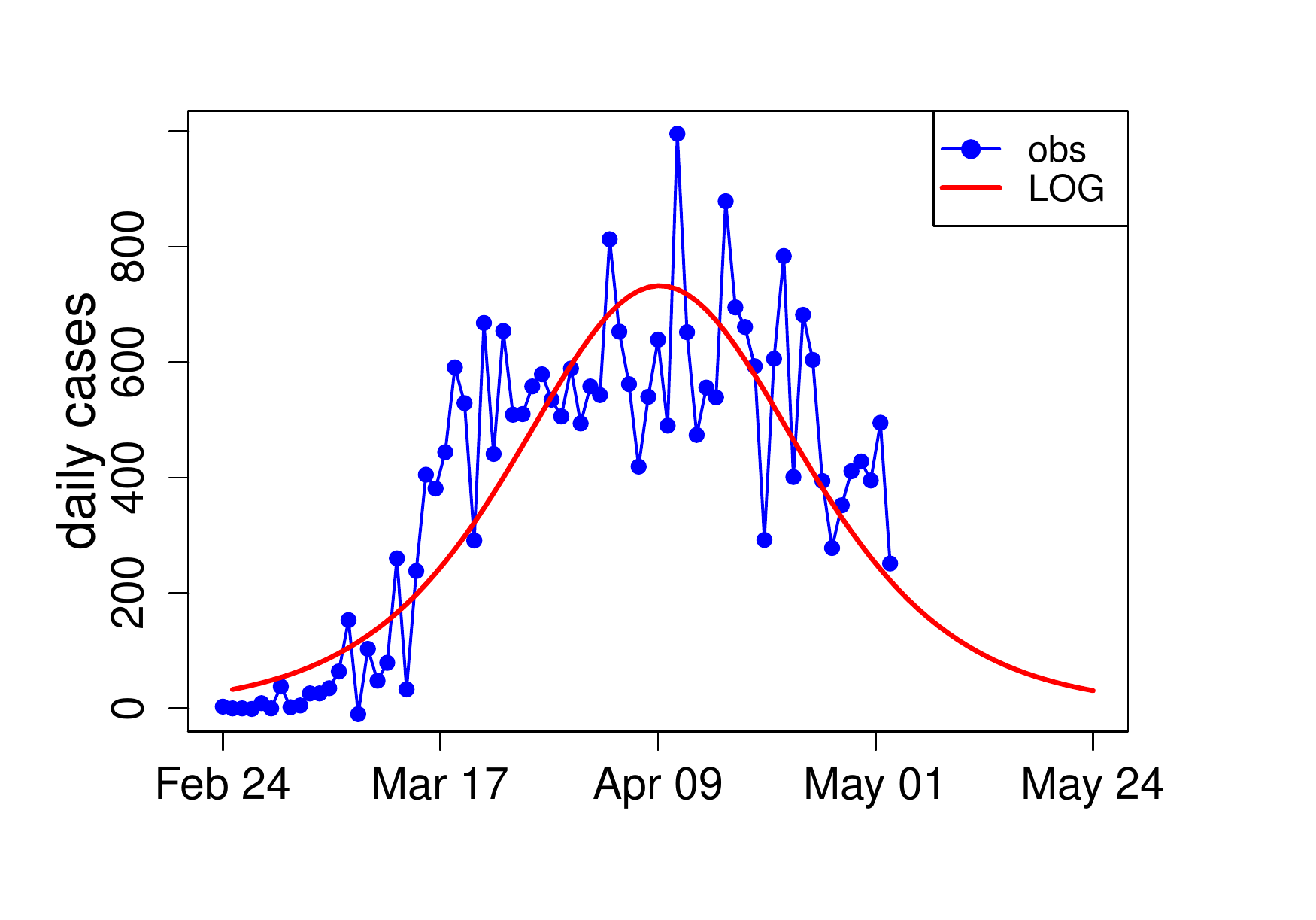} &
\hspace{-10mm} \includegraphics[width=73mm]{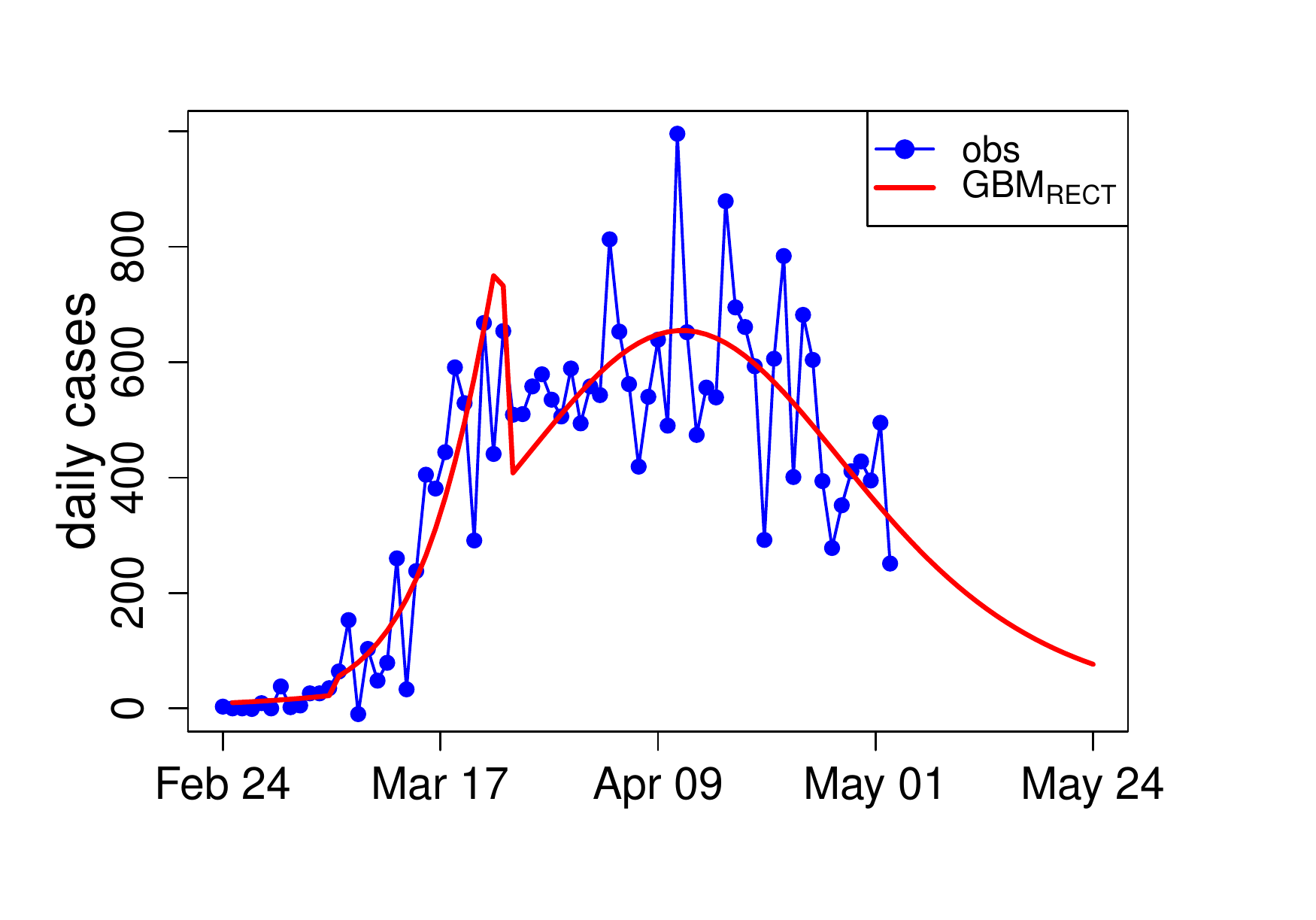}\\
\hspace{-8mm}(c) &  \hspace{-8mm} (d) \\
\hspace{-8mm} \includegraphics[width=73mm]{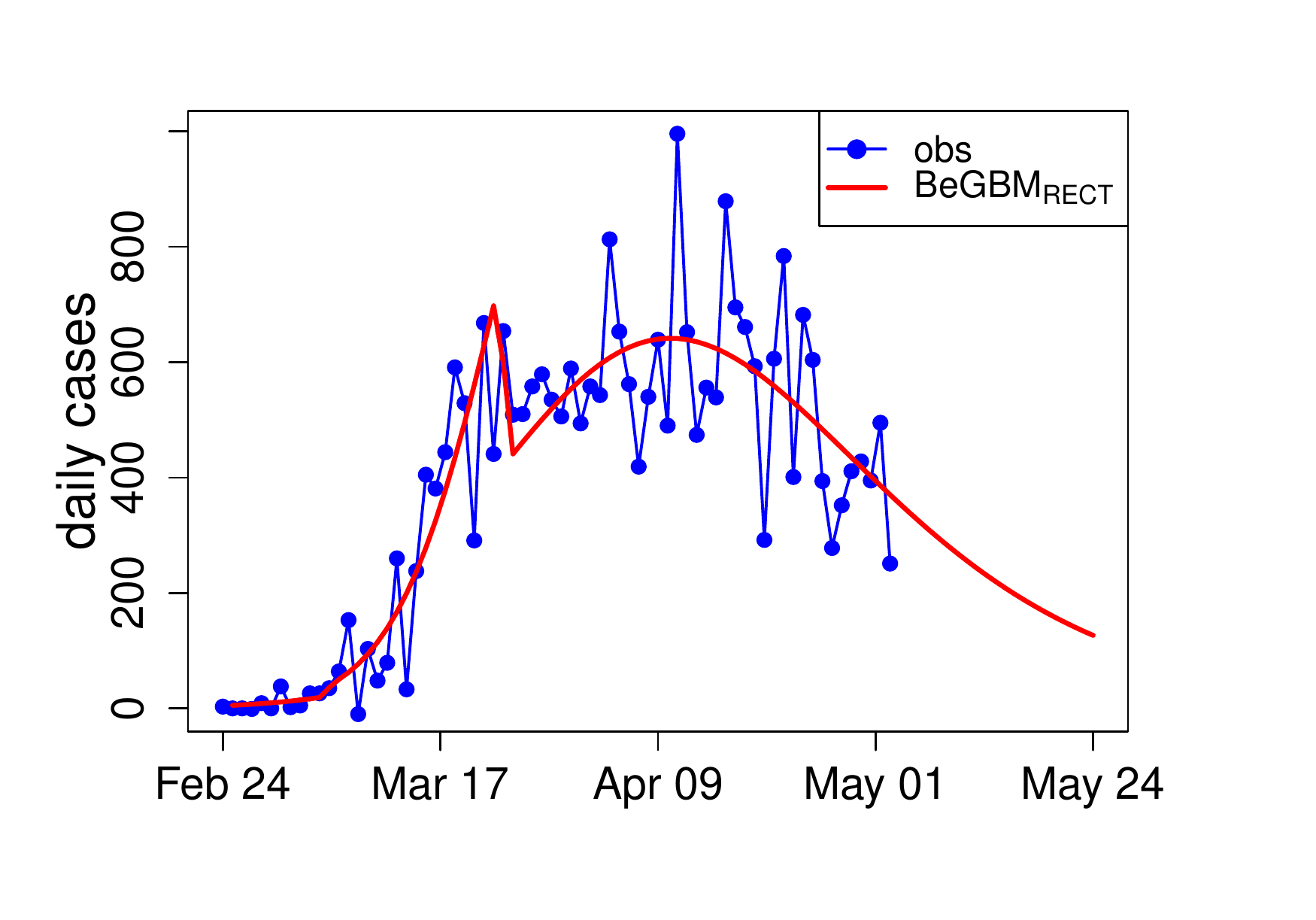} &
\hspace{-8mm}\includegraphics[width=73mm]{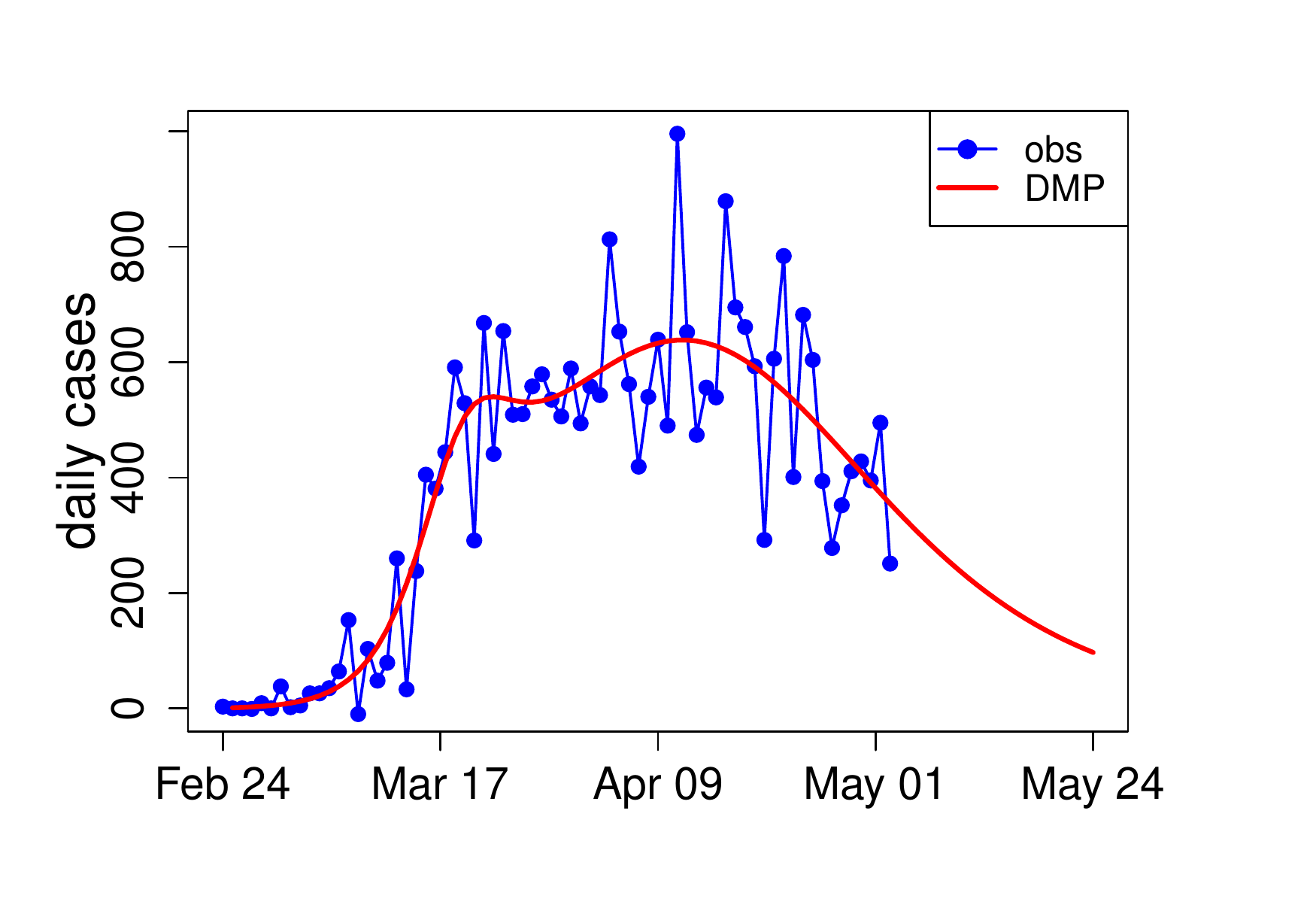}\\
\hspace{-10mm} (e) &  \hspace{-8mm} (f) \\
\hspace{-10mm} \includegraphics[width=73mm]{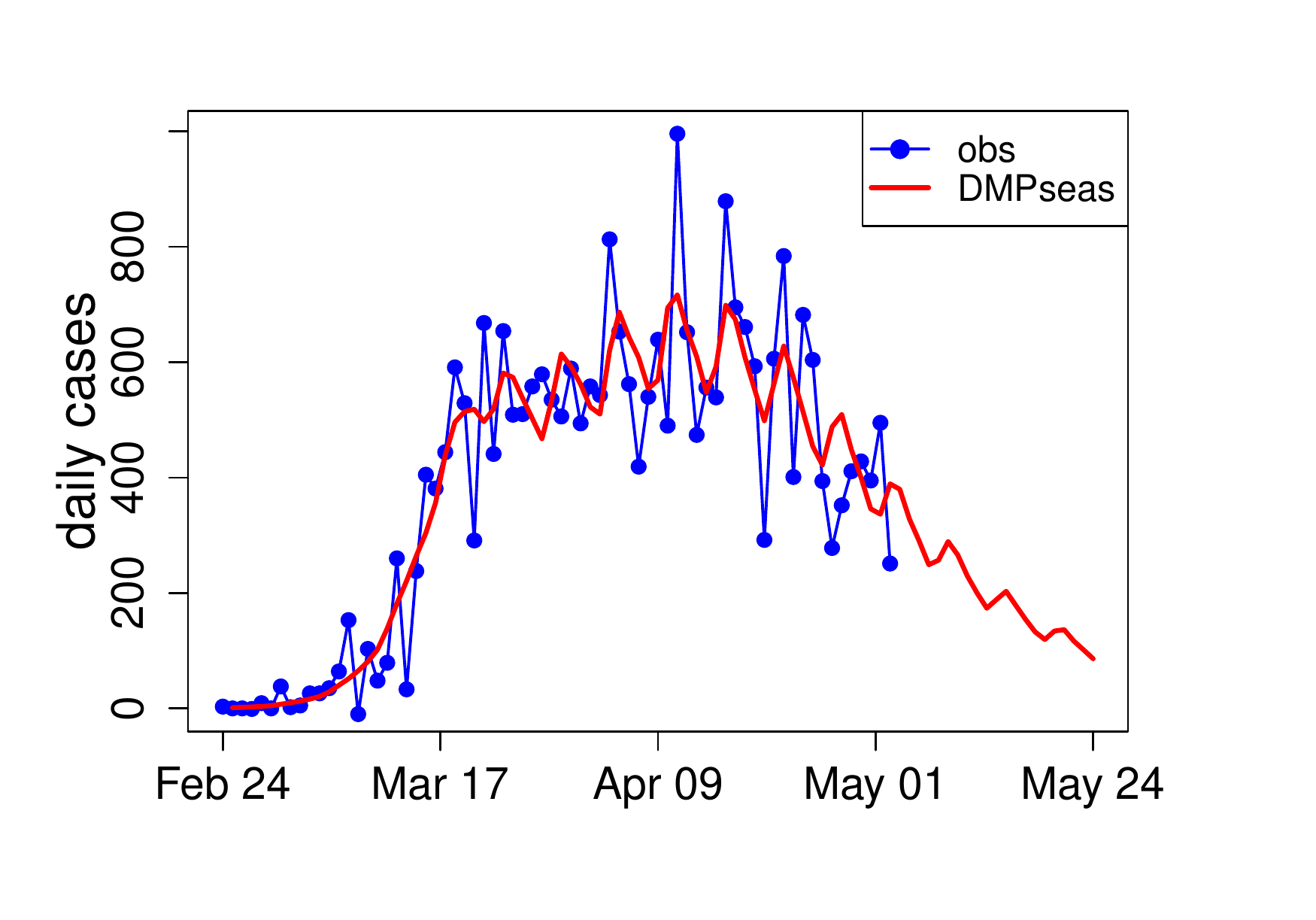} &
 \hspace{-8mm} \includegraphics[width=73mm]{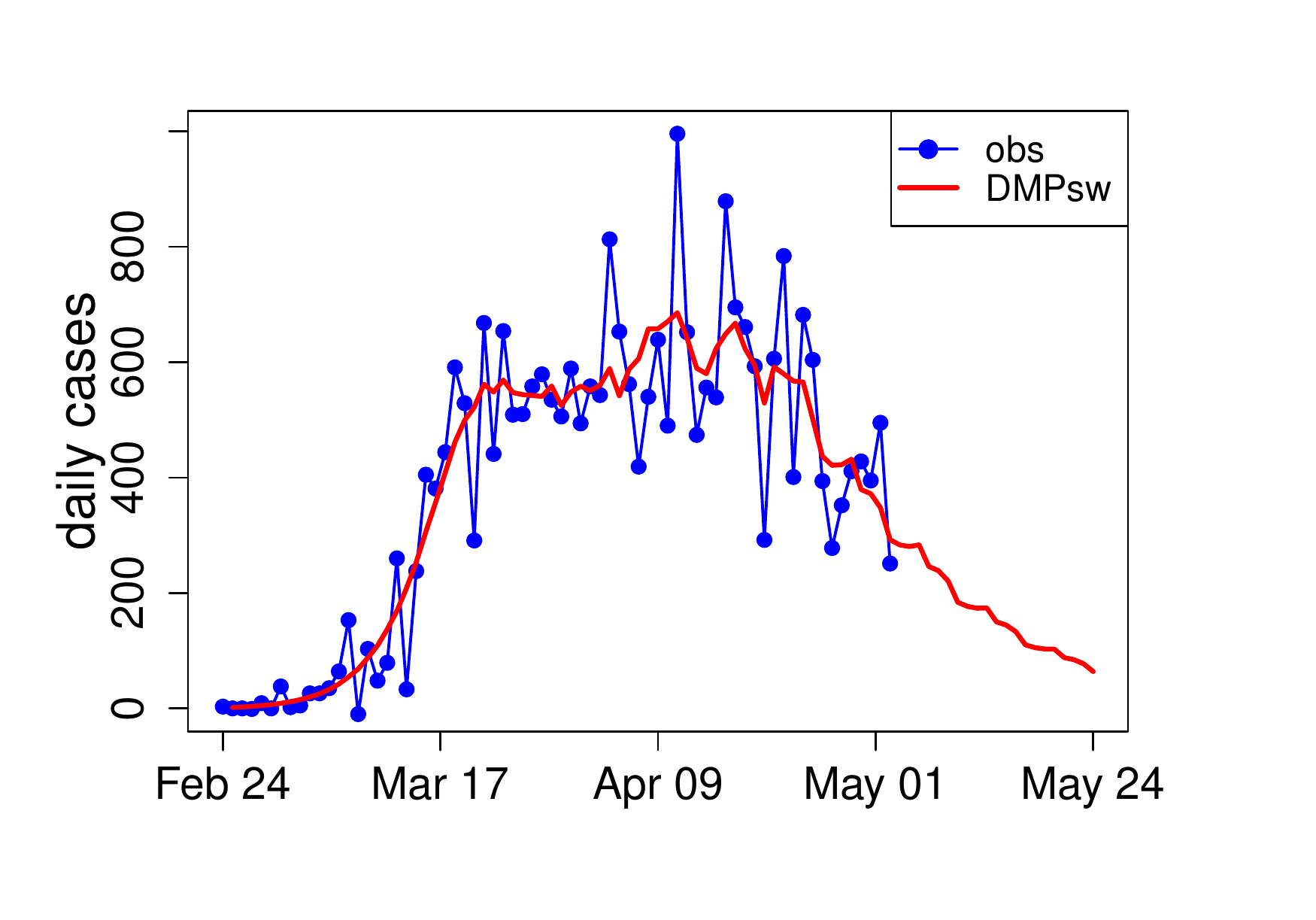}  \\
\end{tabular}
\caption{Piedmont. Observed and fitted values with the alternative models.
{\rm (a)} Logistic                                                 (LOG);
{\rm (b)} GBM with rectangular shock                       (GBM$_{\rm RECT}$);
{\rm (c)} Bemmaor GBM with rectangular shock               (BeGBM$_{\rm RECT}$);
{\rm (d)} Dynamic market potential                   (DMP);
{\rm (e)} Dynamic market potential+seasonal effect (DMPseas);
{\rm (f)} Dynamic market potential+swabs           (DMPsw).}
\label{fig:fit piemonte}
\end{figure}
\clearpage

\subsection{Piedmont}
Piedmont is the Italian region where the peak in diagnosed
cases
occurred
later
(see
Figure \ref{fig:data}),
as,
 at
 the
 beginning
of
April,
one month after the lockdown,
we still observed the highest number of daily
cases.
This could be due to the limited swabbing capacity  in the first part of the epidemic, when
swabs never exceeded 400 per day (see Figure \ref{fig:data_tamponi}(c)).
The results for Piedmont are displayed in Table \ref{tab:R2} ($R^2,$ BIC and $\rho^2$),
in Tables \ref{tab:piemonteTMP},
\ref{tab:piemonteLOG}--\ref{tab:piemonteSIRD} (for parameter estimates for
all the models fitted) and in Figures \ref{fig:data_SIRD}(c) and \ref{fig:fit piemonte},
where observed and fitted daily values are plotted.

Also for this region, the logistic (Figure \ref{fig:fit piemonte}(a))
and SIRD (Figure \ref{fig:data_SIRD}(c)) models are the poorest performing
models in terms of describing  the
asymmetrical
evolution of the epidemic.

The results in Tables \ref{tab:piemonteGBM} and \ref{tab:piemonteBEGBM} show that here
also  
a  positive ($\hat{c}>0$)
rectangular shock is significantly diagnosed at the beginning of the time series,
both in the GBM$_{\rm RECT}$ and the BeGBM$_{\rm RECT}$.
The two models provide the same estimate for the end of the shock on
March 24th ($t\simeq$ 30), exactly as observed for Veneto.
However, after the end of shock, the number of daily cases continued to increase, although at a slower rate
(see Figure \ref{fig:fit piemonte}(b) and (c)).

The DMP model (Figure \ref{fig:fit piemonte}(d) and Table \ref{tab:piemonteDMP})
enables describing, without shocks, the bimodal behaviour of this time
series: the `saddle',
which is the slowdown between two relative peaks,
is exactly positioned immediately after the end of the shocks estimated with
the GBM$_{\rm RECT}$ and the BeGBM$_{\rm RECT}$.
If, on the one hand, the lockdown policy had the effect of reducing the number
of daily cases (saddle after the first peak), on the other hand, the second peak is due to
 the increase in the number of swabs after April 8th, which made it possible to detect more infected
 people.
The DMP for this region performs very well
($R^2$=0.99988), with a small BIC value, also due to the parsimony of a model with only five parameters.

The DMPseas, with a weekly cycle ($\hat{s}=7.01$ days)
(Table \ref{tab:piemonteSTAG}), describes the frequency of the fluctuations up to the second half of
April,
with an insufficient amplitude throughout the entire observed period.
For this region, however, we observe the largest $R^2$ value among all the previous models,
$R^2$=0.999895, although the BIC value is larger than that observed for the simpler DMP.

The DMPsw (Figure \ref{fig:fit piemonte}(f) and Table \ref{tab:piemonteTMP}) provides the
largest values for $R^2,$ 0.999905,
 and $\rho^2,$ 0.843469. The BIC value for this model, 658.7328, supports it with respect
 to all other examined models. The width of the fluctuations in the observed series is not,
 however,
 fully described by this model (Figure \ref{fig:fit piemonte}(f)). This is also apparent
 from the value of
 $\hat{\xi}$=0.13 (Table \ref{tab:piemonteTMP}), which is lower compared to the estimates for Veneto (0.469)
 and Lombardy (0.538).

\begin{table}
\centering
\caption{Piedmont.      Estimates, asymptotic standard errors and $95\%$ mCIs for the parameters of the           DMPsw  {\rm (\ref{eq:ggmwt})+(\ref{eq:wtamp})}.}
\label{tab:piemonteTMP}
\begin{tabular}{crrr@{$\!\;\;$}l}
Parameter    &  Estimate  &  Standard Error  &    \multicolumn{2}{c}{Confidence Interval}    \\
\hline
$m$  &  31620.64  &  323.5048  &  (30974.37, & 32266.92)\\
$p_c$&  0.000056  &  0.000014  &  (0.000028, & 0.000084)\\
$q_c$&  0.329662  &  0.013220  &  (0.303251, & 0.356073)\\
$p$  &  0.003141  &  0.000416  &  (0.002310, & 0.003972)\\
$q$  &  0.068384  &  0.001974  &  (0.064440, & 0.072327)\\
$\xi$&  0.130036  &  0.035910  &  (0.058298, & 0.201774)\\
\hline
 \end{tabular}
\end{table}

The estimates obtained for $m$ in
GBM$_{\rm RECT},$ BeGBM$_{\rm RECT}$, DMP, DMPseas and DMPsw
range from 31621 (DMPsw)
to 34351 (BeGBM$_{\rm RECT}$). Conversely,
we obtained $\hat{m}=28968$ for the LOG and $\hat{N}=28955$
for the SIRD; also for this region, the LOG and  SIRD models predictions are
smaller than for other models.
The total number
of cases until May 3rd was 27430. By comparing this value
to $\hat{m}$=31621 of the DMPsw, we notice that this region, by May 3rd,
experienced 87\% of all expected cases.

\subsection{Tuscany}
\begin{figure}[t]
\centering
\begin{tabular}{cc}
\hspace{-10mm} (a) & \hspace{-8mm} (b) \\
\hspace{-8mm} \includegraphics[width=73mm]{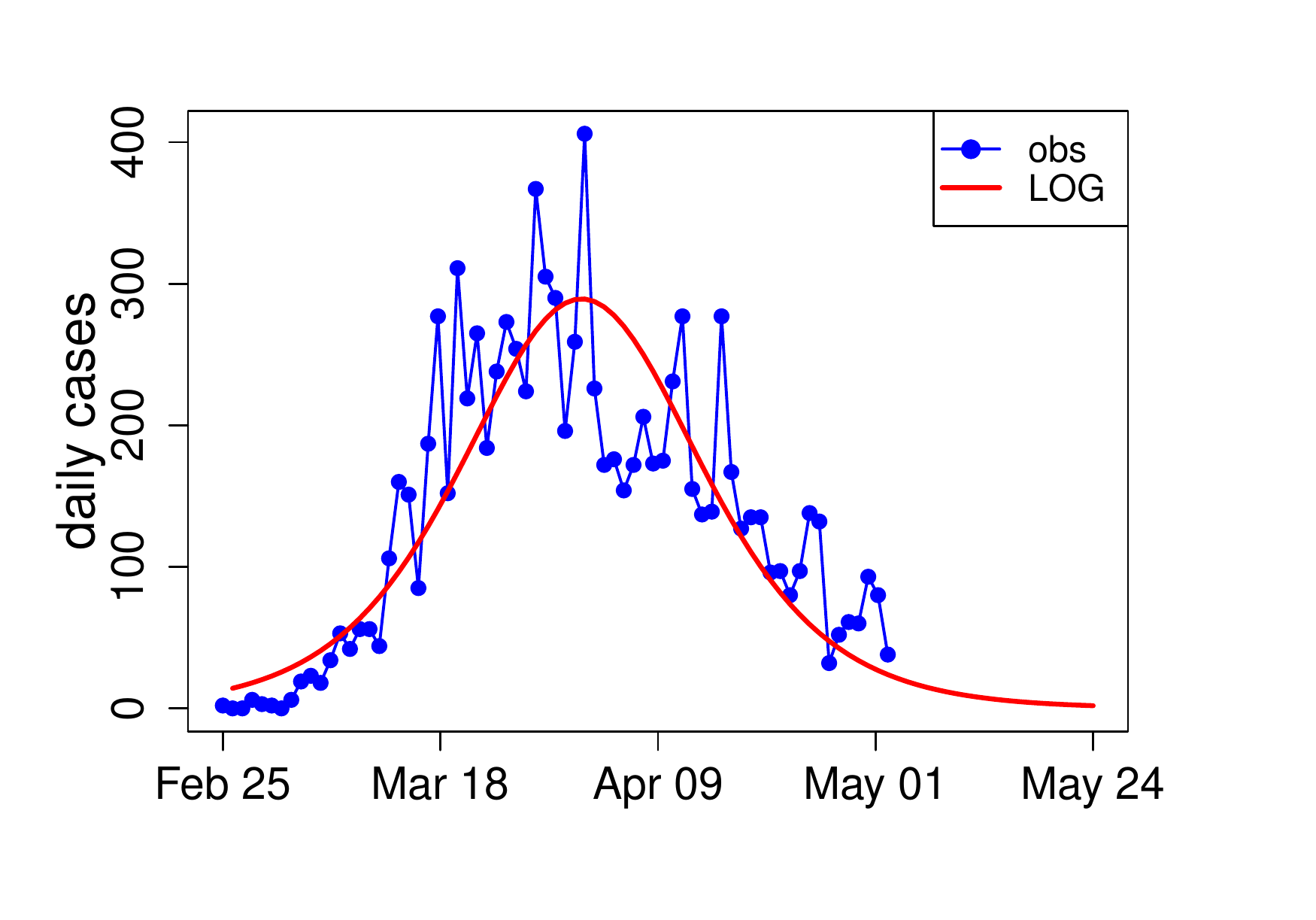} &
\hspace{-10mm} \includegraphics[width=73mm]{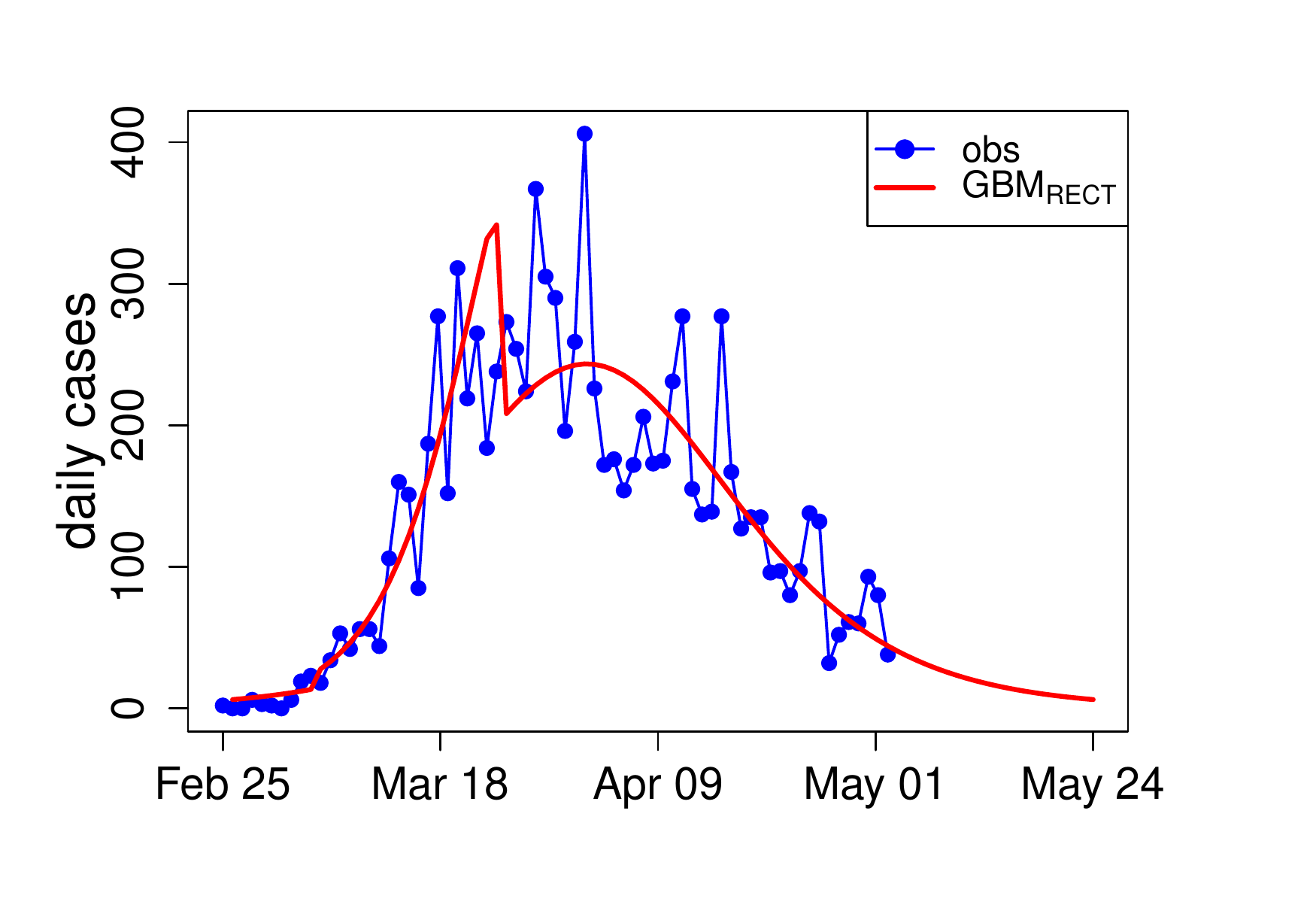}\\
\hspace{-8mm}(c) &  \hspace{-8mm} (d) \\
\hspace{-8mm} \includegraphics[width=73mm]{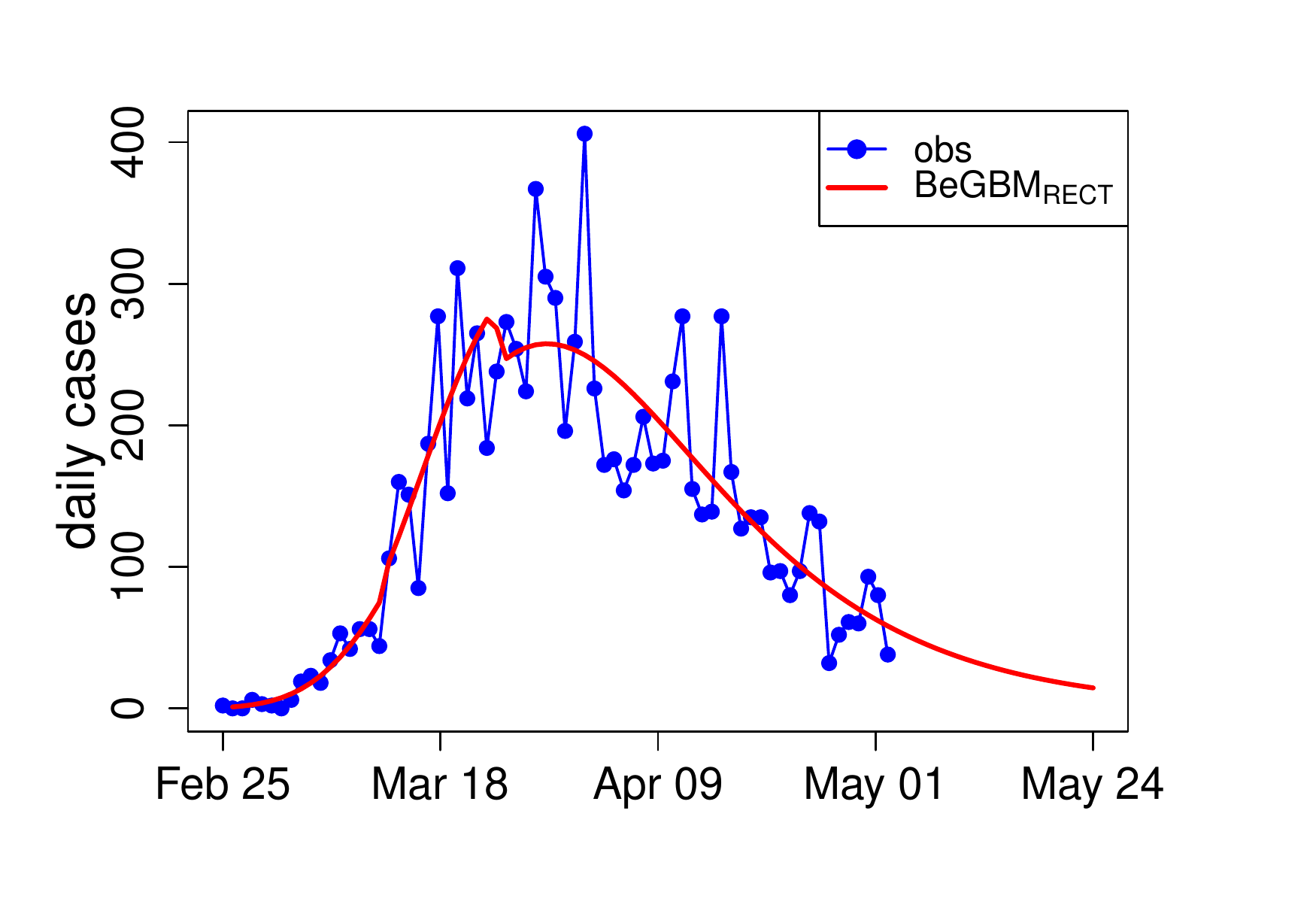} &
\hspace{-8mm}\includegraphics[width=73mm]{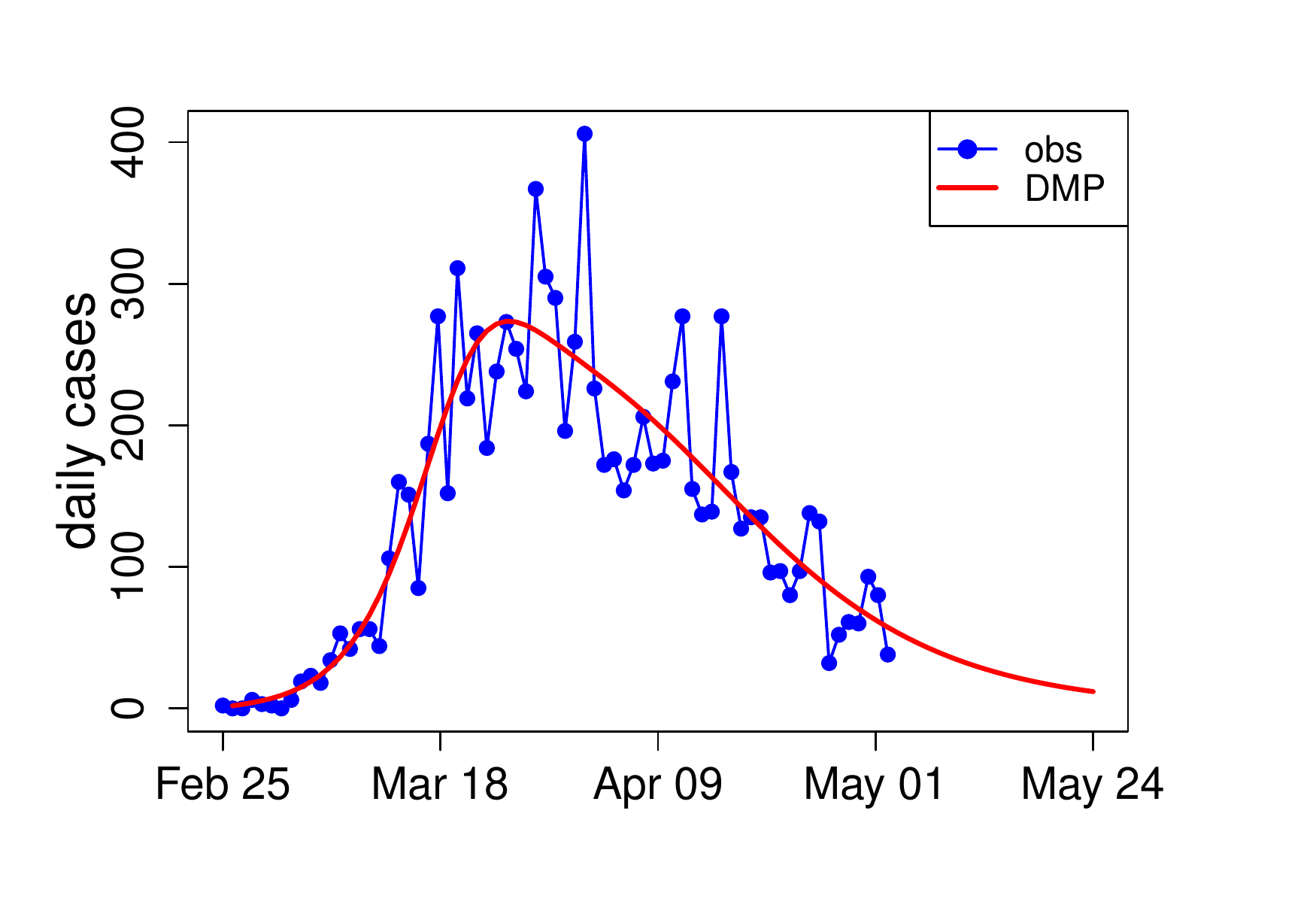}\\
\hspace{-10mm} (e) &  \hspace{-8mm} (f) \\
\hspace{-10mm} \includegraphics[width=73mm]{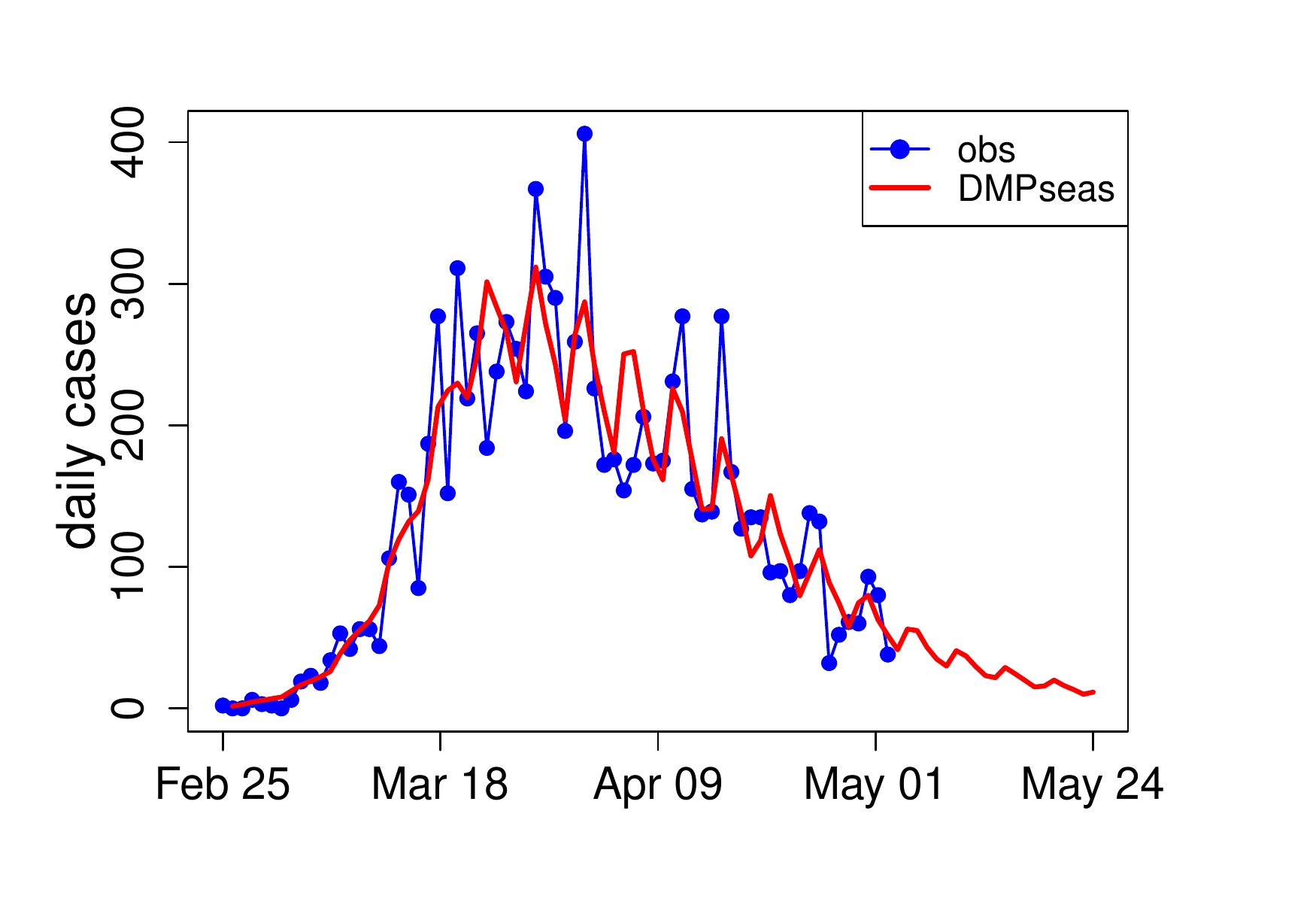} &
 \hspace{-8mm} \includegraphics[width=73mm]{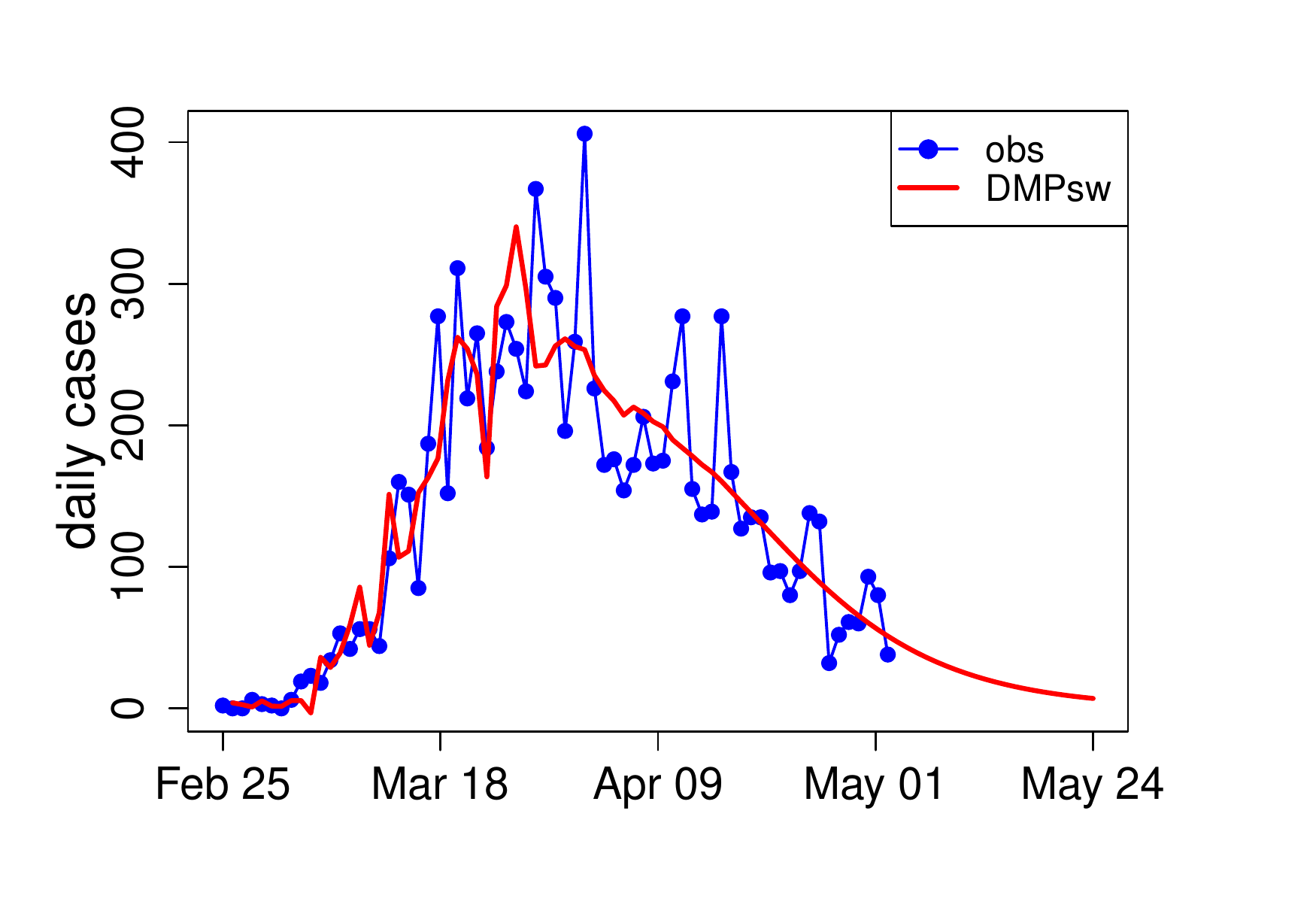}  \\
\end{tabular}
\caption{Tuscany. Observed and fitted values with the alternative models.
{\rm (a)} Logistic                                                 (LOG);
{\rm (b)} GBM with rectangular shock                       (GBM$_{\rm RECT}$);
{\rm (c)} Bemmaor GBM with rectangular shock               (BeGBM$_{\rm RECT}$);
{\rm (d)} Dynamic market potential                   (DMP);
{\rm (e)} Dynamic market potential+seasonal effect (DMPseas);
{\rm (f)} Dynamic market potential+swabs           (DMPsw).}
\label{fig:fit toscana}
\end{figure}

The results for Tuscany are displayed in Table \ref{tab:R2} ($R^2,$ BIC and $\rho^2$),
in Tables \ref{tab:toscanaSTAG}, \ref{tab:toscanaTMP},
\ref{tab:toscanaLOG}--\ref{tab:toscanaSIRD} (for parameter estimates for
all the models fitted) and in Figures \ref{fig:data_SIRD}(d) and \ref{fig:fit toscana},
where observed and fitted daily values are plotted.

Also for this region, the logistic (Figure \ref{fig:fit toscana}(a))
and SIRD (Figure \ref{fig:data_SIRD}(d)) models are the worst performing
models in describing  the
asymmetrical
evolution of the epidemic.

The results in Tables \ref{tab:toscanaGBM} and \ref{tab:toscanaBEGBM} show that
a  positive ($\hat{c}>0$)
rectangular shock is significantly diagnosed at the beginning of the time series,
both in the GBM$_{\rm RECT}$ and the BeGBM$_{\rm RECT}$.
The two models provide the same estimate for the end of the shock on
March 24th ($t\simeq$ 29), exactly as observed for Veneto and Piedmont. Notice that here the data start on February 25th because Toscana did not report cases earlier than that.
Differently from other regions, however, for the BeGBM$_{\rm RECT}$ the path is apparently less  perturbed
by the shock
(see Figure \ref{fig:fit toscana}(c)).

The DMP model (Figure \ref{fig:fit toscana}(d) and Table \ref{tab:toscanaDMP})
enables effectively describing the asymmetric behaviour of this time series without shocks.
The $R^2$ is
very high (0.999725), with a small BIC value equal to 570.1541, also
due to the parsimony of a model with five parameters
only.

The DMPseas, with a weekly cycle ($\hat{s}=6.982$ days)
(Table \ref{tab:toscanaSTAG}), well describes the frequency of the fluctuations
and, differently from other regions, this model is also able to describe their width
 (Figure \ref{fig:fit toscana}(e)).
The $R^2$ is equal to \hfill 0.999792, \hfill although \hfill the \hfill BIC \hfill value \hfill is \hfill larger \hfill than \hfill that \hfill observed \hfill for\hfill
the
\clearpage  
\noindent
simpler DMP.

The DMPsw (Table \ref{tab:toscanaTMP}) returns a value
for $R^2,$ 0.999796, slightly larger than that
observed for the DMPseas.
From Figure \ref{fig:fit toscana}(f), however, we can see that, after the peak,
fitted values are almost unaffected by changes in the
number of
daily swabs,  $B(t),$
although this  time   series
shows important  variations
in time
(see Figure \ref{fig:data_tamponi}(d)), and $\hat{\xi}$ has a large value (0.78) compared
to other regions.
 Both the confirmed cases and swab time series  exhibit  a  weekly
  pattern,
 probably due to the organisation of the laboratories, but since April, the data
 do not appear to be fully synchronized.
This consideration probably explains why the $\rho^2$ value for the DMPsw model (0.778397)
is lower than observed for the DMPseas (0.842169). The latter model better recognizes the weekly fluctuations in cases, even if the model is
less parsimonious.
\begin{table}
\centering
\caption{Tuscany.      Estimates, asymptotic standard errors and $95\%$ mCIs for the parameters of the           DMPseas  {\rm (\ref{eq:ggmwt})+(\ref{eq:wstag})}.}
\label{tab:toscanaSTAG}
\begin{tabular}{crrr@{$\!\;\;$}l}
Parameter    &  Estimate  &  Standard Error  &    \multicolumn{2}{c}{Confidence Interval}    \\
\hline
$m$       &  10312.24  &  77.9095  &     (10156.45, & 10468.03)\\
$p_c$     &  0.000360  &  0.000069  &    (0.000223, & 0.000497)\\
$q_c$     &  0.255746  &  0.011751  &    (0.232249, & 0.279243)\\
$p$       &  0.004071  &  0.000280  &    (0.003510, & 0.004631)\\
$q$       &  0.076667  &  0.002802  &    (0.071064, & 0.082269)\\
$s$       &  6.981769  &  0.001244  &    (6.979281, & 6.984258)\\
$\alpha_1$&  0.063085  &  0.126395  &   (-0.189658, & 0.315828)\\
$\alpha_2$& -0.150143  &  0.063309  &   (-0.276736, & -0.0235492)\\
\hline
 \end{tabular}
\end{table}

\begin{table}[t]
\centering
\caption{Tuscany.      Estimates, asymptotic standard errors and $95\%$ mCIs for the parameters of the           DMPsw  {\rm (\ref{eq:ggmwt})+(\ref{eq:wtamp})}.}
\label{tab:toscanaTMP}
\begin{tabular}{crrr@{$\!\;\;$}l}
Parameter    &  Estimate  &  Standard Error  &    \multicolumn{2}{c}{Confidence Interval}    \\
\hline
$m$  &  10078.90  &  73.32550  &     (9932.373, & 10225.43)\\
$p_c$&  0.000823  &  0.000094  &     (0.000636, & 0.001010)\\
$q_c$&  0.100709  &  0.003683  &     (0.093348, & 0.108069)\\
$p$  &  0.218881  &  0.019900  &     (0.179115, & 0.258647)\\
$q$  & -0.021620  &  0.040863  &    (-0.103278, & 0.060037)\\
$\xi$&  0.780094  &  0.006011  &     (0.768081, & 0.792107)\\
\hline
 \end{tabular}
\end{table}

The estimates obtained for $m$ in
GBM$_{\rm RECT},$ BeGBM$_{\rm RECT}$, DMP, DMPseas and DMPsw
range from 9948 (GBM$_{\rm RECT}$)
to 10416 (BeGBM$_{\rm RECT}$). The total number
of cases until May 3rd was 9563. By comparing this value
to $\hat{m}$=10312 of the DMPseas, we notice that this region, by May 3rd,
experienced 92.7\% of all expected cases.

\begin{figure}[t]
\centering
\begin{tabular}{cc}
\hspace{-10mm} (a) & \hspace{-8mm} (b) \\
\hspace{-8mm} \includegraphics[width=73mm]{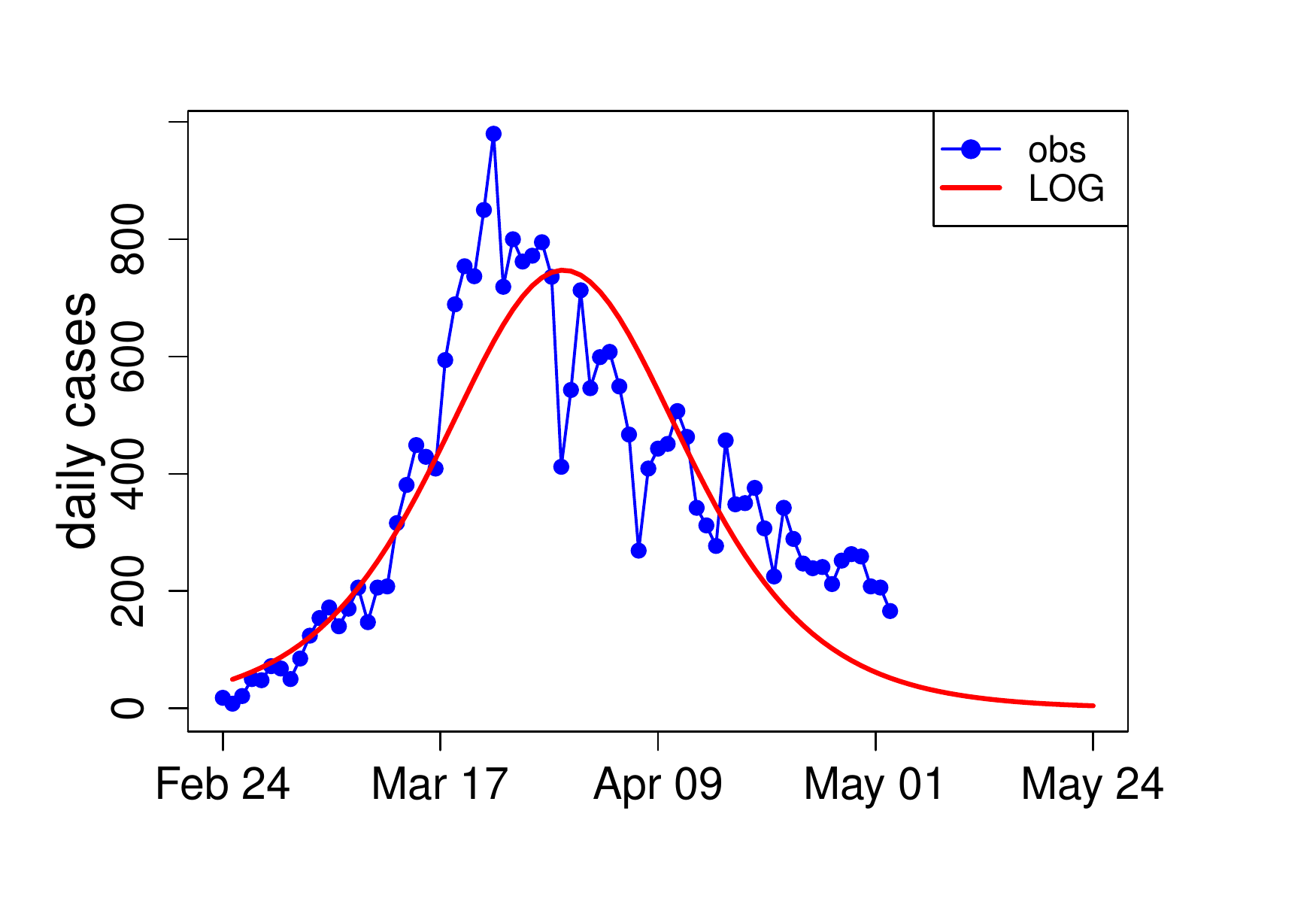} &
\hspace{-10mm} \includegraphics[width=73mm]{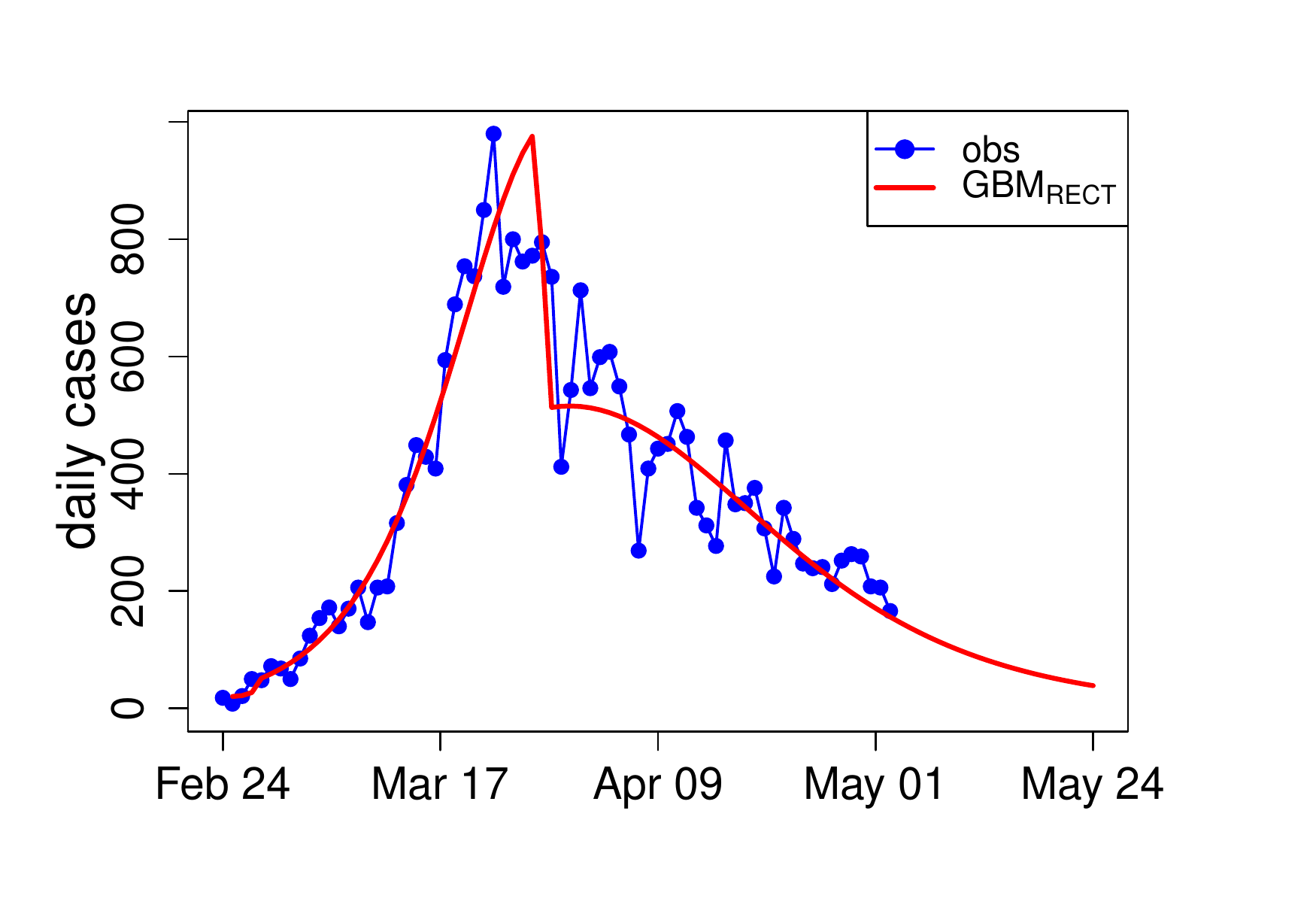}\\
\hspace{-8mm}(c) &  \hspace{-8mm} (d) \\
\hspace{-8mm} \includegraphics[width=73mm]{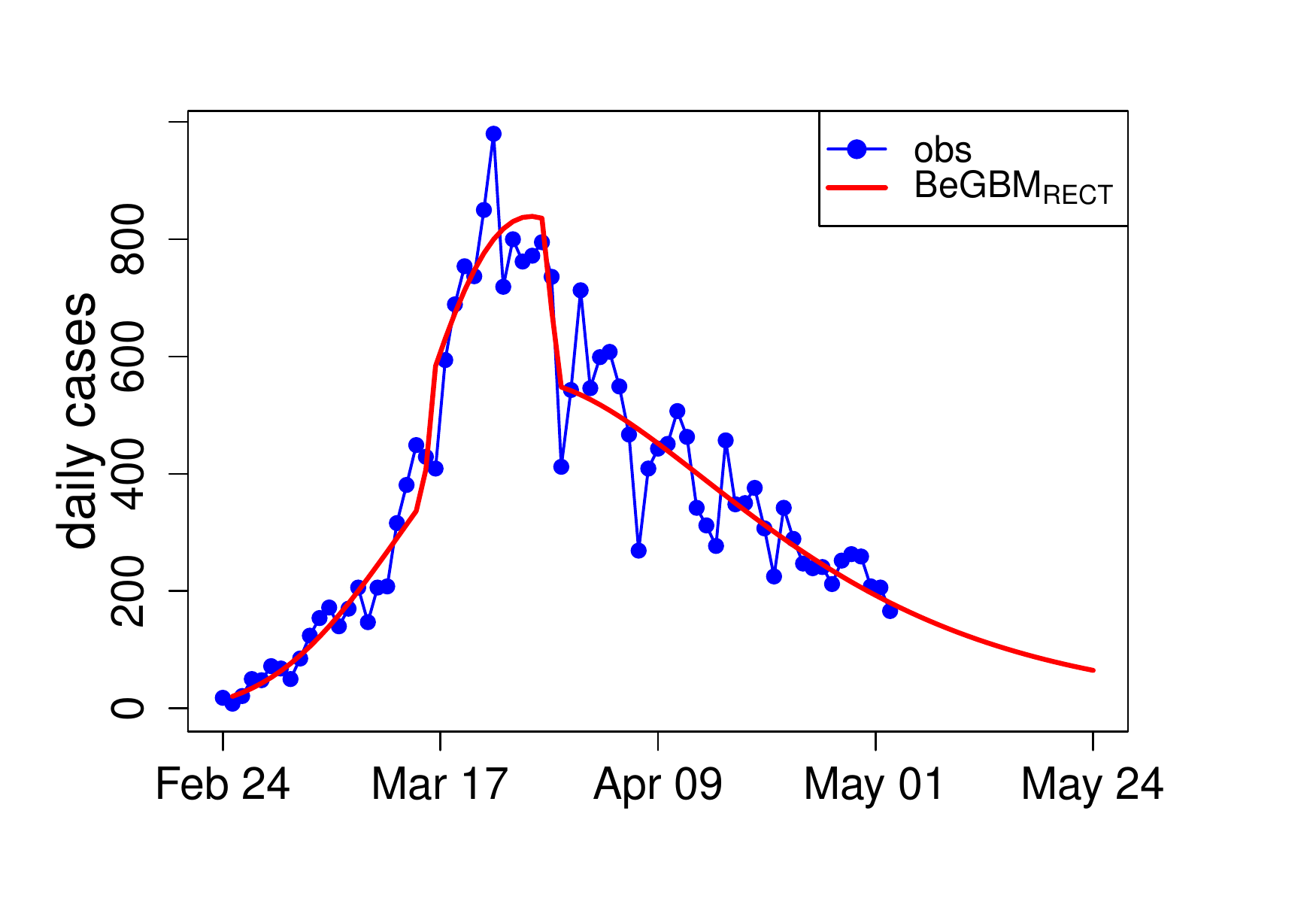} &
\hspace{-8mm}\includegraphics[width=73mm]{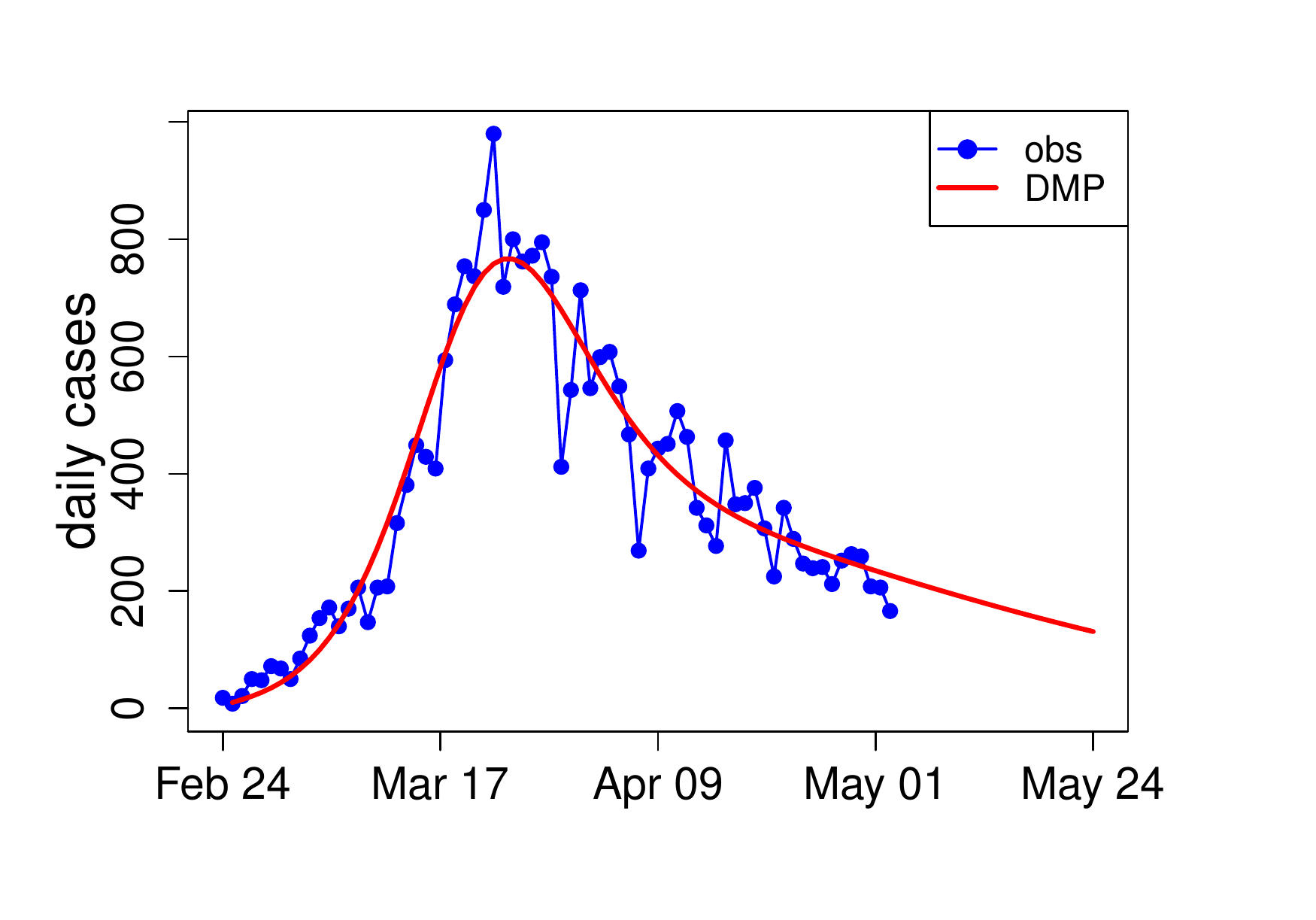}\\
\hspace{-10mm} (e) &  \hspace{-8mm} (f) \\
\hspace{-10mm} \includegraphics[width=73mm]{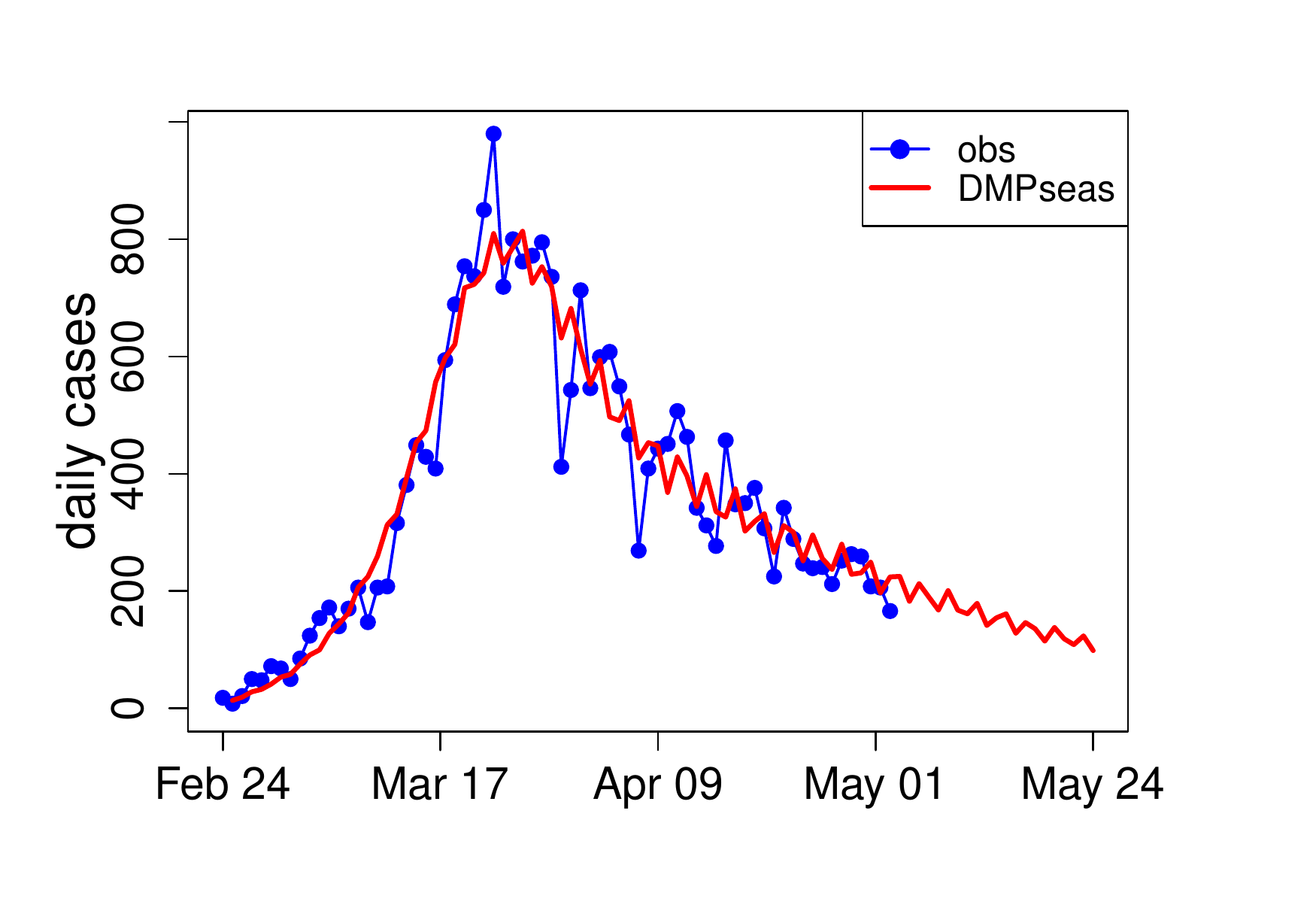} &
 \hspace{-8mm} \includegraphics[width=73mm]{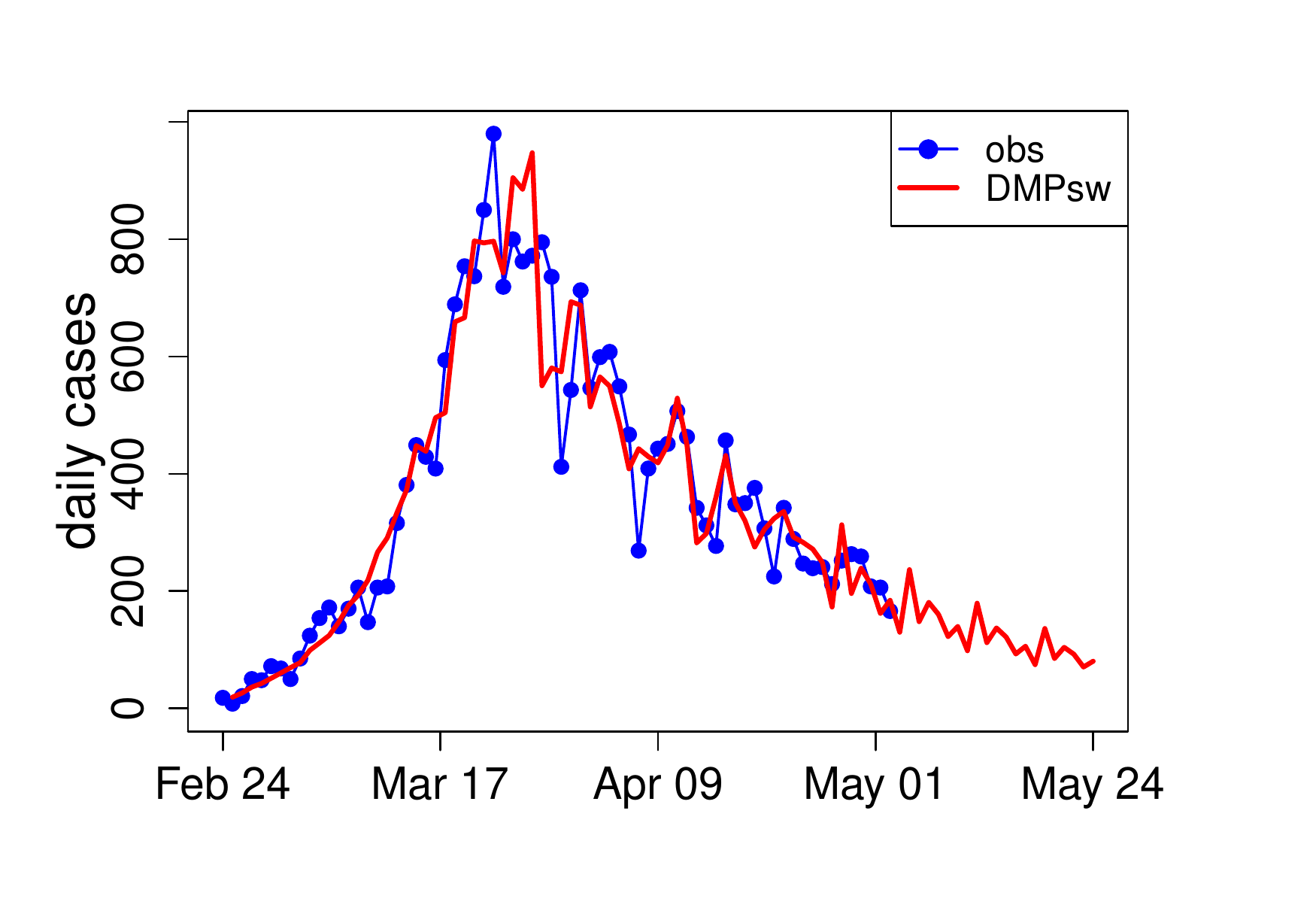}  \\
\end{tabular}
\caption{Emilia--Romagna. Observed and fitted values with the alternative models.
{\rm (a)} Logistic                                                 (LOG);
{\rm (b)} GBM with rectangular shock                       (GBM$_{\rm RECT}$);
{\rm (c)} Bemmaor GBM with rectangular shock               (BeGBM$_{\rm RECT}$);
{\rm (d)} Dynamic market potential                   (DMP);
{\rm (e)} Dynamic market potential+seasonal effect (DMPseas);
{\rm (f)} Dynamic market potential+swabs           (DMPsw).}
\label{fig:fit emilia}
\end{figure}
\clearpage

\subsection{Emilia--Romagna}
The results for Emilia--Romagna are displayed in Table \ref{tab:R2} ($R^2,$ BIC and $\rho^2$),
in Tables  \ref{tab:emiliaTMP},
\ref{tab:emiliaLOG}--\ref{tab:emiliaSIRD} (for parameter estimates for
all the models fitted) and
in
Figures \ref{fig:data_SIRD}(e)
and \ref{fig:fit emilia},
where
observed
and
fitted
daily values are plotted.

Also for this region, the logistic (Figure \ref{fig:fit emilia}(a))
and  SIRD (Figure \ref{fig:data_SIRD}(e)) models are the worst performing
models for describing  the
asymmetrical
evolution of the epidemic.

The results in Tables \ref{tab:emiliaGBM} and \ref{tab:emiliaBEGBM} show that
also here 
a  positive ($\hat{c}>0$)
rectangular shock is significantly diagnosed at the beginning of the time series,
both in the GBM$_{\rm RECT}$ and the BeGBM$_{\rm RECT}$.
The two models provide the same estimate  for the  end of  the shock
on  March 28th  ($t\simeq$ 34), as observed for Lombardy,
which is later than for Veneto, Piedmont
and Tuscany. If we observe Figures  \ref{fig:fit emilia}(b) and (c), we notice, however, that
the fit around the peak is not completely satisfactory, as a small
decrease in the number of confirmed cases actually
occurred  a few days earlier than predicted by both  models.
The $R^2$ for BeGBM$_{\rm RECT}$ is very high (0.999923).

The DMP model (Figure \ref{fig:fit emilia}(d) and Table \ref{tab:emiliaDMP})
allows for a partially satisfactory description without shocks of the asymmetric behaviour of this time series.
The $R^2$ is equal to  0.999776, and its BIC value and $\rho^2$ are worse than for
the BeGBM$_{\rm RECT}$ (713.4355 and 0.9077, respectively).

The DMPseas, with a weekly cycle ($\hat{s}=7.004$ days)
(Table \ref{tab:emiliaSTAG}), is not  able to describe the fluctuations
 (Figure \ref{fig:fit emilia}(e)).
The $R^2$ is equal to 0.999862, and the BIC value is larger than observed for
the BeGBM$_{\rm RECT}$.

The DMPsw (Table \ref{tab:toscanaTMP}) shows a value
for $R^2,$ 0.999925, that is slightly larger than that observed for the BeGBM$_{\rm RECT}$.
From Figure \ref{fig:fit toscana}(f),  we can see that the
fitted values follow the observed data very well, except from the values around the peak.
This behaviour, already noticed for the GBM$_{\rm RECT},$ is responsible
for the low $\rho^2$ value, which is equal to 0.904318, lower than that
observed for the DMPseas (0.921988).\footnote{If we remove the observations from
$t$=27 to $t$=33 (around the peak) from the evaluation of
$\rho^2$ for all the considered models,  we obtain the following values:
0.715193 {for the} LOG, 0.879792 {for the} GBM$_{\rm RECT}$, 0.895025 {for the}
 BeGBM$_{\rm RECT}$, 0.883395
 {for the} DMP,
0.895168
{for the} DMPseas {and}  0.906916 {for the} DMPsw {model}.
}

\begin{table}
\centering
\caption{Emilia--Romagna.      Estimates, asymptotic standard errors and $95\%$ mCIs for the parameters of the           DMPsw  {\rm (\ref{eq:ggmwt})+(\ref{eq:wtamp})}.}
\label{tab:emiliaTMP}
\begin{tabular}{crrr@{$\!\;\;$}l}
Parameter    &  Estimate  &  Standard Error  &    \multicolumn{2}{c}{Confidence Interval}    \\
\hline
$m$  &  30632.96  &  896.4504  &  (28842.10, & 32423.83)\\
$p_c$&  0.000598  &  0.000097  &  (0.000404, & 0.000791)\\
$q_c$&  0.196219  &  0.007415  &  (0.181406, & 0.211033)\\
$p$  &  0.027356  &  0.003847  &  (0.019671, & 0.035041)\\
$q$  & -0.000321  &  0.010265  & (-0.020827, & 0.020185)\\
$\xi$&  0.408953  &  0.058755  &  (0.291576, & 0.526330)\\
\hline
 \end{tabular}
\end{table}

The estimates obtained for $m$ in the LOG model (25140) and  $N$ in the
SIRD model (24982) are even lower than the final
observation, 26016, representing the total number
of cases until May 3rd.
The estimates obtained for $m$ in
GBM$_{\rm RECT},$ BeGBM$_{\rm RECT}$, DMP, DMPseas and DMPsw
range from 28094 (GBM$_{\rm RECT}$)
to 33428 (DMP). If we consider $\hat{m}$=30633 of the DMPsw,
we notice that this region, by May 3rd,
experienced 84.9\% of all expected cases.

\section{Concluding remarks} \label{sec:conclusions}
The aim of this study was to propose a new model to describe
the pattern of COVID-19 cases in the five most affected Italian regions.
The new model and alternative existing nonlinear model structures are fitted
to the available data.

Our results suggest that the commonly
used models, that is, the logistic and SIRD models, are not flexible.
Not only are they incapable of describing fluctuations; they also fail
to follow the asymmetric trend typical of all the regions: the increase in daily
cases has been faster than the decrease observed in the second part of the outbreak.

In all the analyzed regions, both the GBM$_{\rm RECT}$ and the  BeGBM$_{\rm RECT}$
highlight that a positive shock increased the number of daily cases in a period starting about two weeks
after the first cases and ending around March 24th.
From these results, we deduce that the lockdown policy established by the
Italian government on March 8th played a fundamental role in reducing the spread
of the virus and significantly decreasing daily cases two weeks after the start of the
lockdown.

The models that are available in the literature perform quite well in describing the main trend of daily cases.
However, observations also show significant fluctuations.
As highlighted in the introductory section, daily changes have been reported
by the media and have been the focus throughout the most critical weeks of the outbreak.
The available data reveal that the pattern of analyzed swabs is often
concordant with the pattern of confirmed cases. This is not surprising, but the models
available in the literature cannot exploit this information. The model
proposed here, starting from a trend described by a dynamic market potential
diffusion model, makes it possible to perturb the trend through an intervention function
depending on the number of analyzed swabs. The larger the number of swabs
with respect to the average, the larger the number of predicted daily
cases. The proposed model, which is highly parsimonious, is able to describe
the daily fluctuations in cases very well and
proved to be the
best of the models analysed here for four of the five regions (Veneto,
Lombardy, Piedmont and Emilia--Romagna). For the fifth region, Tuscany, the pattern of
daily cases exhibits a weekly pattern, but it does not correspond to the pattern of the
processed swabs. For this region, DMPseas performs better in terms of describing the observed data.

Ahead forecasts have also been evaluated for a period of three weeks.
Forecasting is not the aim of this study since the final observation
corresponds to the last day of complete lockdown. We can, however,
use our forecasts as a benchmark corresponding to the evaluation of the trend
under lockdown for comparison with actual observations pertaining to
the so-called Phase 2, where many restrictions have been removed.
Note that the Italian government decided to start Phase 2 simultaneously
for all the regions, even though there still were differences among them.
By comparing the final cumulative value of number of cases with the estimated final
number of infected patients in each region at the end of the outbreak, we observed that,
while Veneto and Tuscany reached about 92\% of the total number number of the expected cases by May 3rd,
Piedmont, Emilia--Romagna and Lombardy, in particular,  were still facing a more critical
 situation, having experienced, respectively, only
87\%, 85\% and 82\% of all expected cases.

The proposed structure for the intervention function is quite intuitive, and we highlight
that the proposed model could also be used to examine the effect on confirmed
cases of different swabbing strategies by modifying the number of
swabs in the intervention function.
Alternative formulations, with a changepoint to allow for
a different effect of standardised swabs before and after the changepoint,
have also been estimated, but the improvement with respect to the proposed DMPsw model was  negligible.



\clearpage

\section*{Appendix}

\setcounter{table}{0}
\renewcommand{\theequation}{A.\arabic{equation}}
\renewcommand{\thetable}{A.\arabic{table}}
\renewcommand{\thefigure}{A.\arabic{figure}}

This Appendix lists all the parameter estimates for models not included in the main text.

\begin{table}[h]
\centering
\caption{Veneto.      Estimates, asymptotic standard errors and $95\%$ mCIs for the parameters of the           LOG model   {\rm (\ref{eq:log})}  }
\label{tab:venetoLOG}
\begin{tabular}{crrr@{$\!\;\;$}l}
Parameter    &  Estimate  &  Standard Error  &    \multicolumn{2}{c}{Confidence Interval}    \\
\hline
$m$&  18270.81  &  157.9788  &  (17955.73, & 18585.89)\\
$\lambda$&  40.55856  &  0.268768  &  (40.02252, & 41.09460)\\
$\eta$&  8.908960  &  0.198532  &  (8.513001, & 9.304919)\\
\hline
 \end{tabular}
\end{table}

\begin{table}[h]
\centering
\caption{Veneto.      Estimates, asymptotic standard errors and $95\%$ mCIs for the parameters of the           GBM$_{\rm RECT}$ {\rm (\ref{eq:gbm})+(\ref{eq:rect}).}    }
\label{tab:venetoGBM}
\begin{tabular}{crrr@{$\!\;\;$}l}
Parameter    &  Estimate  &  Standard Error  &    \multicolumn{2}{c}{Confidence Interval}    \\
\hline
$m$&  19432.00  &  84.0108  &  (19264.31, & 19599.69)\\
$p$&  0.000786  &  0.00010  &  (0.000586, & 0.000985)\\
$q$&  0.086643  &  0.00111  &  (0.084428, & 0.088858)\\
$c$&  0.678511  &  0.03783  &  (0.603001, & 0.754020)\\
$a$&  14.37807  &  1.89135  &  (10.60292, & 18.15322)\\
$b$&  32.88902  &  0.32947  &  (32.23139, & 33.54665)\\
\hline
 \end{tabular}
\end{table}

\begin{table}[h]
\centering
\caption{Veneto.      Estimates, asymptotic standard errors and $95\%$ mCIs for the parameters of the           BeGBM$_{\rm RECT}$ {\rm (\ref{eq:bemrect})+(\ref{eq:rect}).}  }
\label{tab:venetoBEGBM}
\begin{tabular}{crrr@{$\!\;\;$}l}
Parameter    &  Estimate  &  Standard Error  &    \multicolumn{2}{c}{Confidence Interval}    \\
\hline
$m$&  20084.50  &  169.5056  &  (19746.07, & 20422.93)\\
$p$&  0.007073  &  0.003246  &  (0.000592, & 0.013555)\\
$q$&  0.063168  &  0.006442  &  (0.050305, & 0.076030)\\
$c$&  0.321913  &  0.066829  &  (0.188485, & 0.455342)\\
$a$&  18.00000  &  2.144826  &  (13.71772, & 22.28228)\\
$b$&  33.65430  &  0.503203  &  (32.64960, & 34.65896)\\
$A$&  2.315594  &  0.617207  &  (1.083300, & 3.547888)\\
\hline
 \end{tabular}
\end{table}

\begin{table}[h]
\centering
\caption{Veneto.      Estimates, asymptotic standard errors and $95\%$ mCIs for the parameters of the           DMP  {\rm (\ref{eq:ggm})}.   }
\label{tab:venetoDMP}
\begin{tabular}{crrr@{$\!\;\;$}l}
Parameter    &  Estimate  &  Standard Error  &    \multicolumn{2}{c}{Confidence Interval}    \\
\hline
$m$  &  20031.83  &  153.4688  &  (19725.59, & 20338.07)\\
$p_c$&  0.000506  &  0.000079  &  (0.000349, & 0.000663)\\
$q_c$&  0.219903  &  0.009433  &  (0.201080, & 0.238726)\\
$p$  &  0.003534  &  0.000236  &  (0.003062, & 0.004005)\\
$q$  &  0.071518  &  0.002504  &  (0.066522, & 0.076514)\\
\hline
 \end{tabular}
\end{table}

\begin{table}[h]
\centering
\caption{Veneto.      Estimates, asymptotic standard errors and $95\%$ mCIs for the parameters of the           DMPseas  {\rm (\ref{eq:ggmwt})+(\ref{eq:wstag})}.}
\label{tab:venetoSTAG}
\begin{tabular}{crrr@{$\!\;\;$}l}
Parameter    &  Estimate  &  Standard Error  &    \multicolumn{2}{c}{Confidence Interval}    \\
\hline
$m$       &  20035.56  &  155.7882  &  (19724.43, & 20346.69)\\
$p_c$     &  0.000503  &  0.000079  &  (0.000345, & 0.000661)\\
$q_c$     &  0.220037  &  0.009527  &  (0.201010, & 0.239064)\\
$p$       &  0.003513  &  0.000239  &  (0.003035, & 0.003991)\\
$q$       &  0.071512  &  0.002534  &  (0.066451, & 0.076574)\\
$s$       &  7.003457  &  0.002538  &  (6.998389, & 7.008525)\\
$\alpha_1$&  0.010361  &  0.116495  & (-0.222295, & 0.243017)\\
$\alpha_2$&  0.067384  &  0.058376  & (-0.049201, & 0.183969)\\
\hline
 \end{tabular}
\end{table}


\begin{table}											
\centering											
\caption{Veneto.			Estimates,  asymptotic  standard errors, and  $95\%$ profile likelihood mCIs for the parameters of the 					SIRD model		 {\rm (\ref{eq:SIRD}).}	}
\label{tab:venetoSIRD}											
\begin{tabular}{crrr@{$\!\;\;$}l}											
Parameter		&	Estimate	&	Standard Error	&		\multicolumn{2}{c}{Confidence Interval}			\\
\hline											
$	logit(\beta)	$ &	-1.627417	&	0.006810	&	(-1.665249,& -1.592545)\\
$	logit(\gamma)	$ &	-3.972993	&	0.013783	&	(-4.033721,& -3.913078)\\
$	logit(\delta)	$ &	-5.547785	&	0.009624	&	(-5.566849,& -5.528453)\\
$	\ln(N) $ &	 9.774865	&	0.007822	&	(9.748010,&  9.802344)\\
$	\ln(I_0)$ &	 4.536530	&	0.030909	&	(4.365557,&  4.719183)\\
\hline											
\end{tabular}																					
\end{table}


\begin{table}
\centering
\caption{Lombardy.      Estimates, asymptotic standard errors and $95\%$ mCIs for the parameters of the           LOG model   {\rm (\ref{eq:log})}  }
\label{tab:lombardiaLOG}
\begin{tabular}{crrr@{$\!\;\;$}l}
Parameter    &  Estimate  &  Standard Error  &    \multicolumn{2}{c}{Confidence Interval}    \\
\hline
$m$&  75187.57  &  910.6877  &  (73371.26, & 77003.88)\\
$\lambda$&  37.93709  &  0.400709  &  (37.13791, & 38.73628)\\
$\eta$&  9.313397  &  0.301700  &  (8.711676, & 9.915118)\\
\hline
 \end{tabular}
\end{table}

\begin{table}
\centering
\caption{Lombardy.      Estimates, asymptotic standard errors and $95\%$ mCIs for the parameters of the           GBM$_{\rm RECT}$ {\rm (\ref{eq:gbm})+(\ref{eq:rect}).}    }
\label{tab:lombardiaGBM}
\begin{tabular}{crrr@{$\!\;\;$}l}
Parameter    &  Estimate  &  Standard Error  &    \multicolumn{2}{c}{Confidence Interval}    \\
\hline
$m$&  87337.71  &  897.5725  &  (85546.15, & 89129.27)\\
$p$&  0.001095  &  0.000195  &  (0.000706, & 0.001485)\\
$q$&  0.061445  &  0.001594  &  (0.058263, & 0.064627)\\
$c$&  1.080260  &  0.059873  &  (0.960753, & 1.199768)\\
$a$&  11.30640  &  1.993151  &  (7.328054, & 15.28474)\\
$b$&  33.89623  &  0.292434  &  (33.31253, & 34.47994)\\
\hline
 \end{tabular}
\end{table}

\begin{table}
\centering
\caption{Lombardy.      Estimates, asymptotic standard errors and $95\%$ mCIs for the parameters of the           BeGBM$_{\rm RECT}$ {\rm (\ref{eq:bemrect})+(\ref{eq:rect}).}  }
\label{tab:lombardiaBEGBM}
\begin{tabular}{crrr@{$\!\;\;$}l}
Parameter    &  Estimate  &  Standard Error  &    \multicolumn{2}{c}{Confidence Interval}    \\
\hline
$m$&  97365.72                 &  1037.772                & (95293.75,                & 99437.70)\\
$p$&  0.038627                 &  0.000925                & (0.036781,                & 0.040473)\\
$q$&  2.5$\!\times\! 10^{-6}$  &  1.2$\!\times\! 10^{-7}$ & (2.2$\!\times\! 10^{-6}$, & 2.7$\times\! 10^{-6}$)\\
$c$&  0.554140                 &  0.027383                & (0.499468,                & 0.608813)\\
$a$&  13.87466                 &  1.035562                & (11.80710,                & 15.94221)\\
$b$&  36.77908                 &  0.245369                & (36.28918,                & 37.26897)\\
$A$&  79263.48                 &  0.001628                & (79263.48,                & 79263.48)\\
\hline
 \end{tabular}
\end{table}

\begin{table}
\centering
\caption{Lombardy.      Estimates, asymptotic standard errors and $95\%$ mCIs for the parameters of the           DMP  {\rm (\ref{eq:ggm})}.   }
\label{tab:lombardiaDMP}
\begin{tabular}{crrr@{$\!\;\;$}l}
Parameter    &  Estimate  &  Standard Error  &    \multicolumn{2}{c}{Confidence Interval}    \\
\hline
$m$  &  95623.52  &  2549.990  &     (90535.09, & 100711.9)\\
$p_c$&  0.002288  &  0.000110  &     (0.002069, & 0.002508)\\
$q_c$&  0.050407  &  0.003498  &     (0.043427, & 0.057388)\\
$p$  &  0.002026  &  0.000116  &     (0.001795, & 0.002258)\\
$q$  &  0.172654  &  0.004251  &     (0.164171, & 0.181138)\\
\hline
 \end{tabular}
\end{table}

\begin{table}
\centering
\caption{Lombardy.      Estimates, asymptotic standard errors and $95\%$ mCIs for the parameters of the           DMPseas  {\rm (\ref{eq:ggmwt})+(\ref{eq:wstag})}.}
\label{tab:lombardiaSTAG}
\begin{tabular}{crrr@{$\!\;\;$}l}
Parameter    &  Estimate  &  Standard Error  &    \multicolumn{2}{c}{Confidence Interval}    \\
\hline
$m$       &  98722.82  &  2341.256  &     (94047.01, & 103398.6)\\
$p_c$     &  0.000322  &  0.000029  &     (0.000264, & 0.000380)\\
$q_c$     &  0.231578  &  0.004859  &     (0.221874, & 0.241282)\\
$p$       &  0.008156  &  0.000262  &     (0.007632, & 0.008681)\\
$q$       &  0.032693  &  0.002793  &     (0.027116, & 0.038270)\\
$s$       &  7.004860  &  0.001356  &     (7.002152, & 7.007570)\\
$\alpha_1$&  0.017993  &  0.121406  &    (-0.224471, & 0.260458)\\
$\alpha_2$&  0.125291  &  0.060556  &     (0.004353, & 0.246230)\\
\hline
 \end{tabular}
\end{table}


\begin{table}											
\centering											
\caption{Lombardy.			Estimates,  asymptotic  standard errors, and  $95\%$ profile likelihood mCIs for the parameters of the 					SIRD model		 {\rm (\ref{eq:SIRD}).}	}
\label{tab:lombardiaSIRD}											
\begin{tabular}{crrr@{$\!\;\;$}l}											
Parameter		&	Estimate	&	Standard Error	&		\multicolumn{2}{c}{Confidence Interval}			\\
\hline	
$  logit(\beta)  $   &  -0.847112  &  0.004575 & (-0.902256,  & -0.783506) \\
$  logit(\gamma)  $  &  -3.566582  &  0.019616 & (-3.641281,  & -3.495745) \\
$  logit(\delta)  $  &  -4.107745  &  0.013836 & (-4.182120,  & -4.037120) \\
$  \ln(N) $          &  10.705627  &  0.015777 & (10.68429,  &  10.72949)  \\
$  \ln(I_0)$         &   4.221095  &  0.019883 & (3.941377,  &  4.461812)  \\
\hline											
\end{tabular}																					
\end{table}


\begin{table}
\centering
\caption{Piedmont.      Estimates, asymptotic standard errors and $95\%$ mCIs for the parameters of the           LOG model   {\rm (\ref{eq:log})}  }
\label{tab:piemonteLOG}
\begin{tabular}{crrr@{$\!\;\;$}l}
Parameter    &  Estimate  &  Standard Error  &    \multicolumn{2}{c}{Confidence Interval}    \\
\hline
$m$&  28967.83  &  459.1435  &  (28051.38, & 29884.29)\\
$\lambda$&  45.67634  &  0.450381  &  (44.77738, & 46.57531)\\
$\eta$&  9.883935  &  0.267825  &  (9.349354, & 10.41852)\\
\hline
 \end{tabular}
\end{table}

\begin{table}
\centering
\caption{Piedmont.      Estimates, asymptotic standard errors and $95\%$ mCIs for the parameters of the           GBM$_{\rm RECT}$ {\rm (\ref{eq:gbm})+(\ref{eq:rect}).}    }
\label{tab:piemonteGBM}
\begin{tabular}{crrr@{$\!\;\;$}l}
Parameter    &  Estimate  &  Standard Error  &    \multicolumn{2}{c}{Confidence Interval}    \\
\hline
$m$&  31941.45  &  217.2889  &  (31507.37, & 32375.54)\\
$p$&  0.000272  &  0.000110  &  (0.000052, & 0.000493)\\
$q$&  0.081491  &  0.000917  &  (0.079659, & 0.083323)\\
$c$&  1.223254  &  0.081881  &  (1.059678, & 1.386829)\\
$a$&  12.00000  &  3.486412  &  (5.035091, & 18.96491)\\
$b$&  29.76552  &  0.275815  &  (29.21451, & 30.31652)\\
\hline
 \end{tabular}
\end{table}

\begin{table}
\centering
\caption{Piedmont.      Estimates, asymptotic standard errors and $95\%$ mCIs for the parameters of the           BeGBM$_{\rm RECT}$ {\rm (\ref{eq:bemrect})+(\ref{eq:rect}).}  }
\label{tab:piemonteBEGBM}
\begin{tabular}{crrr@{$\!\;\;$}l}
Parameter    &  Estimate  &  Standard Error  &    \multicolumn{2}{c}{Confidence Interval}    \\
\hline
$m$&  34351.47  &  773.2607  &  (32806.23, & 35896.71)\\
$p$&  0.002377  &  0.001419  & (-0.000460, & 0.005213)\\
$q$&  0.060533  &  0.005707  &  (0.049129, & 0.071937)\\
$c$&  0.830728  &  0.112215  &  (0.606484, & 1.054971)\\
$a$&  11.24809  &  5.389364  &  (0.478300, & 22.01787)\\
$b$&  29.53243  &  0.317684  &  (28.89759, & 30.16727)\\
$A$&  1.910752  &  0.404683  &  (1.102058, & 2.719447)\\
\hline
 \end{tabular}
\end{table}

\begin{table}
\centering
\caption{Piedmont.      Estimates, asymptotic standard errors and $95\%$ mCIs for the parameters of the           DMP  {\rm (\ref{eq:ggm})}.   }
\label{tab:piemonteDMP}
\begin{tabular}{crrr@{$\!\;\;$}l}
Parameter    &  Estimate  &  Standard Error  &    \multicolumn{2}{c}{Confidence Interval}    \\
\hline
$m$  &  32979.56  &  244.2301  &  (32491.80, & 33467.32)\\
$p_c$&  0.000042  &  0.000014  &  (0.000013, & 0.000070)\\
$q_c$&  0.356498  &  0.016712  &  (0.323121, & 0.389875)\\
$p$  &  0.001946  &  0.000043  &  (0.001861, & 0.002031)\\
$q$  &  0.073422  &  0.001021  &  (0.071384, & 0.075461)\\
\hline
 \end{tabular}
\end{table}

\begin{table}
\centering
\caption{Piedmont.      Estimates, asymptotic standard errors and $95\%$ mCIs for the parameters of the           DMPseas  {\rm (\ref{eq:ggmwt})+(\ref{eq:wstag})}.}
\label{tab:piemonteSTAG}
\begin{tabular}{crrr@{$\!\;\;$}l}
Parameter    &  Estimate  &  Standard Error  &    \multicolumn{2}{c}{Confidence Interval}    \\
\hline
$m$       &  33005.60  &  236.0270  &  (32533.80, & 33477.40)\\
$p_c$     &  0.000042  &  0.000014  &  (0.000014, & 0.000069)\\
$q_c$     &  0.356128  &  0.015956  &  (0.324233, & 0.388024)\\
$p$       &  0.001925  &  0.000042  &  (0.001842, & 0.002008)\\
$q$       &  0.073395  &  0.000981  &  (0.071433, & 0.075356)\\
$s$       &  7.009690  &  0.000894  &  (7.007900, & 7.011470)\\
$\alpha_1$&  0.035228  &  0.074368  & (-0.113432, & 0.183887)\\
$\alpha_2$&  0.108353  &  0.037278  &  (0.033835, & 0.182870)\\
\hline
 \end{tabular}
\end{table}

%

\begin{table}											
\centering											
\caption{Piedmont.			Estimates,  asymptotic  standard errors, and  $95\%$ profile likelihood mCIs for the parameters of the 					SIRD model		 {\rm (\ref{eq:SIRD}).}	}
\label{tab:piemonteSIRD}											
\begin{tabular}{crrr@{$\!\;\;$}l}											
Parameter		&	Estimate	&	Standard Error	&		\multicolumn{2}{c}{Confidence Interval}			\\
\hline	
$  logit(\beta)  $   & -1.945829   & 0.007838  & (-1.997635, -1.893736  & ) \\
$  logit(\gamma)  $  & -4.228805   & 0.019824  & (-4.279535, -4.181564 & ) \\
$  logit(\delta)  $  & -4.958486   & 0.015567  & (-4.991483, -4.924573 & ) \\
$  \ln(N) $          & 10.273504   & 0.010933  & (10.23877,   10.30958 & )  \\
$  \ln(I_0)$         &  5.647353   & 0.032546  & (5.444469,   5.839862 & )  \\
\hline											
\end{tabular}																					
\end{table}	


\begin{table}
\centering
\caption{Tuscany.      Estimates, asymptotic standard errors and $95\%$ mCIs for the parameters of the           LOG model   {\rm (\ref{eq:log})}.  }
\label{tab:toscanaLOG}
\begin{tabular}{crrr@{$\!\;\;$}l}
Parameter    &  Estimate  &  Standard Error  &    \multicolumn{2}{c}{Confidence Interval}    \\
\hline
$m$&  9438.855  &  87.63859  &  (9263.879, & 9613.831)\\
$\lambda$&  37.18838  &  0.275616  &  (36.63810, & 37.73867)\\
$\eta$&  8.152733  &  0.209458  &  (7.734544, & 8.570929)\\
\hline
 \end{tabular}
\end{table}

\begin{table}
\centering
\caption{Tuscany.      Estimates, asymptotic standard errors and $95\%$ mCIs for the parameters of the           GBM$_{\rm RECT}$  {\rm (\ref{eq:gbm})+(\ref{eq:rect})}.    }
\label{tab:toscanaGBM}
\begin{tabular}{crrr@{$\!\;\;$}l}
Parameter    &  Estimate  &  Standard Error  &    \multicolumn{2}{c}{Confidence Interval}    \\
\hline
$m$&  9947.536  &  60.19960  &  (9827.236, & 10067.84)\\
$p$&  0.000540  &  0.000325  & (-0.000109, & 0.001190)\\
$q$&  0.096809  &  0.001909  &  (0.092995, & 0.100624)\\
$c$&  0.824558  &  0.087759  &  (0.649185, & 0.999930)\\
$a$&  10.00000  &  6.341729  & (-2.672935, & 22.67294)\\
$b$&  28.88851  &  0.475589  &  (27.93812, & 29.83890)\\
\hline
 \end{tabular}
\end{table}

\begin{table}
\centering
\caption{Tuscany.      Estimates, asymptotic standard errors and $95\%$ mCIs for the parameters of the           BeGBM$_{\rm RECT}$   {\rm (\ref{eq:bemrect})+(\ref{eq:rect})}.   }
\label{tab:toscanaBEGBM}
\begin{tabular}{crrr@{$\!\;\;$}l}
Parameter    &  Estimate  &  Standard Error  &    \multicolumn{2}{c}{Confidence Interval}    \\
\hline
$m$&  10415.94  &  103.1553  &  (10209.74, & 10622.14)\\
$p$&  0.032542  &  0.024590  & (-0.016614, & 0.081697)\\
$q$&  0.037256  &  0.028109  & (-0.018933, & 0.093446)\\
$c$&  0.183372  &  0.079073  &  (0.025308, & 0.341437)\\
$a$&  17.00000  &  3.601708  &  (9.800288, & 24.19971)\\
$b$&  28.62989  &  1.187345  &  (26.25642, & 31.00336)\\
$A$&  9.756513  &  12.00014  & (-14.23142, & 33.74445)\\
\hline
 \end{tabular}
\end{table}

\begin{table}
\centering
\caption{Tuscany.      Estimates, asymptotic standard errors and $95\%$ mCIs for the parameters of the           DMP  {\rm (\ref{eq:ggm})}.   }
\label{tab:toscanaDMP}
\begin{tabular}{crrr@{$\!\;\;$}l}
Parameter    &  Estimate  &  Standard Error  &    \multicolumn{2}{c}{Confidence Interval}    \\
\hline
$m$  &  10314.79  &  79.90432  &  (10155.17, & 10474.42)\\
$p_c$&  0.000357  &  0.000069  &  (0.000218, & 0.000495)\\
$q_c$&  0.256022  &  0.011975  &  (0.232100, & 0.279944)\\
$p$  &  0.004027  &  0.000281  &  (0.003466, & 0.004588)\\
$q$  &  0.076682  &  0.002860  &  (0.070968, & 0.082397)\\
\hline
 \end{tabular}
\end{table}

\begin{table}											
\centering											
\caption{Tuscany.			Estimates,  asymptotic  standard errors, and  $95\%$ profile likelihood mCIs for the parameters of the 					SIRD model		 {\rm (\ref{eq:SIRD}).}	}
\label{tab:toscanaSIRD}											
\begin{tabular}{crrr@{$\!\;\;$}l}											
Parameter		&	Estimate	&	Standard Error	&		\multicolumn{2}{c}{Confidence Interval}			\\
\hline	
$  logit(\beta)  $   &  -1.665359  &  0.007469  & (-1.706638,  & -1.625033) \\
$  logit(\gamma)  $  &  -4.537731  &  0.020362  & (-4.615783,  & -4.464222) \\
$  logit(\delta)  $  &  -5.494510  &  0.012242  & (-5.518653,  & -5.470009) \\
$  \ln(N) $          &   9.121444  &  0.007859  & (9.098362,  &  9.144408)  \\
$  \ln(I_0)$         &   4.003009  &  0.032816  & (3.821543,  &  4.185668)  \\
\hline											
\end{tabular}																					
\end{table}


\begin{table}
\centering
\caption{Emilia--Romagna.      Estimates, asymptotic standard errors and $95\%$ mCIs for the parameters of the           LOG model   {\rm (\ref{eq:log})}  }
\label{tab:emiliaLOG}
\begin{tabular}{crrr@{$\!\;\;$}l}
Parameter    &  Estimate  &  Standard Error  &    \multicolumn{2}{c}{Confidence Interval}    \\
\hline
$m$&  25140.07  &  239.2999  &  (24662.43, & 25617.72)\\
$\lambda$&  35.70744  &  0.299284  &  (35.11007, & 36.30482)\\
$\eta$&  8.407013  &  0.231781  &  (7.944376, & 8.869650)\\
\hline
 \end{tabular}
\end{table}

\begin{table}
\centering
\caption{Emilia--Romagna.      Estimates, asymptotic standard errors and $95\%$ mCIs for the parameters of the           GBM$_{\rm RECT}$ {\rm (\ref{eq:gbm})+(\ref{eq:rect}).}    }
\label{tab:emiliaGBM}
\begin{tabular}{crrr@{$\!\;\;$}l}
Parameter    &  Estimate  &  Standard Error  &    \multicolumn{2}{c}{Confidence Interval}    \\
\hline
$m$&  28094.29  &  167.1121 &   (27760.44, & 28428.13)\\
$p$&  0.000649  &  0.000170  &  (0.000310, & 0.000988)\\
$q$&  0.072118  &  0.001362  &  (0.069398, & 0.074839)\\
$c$&  0.953871  &  0.043847  &  (0.866278, & 1.041465)\\
$a$&  3.846772  &  3.004815  & (-2.156035, & 9.849580)\\
$b$&  33.56896  &  0.208713  &  (33.15201, & 33.98591)\\
\hline
 \end{tabular}
\end{table}

\begin{table}
\centering
\caption{Emilia--Romagna.      Estimates, asymptotic standard errors and $95\%$ mCIs for the parameters of the           BeGBM$_{\rm RECT}$ {\rm (\ref{eq:bemrect})+(\ref{eq:rect}).}  }
\label{tab:emiliaBEGBM}
\begin{tabular}{crrr@{$\!\;\;$}l}
Parameter    &  Estimate  &  Standard Error  &    \multicolumn{2}{c}{Confidence Interval}    \\
\hline
$m$&  29493.66  &  202.3006  &  (29089.39, & 29897.92)\\
$p$&  0.030181  &  0.006000  &  (0.018191, & 0.042172)\\
$q$&  0.022991  &  0.007126  &  (0.008750, & 0.037233)\\
$c$&  0.498745  &  0.024778  &  (0.449231, & 0.548259)\\
$a$&  21.73994  &  0.428230  &  (20.88419, & 22.59569)\\
$b$&  34.45871  &  0.245910  &  (33.96729, & 34.95012)\\
$A$&  8.480216  &  3.603653  &  (1.278889, & 15.68154)\\
\hline
 \end{tabular}
\end{table}

\begin{table}
\centering
\caption{Emilia--Romagna.      Estimates, asymptotic standard errors and $95\%$ mCIs for the parameters of the           DMP  {\rm (\ref{eq:ggm})}.   }
\label{tab:emiliaDMP}
\begin{tabular}{crrr@{$\!\;\;$}l}
Parameter    &  Estimate  &  Standard Error  &    \multicolumn{2}{c}{Confidence Interval}    \\
\hline
$m$  &  33428.13  &  2289.969  &  (28854.75, & 38001.51)\\
$p_c$&  0.003860  &  0.000198  &  (0.003465, & 0.004256)\\
$q_c$&  0.037270  &  0.007561  &  (0.022170, & 0.052370)\\
$p$  &  0.002106  &  0.000113  &  (0.001881, & 0.002331)\\
$q$  &  0.161587  &  0.003967  &  (0.153665, & 0.169510)\\
\hline
 \end{tabular}
\end{table}

 \clearpage
\begin{table}
\centering
\caption{Emilia--Romagna.      Estimates, asymptotic standard errors and $95\%$ mCIs for the parameters of the           DMPseas  {\rm (\ref{eq:ggmwt})+(\ref{eq:wstag})}.}
\label{tab:emiliaSTAG}
\begin{tabular}{crrr@{$\!\;\;$}l}
Parameter    &  Estimate  &  Standard Error  &    \multicolumn{2}{c}{Confidence Interval}    \\
\hline
$m$       &  32126.98  &  880.8021  &  (30366.29, & 33887.68)\\
$p_c$     &  0.000338  &  0.000029  &  (0.000281, & 0.000395)\\
$q_c$     &  0.221433  &  0.003769  &  (0.213899, & 0.228967)\\
$p$       &  0.010660  &  0.000521  &  (0.009619, & 0.011700)\\
$q$       &  0.030346  &  0.004154  &  (0.022043, & 0.038650)\\
$s$       &  7.004196  &  0.001743  &  (7.000712, & 7.007680)\\
$\alpha_1$&  0.111676  &  0.132367  & (-0.152921, & 0.376274)\\
$\alpha_2$& -0.010882  &  0.066052  & (-0.142918, & 0.121154)\\
\hline
 \end{tabular}
\end{table}


\begin{table}											
\centering											
\caption{Emilia--Romagna.			Estimates,  asymptotic  standard errors, and  $95\%$ profile likelihood mCIs for the parameters of the 					SIRD model		 {\rm (\ref{eq:SIRD}).}	}
\label{tab:emiliaSIRD}											
\begin{tabular}{crrr@{$\!\;\;$}l}											
Parameter		&	Estimate	&	Standard Error	&		\multicolumn{2}{c}{Confidence Interval}			\\
\hline	
$  logit(\beta)  $   &  -1.622630  &  0.008520  &  (-1.66412,& -1.582144)\\
$  logit(\gamma)  $  &  -3.994620  &  0.016080  &  (-4.05168,& -3.939858)\\
$  logit(\delta)  $  &  -4.846019  &  0.020229  &  (-4.88889,& -4.804116)\\
$  \ln(N) $          &   10.12591  &  0.009268  &  (10.09949,& 10.15281)\\
$  \ln(I_0)$         &   5.388645  &  0.035762  &  (5.21357,&  5.562004)\\
\hline											
\end{tabular}																					
\end{table}


\end{document}